\documentclass[12pt
]{article}

\usepackage{graphicx}
\usepackage{amsfonts}
\usepackage{amsmath}
\usepackage[usenames]{color}
\usepackage{amssymb}
\usepackage[mathscr]{eucal} 
\usepackage{longtable}

\usepackage{color}
\input{colordvi.tex}

\def\kesu#1{}

\def\Z{\mathbb{Z}}

\def\R{\mathbb{R}}
\def\C{\mathbb{C}}
\def\P{\mathbb{P}}

\begin{document}
\baselineskip 0.6cm
\newcommand{\gsim}{ \mathop{}_{\textstyle \sim}^{\textstyle >} }
\newcommand{\lsim}{ \mathop{}_{\textstyle \sim}^{\textstyle <} }
\newcommand{\vev}[1]{ \left\langle {#1} \right\rangle }
\newcommand{\bra}[1]{ \langle {#1} | }
\newcommand{\ket}[1]{ | {#1} \rangle }
\newcommand{\Dsl}{\mbox{\ooalign{\hfil/\hfil\crcr$D$}}}
\newcommand{\nequiv}{\mbox{\ooalign{\hfil/\hfil\crcr$\equiv$}}}
\newcommand{\nsupset}{\mbox{\ooalign{\hfil/\hfil\crcr$\supset$}}}
\newcommand{\nni}{\mbox{\ooalign{\hfil/\hfil\crcr$\ni$}}}
\newcommand{\EV}{ {\rm eV} }
\newcommand{\KEV}{ {\rm keV} }
\newcommand{\MEV}{ {\rm MeV} }
\newcommand{\GEV}{ {\rm GeV} }
\newcommand{\TEV}{ {\rm TeV} }

\def\diag{\mathop{\rm diag}\nolimits}
\def\tr{\mathop{\rm tr}}

\def\Spin{\mathop{\rm Spin}}
\def\SO{\mathop{\rm SO}}
\def\O{\mathop{\rm O}}
\def\SU{\mathop{\rm SU}}
\def\U{\mathop{\rm U}}
\def\Sp{\mathop{\rm Sp}}
\def\SL{\mathop{\rm SL}}

\def\change#1#2{{\color{blue}#1}{\color{red} [#2]}\color{black}\hbox{}}


\begin{titlepage}
  
\begin{flushright}
  UT-10-05 \\
  IPMU10-0067 \\
NSF-KITP-10-041
\end{flushright}
  
\vskip 1cm
\begin{center}
  
{\large \bf More on Dimension-4 Proton Decay Problem in F-theory}

{\bf --- Spectral Surface, Discriminant Locus and Monodromy ---}
  
\vskip 1.2cm
  
Hirotaka Hayashi$^1$, Teruhiko Kawano$^1$, Yoichi Tsuchiya$^1$ and 
Taizan Watari$^{2}$

\vskip 0.4cm
    

{\it
  $^1$Department of Physics, University of Tokyo, Tokyo 113-0033, Japan  
  \\[2mm]
  
 $^2$Institute for the Physics and Mathematics of the Universe, University of Tokyo, Kashiwano-ha 5-1-5, 277-8583, Japan  
  }
\vskip 1.5cm
  
\abstract{Factorized spectral surface scenario has been considered as one 
of solutions to the dimension-4 proton decay problem in supersymmetric 
compactifications of F-theory. 
It has been 
formulated in language of gauge theory on 7+1 dimensions, but the 
gauge theories descriptions can capture physics of geometry of 
F-theory compactification only approximately at best. 
Given the severe constraint on the renormalizable couplings 
that lead to proton decay, it is worth studying without an approximation 
whether or not the proton decay operators are removed completely 
in this scenario. 
We clarify how the behavior of 
spectral surface and discriminant locus are related, study monodromy of 
2-cycles in a Calabi--Yau 4-fold geometry, and find that the proton 
decay operators are likely to be generated in a simple factorization limit 
of the spectral surface. A list of loopholes in this study, and hence 
a list of chances to save the factorized spectral surface scenario, 
is also presented. } 
  
\end{center}
\end{titlepage}



\section{Introduction}

Considerable progress has been made in the last two years 
in understanding flavor structure in F-theory compactifications. 
Supersymmetric compactifications of F-theory to 3+1 
dimensional space-time is described primarily by a set of data 
$(X_4, G^{(4)})$, where $X_4$ is an elliptic fibered Calabi--Yau 
4-fold and $G^{(4)}$ a 4-form flux on it. Non-Abelian gauge theories 
like SU(5)$_{\rm GUT}$ unified theories can arise from singular geometry 
of $X_4$.
It is difficult to extract physics directly from singular geometry, but 
low-energy effective theory associated with the singular local 
geometry of $X_4$ can be described by using non-Abelian gauge theory 
in 7+1 dimensions \cite{KV} to some level of approximation. 
It has been the key in the progress in the understanding of 
flavor structure in F-theory to use this gauge-theory effective 
description \cite{DW-1, BHV-1}. 
See also \cite{Hayashi-2, DW-3, Tsuchiya, Harvard-nonC, Conlon, Hayashi-3App}. 

In supersymmetric compactification for realistic models, 
dimension-4 proton decay operators 
\begin{equation}
 \Delta W \ni \bar{\bf 5}_M \cdot {\bf 10} \cdot \bar{\bf 5}_M 
\label{eq:dim-4}
\end{equation}
have to be brought under control. An obvious solution is 
\begin{itemize}
\item [(i)] to consider a compactification with $(X_4, G^{(4)})$ that has 
  a $\Z_2$ symmetry \cite{Tsuchiya}.
\end{itemize}
The $\Z_2$ symmetry will become an $R$-parity (or equivalently 
a matter parity) in the low-energy effective theory below the 
Kaluza--Klein scale. Other solutions to the problem include 
\begin{itemize}
 \item [(ii)] rank-5 GUT scenario, that is, to consider $\SU(6)$ or 
 $\SO(10)$ GUT models \cite{TW-1}, 
 \item [(iii)] factorized spectral surface scenario 
  \cite{Tsuchiya, Caltech-0906} 
     (see also \cite{BHV-2, DW-2, Harvard-nu, Harvard-E8}), and
 \item [(iv)] spontaneous $R$-parity violation scenario \cite{TW-1, Tsuchiya} 
(see also \cite{Blhg-0908}).
\end{itemize}
The scenario (ii) is a special case of (iii), and is also a special case 
of (iv). 

The factorized spectral surface scenario (iii), however, is not 
without a theoretical concern \cite{Tsuchiya}. The dimension-4 proton 
decay problem is so severe that the coupling (\ref{eq:dim-4}) should 
be extremely small even if they are present. 
The question is whether we have a theoretical framework in which 
the scenario can be formulated rigorously and even an extremely small 
contribution to (\ref{eq:dim-4}) can be studied. 
At least so far, the answer is no. 
It is, thus, important to study to what extent this scenario works,   
and that is what we do in this article.  

Heart of the idea of the factorized spectral surface scenario is to
consider a factorization limit of the spectral surface so that 
\begin{itemize}
 \item matter fields $\bar{\bf 5}_M$ and Higgs fields $\bar{\bf 5}_H$
       are associated with irreducible components of the spectral
       surface for fields in the $\bar{\bf 5}$ representation 
       of $\SU(5)_{\rm GUT}$, and 
 \item an unbroken U(1) symmetry remains in the low-energy effective
       theory below the Kaluza--Klein scale, and the 
       operator (\ref{eq:dim-4}) is ruled out because of the symmetry.
\end{itemize}
This scenario has been formulated by using the gauge theory description 
in 7+1 dimensions. 
Because of the approximate nature of the gauge theory description, 
factorization is not rigorously well-defined; there are also some 
corrections that have not been captured in the gauge theory
description on 7+1 dimensions \cite{Tsuchiya}. We will explain 
these limitations of the gauge theory description on 7+1 dimensions 
more in detail in section \ref{sec:effect_beyond_gauge_theory}, 
although these materials have almost been spelled out in \cite{Tsuchiya,
Caltech-0906} (see also \cite{DW-3}).

In section \ref{sec:mod}, we study monodromy of 2-cycles in $X_4$
instead of a gauge theory on 7+1 dimensions, in order to find out 
whether an unbroken U(1) symmetry remains in the low-energy 
effective theory. We conclude in section \ref{sec:physics} 
that the proton decay operators (\ref{eq:dim-4}) are expected to be 
generated in the factorized spectral surface scenario, but we also 
list some loopholes in the argument. 
The appendix \ref{sec:calculation} explains calculation of the 
monodromy of 2-cycles using language of string junction in detail; 
all the necessary techniques in the appendix \ref{sec:calculation}, 
however, are already available in the literature (e.g., \cite{jcn-KcMd}).
The monodromy matrices obtained in this way in F-theory language 
can be understood much more transparently in dual Heterotic language;  
this is the subject of the appendix \ref{sec:Het}.

\section{Effects Beyond Gauge Theory in 7+1 Dimension}
\label{sec:effect_beyond_gauge_theory}

The gauge theory description on 7+1 dimensions in F-theory captures 
geometry of a local neighbourhood of a discriminant locus of an elliptic 
fibration \cite{KV}. 
When a geometry of an elliptically fibered Calabi--Yau 4-fold 
$\pi_X: X_4 \rightarrow B_3$ is locally approximately an ALE space of 
A--D--E type in the direction transverse to the discriminant locus, 
gauge theory of the same A--D--E type describes the physics associated with 
this local geometry. 
Thus, by construction, the gauge theory description has limited power 
in capturing the physics of geometry in the transverse direction; 
only physics associated with {\it non-compact} (local) geometry that is 
{\it approximately} ALE fibration can be captured {\it approximately}. 
When it comes to the dimension-4 proton decay problem, this approximate 
nature of the gauge theory description of F-theory can be a problem. 

To see this more explicitly, let us consider a local geometry given by the 
generalized Weierstrass form 
\begin{eqnarray}
y^2 & = & x^3 - A_1 y x + A_2 x^2 - A_3 y + A_4 x + A_6, 
\label{eq:local-defeq}
\end{eqnarray}
where $(x,y)$ are coordinates of the elliptic fiber of $X_4$, and 
$A_{1,\cdots,4,6}$ are regarded as functions locally on $B_3$.
For an ${\rm SU}(5)_{\rm GUT}$ GUT model \cite{6authors}, we need to 
consider a case where $A_{1,\cdots, 4,6}$ are in the form of 
\begin{equation}
 A_k = (-)^k \left[ 
  z^{k-1} a_{6-k} + z^k a'_{6-k} + z^{k+1} a''_{6-k} + \cdots \right],
\label{eq:Ak-expnsn}
\end{equation} 
where $z$ is a local coordinate of a base manifold $B_3$;
the $z=0$ locus becomes an irreducible component $S_{GUT}$ of the 
discriminant locus $\Delta$. $a_{0,2,3,4,5}$ and $a'_{0,2,3,4,5}$ are 
coefficients that depend on two local coordinates on $S_{GUT}$.
Consider, for example, a case\footnote{This can be interpreted as 
considering a region of $S_{GUT}$ where the condition 
(\ref{eq:eK-scale}) is satisfied.}\raisebox{4pt}{,}\footnote{
Additional scaling may hold in some regions of $S_{GUT}$.
For example, along the matter curve $a_5 = 0$, 
\begin{equation}
a_{5,*}=\lambda \tilde{a}_{5,*}, \qquad \tilde{a}_{5,*} \sim {\cal O}(1)
\end{equation}
is satisfied for $\lambda \ll 1$. One of the four 2-cycles corresponding 
to the simple roots of the structure group SU(5)$_{\rm str}$ becomes 
much smaller than all the others in this region, and this corresonds to 
a hierarchical symmetry breaking $E_8 \rightarrow D_5 \rightarrow A_4$.
The rank-1 extended gauge theories in \cite{KV} and rank-2 extended 
gauge theories in \cite{BHV-1, Hayashi-2} describe physics of local 
geometries that have such scalings. 
We also introduce a hierarchical symmetry breaking in the choice of 
a base point of monodromy analysis in (\ref{eq:reference_point}) 
in this article.
} 
when all of $a_{2,3,4,5}$ are small in a way indicated by 
\begin{equation}
 a_{5}=\epsilon_{K}^{5}a_{5,*},\;\;a_{4}=\epsilon_{K}^{4}a_{4,*},\;\;
     a_{3}=\epsilon_{K}^{3}a_{3,*},\;\;a_{2}=\epsilon_{K}^{2}a_{2,*}
\label{eq:eK-scale}
\end{equation}
for $a_{5,*}$, $a_{4,*}$, $a_{3,*}$, $a_{2,*}$ of all $\sim {\cal O}(1)$. 
$\epsilon_K$ is a small parameter that is different from zero. 
We know that this geometry nearly has an $E_8$ singularity around $z=0$; 
to be more precise, there are four vanishing 2-cycles right at $z=0$, 
and there are four small 2-cycles near $z=0$. One can focus on 
the geometry of these small 2-cycles by choosing a new set of 
coordinates $(x', y', z')$:
\begin{equation}
x=\epsilon_{K}^{10}x^{\prime},\;\; 
y=\epsilon_{K}^{15}y^{\prime},\;\;
z=\epsilon_{K}^{6}z^{\prime}.
\label{eq:coordinate-scale}
\end{equation}
With these new coordinates and coefficients $a_{r,*} \sim {\cal O}(1)$ 
($r=0,2,3,4,5$), the local defining equation (\ref{eq:local-defeq}) 
becomes exactly the $E_8$ singularity with relevant deformation
parameters \cite{KM}, with correction terms whose coefficients are 
suppressed by positive power of the small number $\epsilon_K$. 
The gauge theory description focuses on the geometry of these small 
2-cycles (i.e., forgets about the rest of the geometry), and further 
make an approximation of ignoring the correction terms 
that are small but present \cite{Hayashi-2} (see also \cite{DW-3, 
Caltech-0906}). Even when the $\epsilon_K$ scaling of the coefficients 
is assumed as above to maximize the number of 
small 2-cycles captured in a gauge theory description, the gauge 
theory description is only an approximate description of the 
compact geometry $X_4$ by construction \cite{Tsuchiya}. 
Let us call this {\bf Problem A}.

We may have another difficulty in justifying the factorized spectral 
surface scenario in the gauge theory description. 
To see this, let us remind ourselves of the following.
Since $\bar{\bf 5}_M$ and $\bar{\bf 5}_H$ multiplets are zero modes
appearing in the low-energy effective theory below the Kaluza--Klein 
scale, the factorization of the spectral surface should be defined 
{\it globally} on the GUT divisor $S_{GUT}$. 
Field theory local model in an open patch $U_a \subset S_{GUT}$, on the other 
hand, captures very small 2-cycles in the ALE fiber in addition to 
the four vanishing 2-cycles in $A_4$ singularity, and hence the rank of 
gauge group in $U_a$ may be different from the rank in another 
open patch $U_b \subset S_{GUT}$ \cite{KV, BHV-1, DW-2, Hayashi-2}; these 
local descriptions with gauge groups of varying rank are glued 
together approximately in overlapping regions of open patches 
$\{ U_a \}_{a\in A}$ to cover the entire ALE-fibered geometry over 
$S_{GUT}$. 
The global factorization of the spectral surface, however, cannot be 
well-defined, when the rank of the spectral surface varies from  
one patch to another. This problem can be avoided if we consider 
a Higgs bundle that has a fixed rank over the entire $S_{GUT}$.
 
Fixed rank Higgs bundle description over the entire $S_{GUT}$, however,  
often breaks down somewhere in $S_{GUT}$. For example, a rank-$k$ spectral 
surface $C$ is defined globally in $K_S$ by 
\begin{equation}
 a_{5-k} \xi^k + \cdots + a_4 \xi + a_5 = 0,
\label{eq:spec-surf-1}
\end{equation}
where $\xi$ is the fiber coordinate of the canonical bundle 
$\pi_{K_S}: K_S \rightarrow S$, 
if $a_r$'s ($5-k \leq r \leq 5$) are global holomorphic sections of line 
bundles ${\cal O}(r K_S + \eta)$ for some divisor $\eta$ in $S_{GUT}$. 
$\pi_{K_S}|_C: C \rightarrow S$ is 
an $k$-fold cover at generic points of $S_{GUT}$, but when the coefficient 
of the highest degree term $a_{5-k}$ vanishes, one of the $k$ solutions 
of (\ref{eq:spec-surf-1}) shoots off to infinity in the fiber direction of 
$\pi_{K_S} : K_S \rightarrow S_{GUT}$. This behavior of the 
spectral surface indicates that one of (very) small 2-cycles becomes 
relatively large there.\footnote{\label{fn:eff-KS}
The vev of the Higgs field $\varphi$ is obtained by 
a holomorphic 4-form on a Calabi--Yau 4-fold on 2-cycles. The $\varphi$ 
vev being large means that the period integral over the corresponding 
2-cycle is large \cite{DW-1, Tsuchiya}. } It is not a 
sensible approximation to keep the physics associated with all the 
2-cycles in (\ref{eq:spec-surf-1}) while ignoring all others 
in a local neighbourhood around the zero locus of $a_{5-k}$ . 
As long as the divisor $(5-k)K_S + \eta$ is effective,\footnote{
The divisor $5K_S+\eta$ needs to be effective so that the matter curve 
for the SU(5)$_{\rm GUT}$-{\bf 10} representation field is effective. 
If $-K_S$ is effective, as in del Pezzo surfaces and Hirzebruch surfaces, 
then $(5-k)K_S + \eta$ is also effective. Thus, this problem is 
unavoidable for surfaces $S_{GUT}$ with effective $-K_S$.
See section \ref{sec:physics} for comments on $S_{GUT}$ with 
effective $K_S$ (rather than effective $-K_S$).} 
this does happen somewhere in $S_{GUT}$. 
The problem in this neighbourhood can be fixed (c.f. \cite{BHV-1}) by adopting a Higgs 
bundle that is one-rank higher, a rank-$(k+1)$ Higgs bundle, as long as 
$a_{4-k}$ is not identically zero.\footnote{We will come back to a loophole 
here in section \ref{subsec:loophole}.} We still encounter the same problem 
around the zero locus of $a_{4-k}$. One could still crank the rank 
of gauge group up, once again. But $E_8$ gauge group is maximal in 
the E series of the A--D--E classification, and we come to a dead end. 

We could consider a Calabi--Yau 4-fold $X_4$ for F-theory compactification 
which is approximately an ALE fibration of $E_8$ type over $S_{GUT}$ in the 
neighbourhood along $S_{GUT}$. 
This is done by choosing the coefficients indicated as in
(\ref{eq:eK-scale}) \cite{DW-3}.
When we take GUT gauge group as SU(5)$_{\rm GUT}$, we have a 5-fold 
spectral cover,
\begin{equation}
a_{0}\xi^{5}+a_{2}\xi^{3}+a_{3}\xi^{2}+a_{4}\xi+a_{5}=0.
\label{eq:5-fold}
\end{equation}
Factorization conditions can be imposed on (\ref{eq:5-fold}) in an 
$E_8$ gauge theory defined globally on $S_{GUT}$ \cite{Caltech-0906}.
That is the best we hope to do within gauge theories on 7+1 dimensions.
In a region of $S_{GUT}$ around a point where $a_0$ vanishes, however, 
two roots of (\ref{eq:5-fold}) become large, as 
\begin{equation}
\xi \sim \pm i \sqrt{(a_2/a_0)}. 
\label{eq:shoot-spec-surf}
\end{equation}
The $E_8$ gauge theory is not a sensible approximation in the local 
neighbourhood of the $a_0 = 0$ locus, because the corresponding two 
2-cycles are no longer relatively small.  
In a region slightly away from the $a_0 = 0$ locus, one can see 
that the two roots (and hence the Higgs field vevs $\varphi$ in 
two diagonal entries) are exchanged when the phase of $a_0$ changes 
by $2\pi$. This phenomenon indicates that the two 2-cycles not just become
large, but also are twisted by a monodromy around the $a_0=0$ locus. 
2-cycles that are not captured by the $E_8$ gauge theory may also 
be involved in this monodromy, because there is no clear separation 
between the 2-cycles within $E_8$ and those that are not around 
the $a_0=0$ locus. 
In order to study geometry and physical consequences associated with 
this behavior, one has to go beyond $E_8$ gauge theory on 7+1 dimensions.
We call this {\bf Problem B}.

Clearly we need a theoretical idea how to study whether the operator
(\ref{eq:dim-4}) is generated or not; a new idea should keep the 
successful aspects of the factorized spectral surface scenario, while 
it should not rely on the gauge theory descriptions in 7+1 dimensions. 
While spectral surfaces can be defined only within the gauge theory 
descriptions, the essence of the scenario is to keep an unbroken 
U(1) symmetry in the low-energy effective theory below the 
Kaluza--Klein scale.
We can thus focus on a question whether an unbroken U(1) symmetry 
is maintained (or how to find one) in F-theory compactifications.
Noting that topological 2-forms of certain type in $X_4$ yield U(1) 
vector fields in 3+1 dimensions \cite{MV2, DRS},\footnote{
These vector fields can be massive, yet global U(1) symmetries may remain 
in the effective theory. That is the situation we are interested in.} 
and that the vector fields are always accompanied by 
U(1) symmetries\footnote{The similar idea that an invariant 2-cycle 
under the monodoromy group gives rise to an unbroken U(1) symmetry, has been 
applied to the system of D5-branes wrapped on 2-cycles in an ALE fibered 
geometry \cite{CKV}.}, 
we see that a solution\footnote{One might alternatively think of factorizing 
the discriminant
as a generalization of factorizing the spectral surface. However, this
idea does not work from the very beginning. 
Global factorization of the {\it spectral surface} makes sense as a solution 
to the dimension-4 proton decay problem, 
because the spectral surface for the $\SU(5)_{\rm GUT}$-$\bar{\bf 5}$ 
representation is factorized into 2 irreducible branches around 
points of down-type Yukawa coupling ($D_6$ singularity enhancement 
points).
As studied in section 4.3 of \cite{Hayashi-2}, however, 
the {\it discriminant locus} does not have this property. 
See Figure~10.(b) of \cite{Hayashi-2}.
} is to focus on geometry of $X_4$, instead of gauge theory associated 
with the canonical bundle $K_S$ on $S_{GUT}$. 

Since we are interested in U(1) symmetries under which 
$\SU(5)_{\rm GUT}$-charged matter fields on $S_{GUT}$ are charged, 
we are interested in 2-forms of $X_4$ which have components dual to 
2-cycles in the local ALE fiber approximation along $S_{GUT}$ in $X_4$.
It is a topological problem whether or not an extra U(1) symmetry 
remains unbroken. We keep tracks of topological 2-forms / 2-cycles
to address this problem; it is not necessary to restrict our 
attention only to 2-cycles contained in one of the $A$--$D$--$E$ types, 
or to 2-cycles that are relatively small. All the 2-cycles in $X_4$ 
can be treated in an equal footing, and hence, we are free from all 
the problems in the gauge theory description on 7+1 dimensions.
The factorization condition of the spectral surface for an unbroken 
U(1) symmetry is replaced by a condition that there is at least one\footnote{
To be more precise, the condition is the existence of an extra 6-cycle 
[equivalently its Poincare dual] of $X_4$ contributing to $H^{1,1}(X)$ 
other than $B_3$ or those in $\pi_X^* (H^{1,1}(B_3))$ \cite{MV2, DRS}. 
We will come back to this issue in section \ref{subsec:non-K3}.} 
monodromy-invariant 2-cycle over $S_{GUT}$.

\section{Monodromy of 2-Cycles}
\label{sec:mod}

In this section, we study monodromy of 2-cycles in a Calabi--Yau 
4-fold $X_4$ and discuss whether an unbroken U(1) symmetry remains 
or not when the spectral surface satisfies a factorization condition. 
In section \ref{subsec:mod_gauge}, we show that certain subgroup 
of the monodromy corresponds to what we expect from the gauge theory 
descriptions on 7+1 dimensions. This subgroup reduces to a smaller 
one when the spectral surface is in the factorization limit, and there 
is a monodromy invariant 2-cycle [and hence an unbroken U(1)] at this 
level of analysis. This guarantees that we can keep the heart of the 
idea of the factorized spectral surface scenario without relying on 
the gauge theory description.
Section \ref{subsec:str_sing} explains the structure 
of the full monodromy group; the subgroup in section \ref{subsec:mod_gauge} 
is a proper subgroup of the full monodromy group. We study monodromy 
of 2-cycles for some other generators in 
section \ref{subsec:monod_beyond_gauge}, and find that there is no 
monodromy-invariant 2-cycles when the full monodromy group is taken 
into consideration. This means that there is no unbroken U(1) symmetry 
in the effective theory in the factorized limit of the spectral surface. 

\subsection{The model and the notation}

In the study of monodromy of 2-cycles in 
sections \ref{subsec:mod_gauge}--\ref{subsec:monod_beyond_gauge}, we only 
consider a special case where a Calabi--Yau 4-fold $X_4$ is a K3 
fibration over $S_{GUT}$: $\pi'_X: X_4 \rightarrow S_{GUT}$.
One of the advantages of this $X_4$ as a global fibration on $S_{GUT}$ is 
that the U(1) vector fields in 3+1 dimensions, and hence the 
non-$H^2(B_3)$ non-$H^0(B_3; R^2 \pi_{X*} \Z)$ components of $H^2(X_4; \Z)$ can 
be studied by $H^0(S_{GUT}; R^2 \pi'_{X*} \Z)$ \cite{CD, DW-1, Hayashi-1}. 
Global sections of the {\it local} system $R^2 \pi'_{X*} \Z$ correspond 
precisely to the monodromy invariant 2-cycles in the fiber of 
$\pi'_{X}: X_4 \rightarrow S_{GUT}$, and hence the problem can be 
formulated as a local theory on 7+1 dimensions (if not as a local 
{\it gauge} theory on 7+1 dimensions). Although this advantage may appear 
to be available only for a special case, global fibration of 
$X_4$ over $S_{GUT}$, we consider otherwise. Topological 2-cycles 
in a neighbourhood of $S_{GUT}$ can be captured by a local system 
like $R^2 \pi'_{X*} \Z$ at least locally, if not globally.
Since the monodromy among 2-cycles is about a non-trivial local 
behavior of such a local system at special points (which we call monodromy 
locus later in this article), one only has to examine local behavior of such 
local systems to find out whether a U(1) symmetry is projected out 
or not. Thus, although the following presentation in this section 
may appear to rely exclusively on a $X_4$ that is a global fibration 
on $S_{GUT}$, we do not loose generality as a study of F-theory
compactifications. Discussion on non-K3 fibered cases is found in 
section \ref{subsec:non-K3}.

The other advantage is that we know a lot about 2-cycles of 
K3 manifold, so that we can make our presentation very concrete. 
Since the monodromy around $a_0 = 0$ locus might involve not just 
2-cycles within $E_8$ but also other 2-cycles in the direction
transverse to the GUT divisor $S_{GUT}$, we certainly need a geometry 
for analysis where such an ``other 2-cycle'' is identified. 
With a compact K3 manifold in the transverse direction, we have 
all the necessary techniques. On the other hand, we do not loose 
generality by this specific choice, because we are interested 
only in the local behavior of monodromy among 2-cycles in $E_8$ and 
those that are ``adjacent'' to them.

In principle, we could study monodromy of 2-cycles in a configuration 
of real interest, where the $A_4$ singularity is along $S_{GUT}$, and 
a factorization condition is imposed on the 5-fold spectral 
cover (\ref{eq:5-fold}).
Instead of this realistic $E_8 \rightarrow \SU(5)_{\rm GUT}$ symmetry 
breaking model, however, we study monodromy of 2-cycles in an 
$E_8 \rightarrow E_6$ symmetry breaking model. 2+1 factorization
condition can be imposed on the 3-fold spectral cover 
\begin{equation}
a_0 \xi^3 + a_2 \xi + a_3 = 0.
\label{eq:rank3_spect}
\end{equation}
Moreover, since the 2-cycles associated with the hidden sector is 
not essential in our problem, we consider a configuration with an 
unbroken hidden $E_8$ symmetry. 
Namely, we study monodromy of 2-cycles in $X_4$ whose K3 fibration
structure is given by 
\begin{equation}
y^2 = x^3 + (a_2 z^3 + f_0 z^4) x
+ \left( \frac{1}{4} a_3^2 z^4 + a_0 z^5 + g_0 z^6
+ a''_0 z^7 \right) \, .
\label{eq:model}
\end{equation}
Here, $(x,y)$ are coordinates of the elliptic fiber of $X_4$,
$z$ an inhomogeneous coordinate of the base $\mathbb{P}^1$ of K3
and $a_{0,2,3}, a''_0, f_0, g_0$ are sections of some line bundles 
on $S_{GUT}$.

When F-theory is compactified on a K3 manifold that is an elliptic 
fibration $\pi_{K3}: K3 \rightarrow \P^1$ with a section, then there 
are 20 2-cycles dual to $H^1(\P^1; R^1 \pi_{K3 *} \Z)$ and corresponding 
U(1) symmetries in 7+1 dimensions. 
Such topological 2-cycles can be described by string junctions on the 
base $\P^1$. All the necessary details in how to find independent 
20 configuration of such junctions \cite{jcn-KcMd} are reviewed 
in the appendix \ref{sec:calculation}.

In order to calculate monodromy of 2-cycles, or equivalently to see 
how the local system $R^2 \pi'_{X*} \Z$ is twisted, one first has to 
take a base point $b \in S_{GUT}$, and set a basis in the (co)homology 
group of $H_2(\pi^{'-1}_X(b); \Z) \simeq H_2({\rm K3}; \Z)$. 
For a loop $\gamma$ on $S_{GUT}$ that starts and ends on the base 
point $b$, we can trace the basis elements $H_2(\pi^{'-1}_{X}(p); \Z)$ for points 
$p$ on the loop $\gamma$, starting from the base point $b$ until 
we come back to the base point again. The calculation of monodromy 
is practically carried out by tracing string junction configuration 
that correspond to the basis elements of 
$H_2(\pi^{'-1}_X(b); \Z)$. We only have to trace along the loop $\gamma$ 
the motion of discriminant locus points---$[p,q]$ 7-branes---in the 
complex $z$ plane. All the necessary techniques in this monodromy analysis 
is quite standard in Type IIB string theory. We thus present detailed procedure 
of practical calculation of monodromy for only three loops in the
appendix \ref{sec:calculation}; for all other loops, only the results 
are presented in this article. 

\begin{figure}[tb]
\begin{center}
\includegraphics[scale=0.3]{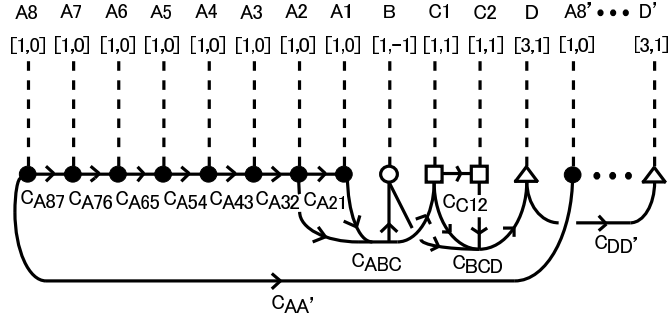}
\caption{The independent 2-cycles of a K3 surface. Dotted lines
 represent the branch cuts of 7-branes. Black blobs represent
 $A$-branes, open circles are B-branes, open boxes correspond to 
C-branes and open triangles are ``D''-branes whose $[p,q]$ charge 
is [3, 1]. $A8^{\prime} \sim D^{\prime}$ 7-branes have the same $[p,q]$ 
charges as those without a $'$, and are ordered from left to right as in
 the same way as $A8 \sim D$ 7-branes. 
7-branes from $A7$ to $C2$ and those from $A7'$ to $C2'$ constitute 
separate sets of $[A^7BC^2]$ 7-brane configuration of $E_8$.
}
\label{fig:2-cycles}
\end{center}
\end{figure}

It is convenient to introduce a notation for a set of independent
2-cycles of the K3 fiber and also to assign names to individual 
$[p,q]$ 7-branes at a base point $b$. To the 24 discriminant 
points, we assign the following names: $A8$--$A1$, $B$, $C1$, $C2$, 
$D$, $A8'$--$A1'$, $B'$, $C1'$, $C2'$ and $D'$. 
The appendix \ref{sec:calculation} explains the choice of the base point 
$b$, location of the the 24 7-branes in the $z$ plane, as well as 
their $[p,q]$ charges for the base point. We assign names (notation) 
such as $C_{A76}$, $C_{A65}$ etc. to string junction configurations and
corresponding 2-cycles of ${\rm K3} = \pi^{'-1}(b)$. All the details 
are written in the appendix \ref{sec:calculation}, but the most relevant 
part of the information is summarized in Figure \ref{fig:2-cycles}.

\begin{figure}[tb]
\begin{center}
\includegraphics[scale=0.3]{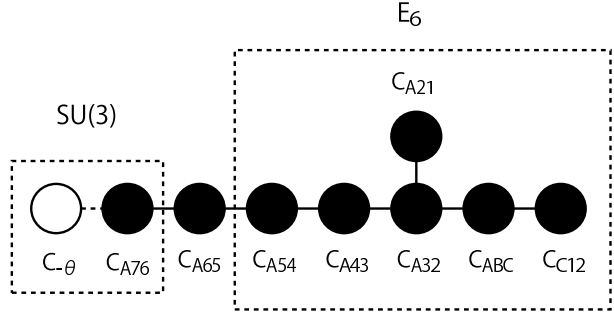}
\caption{2-cycles inside $E_8$ and their intersection. 
Note that $C_{-\theta}$ is not the same as $C_{A87}$; 
see footnote \ref{fn:K3-RES} for more.}
\label{fig:dynkin}
\end{center}
\end{figure}

There is no unique choice of a basis of $H_2({\rm K3}; \Z)$, 
but one of the most convenient choices is to use the 2-cycles 
listed in Table~\ref{tab:2-cycles}.
\begin{table}[tb]
 \begin{center}
  \begin{tabular}{l|l} \hline
 2-cycles inside $E_8$ & $C_{A76}, C_{A65}, C_{A54}, C_{A43}, C_{A32}, 
  C_{A21}, C_{ABC}, C_{C12}$ \\ \hline
 2-cycles outside $E_8$ & $C_{\alpha}^1$, $C_\alpha^2$, $C_{\beta}^1$, 
  $C_\beta^2$ \\ \hline
 those for hidden $E_8$ & $C'_{A76}, C'_{A65}, C'_{A54}, C'_{A43}, 
C'_{A32}, C'_{A21}, C'_{ABC}, C'_{C12}$ \\ \hline 
  \end{tabular}
\caption{A convenient choice of a basis of $H_2(\pi^{'-1}(b); \Z)$. }
\label{tab:2-cycles}
 \end{center}
\end{table}
Among them, $C_\alpha^{1,2}$, $C_\beta^{2}$ need to be defined 
by linear combinations of string junction configurations like those in 
Figure~\ref{fig:2-cycles}; see (\ref{eq:basis1}, \ref{eq:basis2}).
The intersection form becomes (\ref{eq:K3-int-form}).
The intersection form within the eight 2-cycles in the first row of 
Table \ref{tab:2-cycles} is described by the Dynkin diagram of $E_8$ 
shown in Figure~\ref{fig:dynkin}.

As we study monodromy of 2-cycles in (\ref{eq:model}) that has 
unbroken $E_6$ symmetry (split form $E_6$ singularity) \cite{6authors}, 
monodromy matrices must be trivial on a subspace generated purely by 
$C_{A54}, C_{A43}, C_{A32}, C_{A21}, C_{ABC}, C_{C12}$, the simple 
roots of $E_6$. Thus, we can choose $C_{-\theta}$ as a basis element 
instead of $C_{A65}$ as in Table~\ref{tab:2-cycles}, where 
$C_{-\theta}$ is a linear combination purely of the 2-cycles in the 
first row of Table~\ref{tab:2-cycles}, 
\begin{equation}
C_{-\theta} = -(2 C_{A76} + 3 C_{A65} + 4 C_{A54} + 5 C_{A43} + 6 C_{A32} + 4
C_{ABC} + 3 C_{A21} + 2 C_{C12}) \, .
\label{eq:C-theta-def}
\end{equation}
The six 2-cycles of $E_6$ and the two 2-cycles $C_{A76}$ and 
$C_{-\theta}$ for the ``structure group'' $\SU(3)_{\rm str}$ 
are orthogonal in the intersection form. 
Since we restrict ourselves in a region of moduli space where 
the hidden $E_8$ 2-cycles do not participate in the monodromy, 
the monodromy can be represented by $6 \times 6$ matrices 
acting on a vector space generated by $C_{A76}, C_{-\theta}, C_\alpha^1,
C_\alpha^2, C_\beta^1$ and $C_\beta^2$.
   
The calculation of monodromy boils down to following the trace 
of the $[p,q]$ 7-branes along paths in $S_{GUT}$. Thus, an explicit 
expressions for the discriminant of \eqref{eq:model} is very useful:
\begin{equation}
\Delta = z^8 \Delta'
\end{equation}
where $\Delta'$ is given by
\begin{equation}
\begin{split}
\Delta' &= \frac{27}{16} a_3^4
 + \left( \frac{27}{2} a_0 a_3^2 + 4 a_2^3 \right) z
 + \left( \frac{27}{2} a_3^2 g_0 + 27 a_0^2 + 12 a_2^2 f_0 \right) z^2 \\
 & \quad + \left( 54 a_0 g_0 + 12 a_2 f_0^2 + \frac{27}{2} a_3^2 a''_0 \right) z^3
  + (4 f_0^3 + 27 g_0^2 + 54 a_0 a''_0) z^4 \\
 & \quad + 54 a''_0 g_0 z^5 + 27 a^{''2}_0 z^6 \, .
\end{split}
\label{eq:discriminant}
\end{equation}
In this model, eight 7-branes $A5$--$A1$ and $B$, $C1$, $C2$ are at 
an $E_6$ singularity at $z=0$ \cite{Johansen, GZ}, and ten 7-branes 
$A7'$--$A1'$, $B'$, $C1'$and $C2'$ form an $E_8$ singularity at $z = \infty$.
The six other 7-branes, $A6$, $A7$, $A8$, $D$, $A8'$ and $D'$, are 
the zero points of $\Delta'$ above.
Along a path $\gamma$ in $S_{GUT}$, values of all the coefficients 
$a_{0,2,3}$, $f_0$, $g_0$ and $a''_0$ change, and hence the six 7-branes 
change their locations in the $z$-plane. We can thus determine 
the monodromy matrix by following the 7-branes and string junctions.

In sections \ref{subsec:mod_gauge}--\ref{subsec:monod_beyond_gauge}, 
we present monodromy matrices obtained as a result of such a
calculation. Although the monodromy for a loop $\gamma$ in $S_{GUT}$ 
is ultimately what we are interested in, it is a better idea to 
study monodromy of 2-cycles in a little more abstract level;
we will study monodromy of 2-cycles for 
loops in a moduli space of an elliptic K3 manifold.
Since any loops in $S_{GUT}$ are mapped by the choice of sections 
$a_{0,2,3}$, $f_0$ etc. to loops in the moduli space of the elliptic K3 manifold, 
the monodromy group for the loops on $S_{GUT}$ is obtained by a pull-back 
of the monodromy group for the loops on the elliptic K3 moduli space 
by the sections. Such questions as whether an unbroken U(1) symmetry 
exists can be answered at the level of monodromy on the elliptic 
K3 moduli space, independent of specific details of $S_{GUT}$ or 
sections $a_{0,2,3}$ etc. on it.

\subsection{The monodromy in 8D gauge theory region}
\label{subsec:mod_gauge}

In the factorization limit of the spectral surface, the monodromy 
group of 2-cycles captured in the gauge theory description on 
7+1 dimensions is reduced; there is no question about this statement 
in the deformation of $A_{N+M}$ singularity to $A_M$. This statement 
is believed to be true also in deformation of $E_{6,7,8}$ type singularity 
to $A_4$ \cite{Hayashi-2, Tsuchiya} (see also \cite{DW-1}), for example,
but it has not been confirmed directly so far by looking at 2-cycles
in $X_4$. In this section \ref{subsec:mod_gauge}, we study explicitly 
the geometry of 2-cycles in $X_4$ when the spectral surface is at the 
factorization limit, and confirm that the statement above is correct.
This justifies our strategy to replace the factorization limit of 
spectral surfaces by existence of monodromy invariant 2-cycles.

At the same time, we will see that the monodromy of 2-cycles that appear 
in the gauge theory description on 7+1 dimensions is only a proper
subgroup of the entire monodromy group of the whole theory. Thus, it 
will be clear what one overlooks in the gauge theory description on 
7+1 dimensions. 

The family of elliptic K3 manifold (\ref{eq:model}) is parametrized 
by $(a_{0}, a_2, a_3, f_0, g_0, a''_0)$, among which two are redundant 
because of the freedom to rescale the coordinates $(x,y,z)$. 
Instead of using a set of ``gauge invariant'' parameters of this 
moduli space such as $(a_0a''_0/g_0^2)$ and 
$(a_2/a_0)(4f_0^3 + 27 g_0^2)^{1/6}$, we fix a gauge by fixing the
values of $f_0$ (or $g_0$) and $a''_0$.

The full scope of the problem is to consider all possible loops 
in the moduli space of (\ref{eq:model}) and determine the monodromy 
of 2-cycles of the K3 for the loops. We restrict our attention in this 
section, however, to a subset of the moduli space 
\begin{equation}
 a_0 = a_{0,*} \epsilon_\eta, \quad 
 a_2 = a_{2,*} \epsilon_K^2 \epsilon_\eta, \quad
 a_3 = a_{3,*} \epsilon_K^3 \epsilon_\eta, 
\label{eq:scaling-rg}
\end{equation}
where parameters $a_{r, *} \in \C$ and $g_0 \in \C$ are at most 
${\cal O}(1)$; $0 \neq |\epsilon_K| \ll 1$ and 
$0 \neq |\epsilon_\eta| \ll 1$ are fixed numbers, and 
we fix the ``gauge'' by setting $f_0 = -1$ and $a''_0 = \epsilon_\eta$. 
Because $|\epsilon_\eta| \ll 1$, the elliptic K3 manifold is always 
close to the stable degeneration limit \cite{MV2, FMW} within 
the subset specified above. 
We call this subset scaling region. 
It helps in ensuring the validity of the gauge theory description 
in 7+1 dimensions, but not quite, to set $\epsilon_K$ as a small number.
We will see this in detail in the rest of this section in F-theory
language. The appendix \ref{sec:Het} will provide a very clear explanation 
of this in the dual Heterotic language.  
The base point of the moduli space is chosen within this subset, and 
only loops that stay within this subset are considered in this section. 
Although this is only to find part of the monodromy group, we will see 
that this partial result is interesting enough from mathematical and 
physical points of view, and is also sufficient in drawing conclusions 
for practical purposes.

We will call a subset of the scaling region characterized by 
\begin{equation}
 |\epsilon_K|^{2} \ll |a_{0,*}| \sim {\cal O}(1)
\label{eq:gauge-th-rg}
\end{equation}
as an {\it 8D gauge theory region}. We will see in this 
section \ref{subsec:mod_gauge} that all the loops (except the one
mentioned at the end of section \ref{subsec:monod_beyond_gauge}) 
that stay within this `` 8D gauge theory region'' yield monodromies of 
2-cycles that are expected from the gauge theory descriptions on 
7+1 dimensions. 

For any points in the 8D gauge theory region of the moduli space, 
the six 7-branes specified by $\Delta' = 0$ are distributed in 
the complex $z$ plane as follows; there are two 7-branes in the 
region $z \sim {\cal O}(\epsilon_K^6 \epsilon_\eta)$, two 
in the region $z \sim {\cal O}(\epsilon_\eta)$, and two 
in $z \sim {\cal O}(\epsilon_\eta^{-1})$. These three groups 
of 7-branes remain distinct along any loops that stay within 
the 8D gauge theory region of the moduli space. It is, thus, 
appropriate to identify that the two 7-branes in the 
${\cal O}(\epsilon_\eta^{-1})$ region as the 7-branes $A8'$ and $D'$, 
those in the ${\cal O}(\epsilon_\eta)$ region as $A8$ and $D$, and 
those in the ${\cal O}(\epsilon_K^6 \epsilon_\eta)$ region as 
$A6$ and $A7$, respectively. 
Over the entire 8D gauge theory region, the positions 
of the 7-branes A6, A7 are effectively given by the zero points of
\begin{equation}
\Delta'_{\mathrm{lower-quadr}} = \frac{27}{16} a_3^4
 + \left( \frac{27}{2} a_0 a_3^2 + 4 a_2^3 \right) z
 + 27 a_0^2 z^2 \, ,
\label{eq:discriminant_lqp}
\end{equation}
which is approximately the first three terms of the right hand side 
of \eqref{eq:discriminant}.

The string junction configuration (and the corresponding 2-cycles)
$C_{A76}$ and $C_{-\theta}$ correspond to the two roots of 
$\SU(3)_{\rm str}$, the commutant of $E_6$ within $E_8$.
As a point moves along a loop in (the 8D gauge theory region of) the 
moduli space, the coefficients $a_{0,2,3}$ in (\ref{eq:discriminant_lqp}) 
change, and the positions of the $A7$ and $A6$ 7-branes move in the 
complex $z$-plane. Consequently the branch cuts and the string junction
configurations also have to change continuously. Monodromy of the 
two 2-cycles $C_{A76}$ and $C_{-\theta}$ should reproduce what we expect
from the 8D gauge theory descriptions. 

Such loops in the moduli space should avoid a locus where more than 
one 7-branes come on top of one another, because the elliptic K3 ``manifold'' 
becomes singular and it is subtle to talk of ``topological 2-cycles'' there. 
The degeneration of the A6 and A7 7-brane positions are characterized 
by the discriminant locus of the quadratic equation 
$\Delta'_{\mathrm{lower-quadr}} = 0$: 
\begin{equation}
\tilde \Delta_{\mathrm{lower-quadr}} =
 4 a_2^3 (27 a_0 a_3^2 + 4 a_2^3) = 0. 
\label{eq:tilde_discriminant_lqp}
\end{equation}
As one moves on a loop around such a locus of degeneration of the 7-brane
positions, $\tilde{\Delta} = 0$, more than one 7-branes may exchange 
their positions at the end of the loop. Moreover, branch cuts may have to be 
rearranged in case of degeneration of mutually non-local $[p,q]$ 7-branes. 
Thus, the locus of 7-brane degeneration is where monodromy of 2-cycles 
could be generated, and hence we call $\tilde{\Delta} = 0$ locus as 
the monodromy locus. 

The monodromy locus in the 8D gauge theory region is given 
by (\ref{eq:tilde_discriminant_lqp}) 
[see also section \ref{subsec:str_sing}]. 
Interestingly the second factor of \eqref{eq:tilde_discriminant_lqp} is
the same as the ramification locus of the spectral 
surface \eqref{eq:rank3_spect}. 
Since we expect a monodromy among 2-cycles along the ramification 
locus of the spectral surface in the gauge theory description on 
7+1 dimensions, it is more than natural that the same factor 
appears in the monodromy locus that we introduced in the language 
of $X_4$. The first factor of the monodromy locus, $a_2 = 0$ with 
a multiplicity 3, on the other hand, does not appear in the 
gauge theory description. We thus move on to calculate the monodromy 
matrices explicitly to clarify the relation between the two
descriptions. 

\subsubsection{Monodromy without factorization}
\label{sssec:wo-factorization}

We begin with a general choice of complex structure without a
factorization of (\ref{eq:rank3_spect}) in 
section \ref{sssec:wo-factorization}. From the gauge theory 
description on 7+1 dimensions, we expect that the monodromy of 
the 2-cycles $C_{A76}$ and $C_{-\theta}$ is the full Weyl group of 
$\SU(3)_{\rm str}$, $S_3$, and there is no mixing between 
the 2-cycles in the visible $E_8$ and others in $H_2({\rm K3} ; \Z)$.
Moreover, we expect that the monodromy is generated along 
the ramification locus $(27 a_0 a_3^2 + 4 a_2^3) = 0$, but 
nowhere else. These expectations are indeed confirmed by 
explicit calculations of the monodromy, as we explain in 
the following.

Let us pick a base point in the moduli space, and fix it once 
and for all. All the loops start and end at the base point.
We choose it in the 8D gauge theory region: 
\begin{equation}
(a_0,a_2,a_3) = (a_{0,0}, a_{2,0}, a_{3,0}) \equiv 
 (1, \epsilon_K^2, i  \delta \epsilon_K^3) \epsilon_\eta,
\label{eq:reference_point}
\end{equation}
with real positive small numbers $\epsilon_K$ and $\epsilon_\eta$.
$\delta$ is also a small real positive number, which does not play 
an important role in this article. We set $g_0 = g_{0,0} \equiv 1$ 
at the base point. 
Arrangement of the branch cuts and the $[p,q]$ charges of the 
7-branes are described explicitly in the appendix \ref{sec:calculation}
for this choice of the base point. 

Although we could study monodromy of all the loops in the 8D gauge theory 
region of the moduli space parametrized by 
$(a_0, a_2, a_3, g_{0})$, it just introduces a mess that 
is not essential to the problem. 
Instead, let us take a slice of constant $a_0 = a_{0,0}$, $a_3 = a_{3,0}$ and 
$g_{0} = g_{0,0}$ of the moduli space, and consider loops in the 
complex $a_2$-plane. 

On the $a_2$-plane, the monodromy locus 
$\tilde \Delta_{\mathrm{lower-quadr}} = 0$
consists of one point $a_{2-A}$ with a multiplicity 3, and three points 
$a_{2-1}$, $a_{2-2}$ and $a_{2-3}$ with multiplicity 1. 
\begin{figure}[tb]
 \begin{center}
  \includegraphics[scale=0.3]{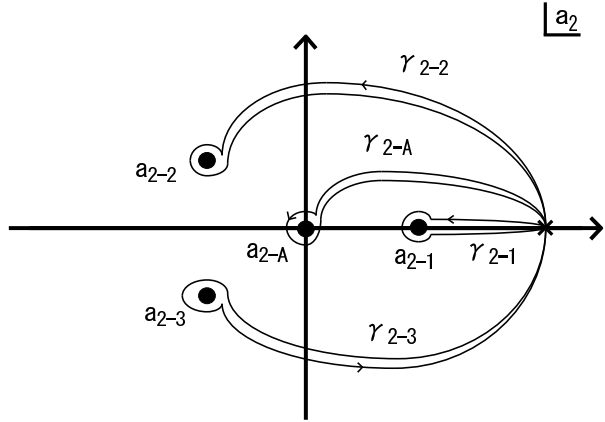}
  \caption{
The monodromy locus $\tilde \Delta_{\mathrm{lower-quadr}} = 0$ 
in (\ref{eq:tilde_discriminant_lqp}) on the $a_2$-plane and independent 
loops around these points. $a_2 = a_{2-A}$ is a triple point and
correspond to the first factor $a_2 = 0$ of 
$\tilde{\Delta}_{\mathrm{lower-quadr}}$. $a_2 = a_{2-1,2,3}$ are 
single points, correspond to the second factor of 
$\tilde{\Delta}_{\mathrm{lower-quadr}}$.
The base point is denoted by a cross mark.}
  \label{fig:a2-plane_e_K}
 \end{center}
\end{figure}
Figure \ref{fig:a2-plane_e_K} shows four independent non-trivial 
loops in the $a_2$-plane. We calculate the monodromy for these
four loops. The calculation itself is straightforward, and we 
just quote the results in this article; calculations for 
$\gamma_{2-A}$ and $\gamma_{2-2}$ here and $\gamma_{0-4}$ that 
appears later, however, are explained explicitly in the appendix 
\ref{sec:calculation} in detail as samples. 

It turns out that the monodromy associated with $\gamma_{2-A}$ 
is trivial.
This is consistent with the expectation that the monodromy is 
generated only at the second factor of the monodromy 
locus (\ref{eq:tilde_discriminant_lqp}), not at the $a_2 = 0$ factor.

Next we consider the loops $\gamma_{2-1}, \gamma_{2-2}$ and $\gamma_{2-3}$.
The monodromy associated with $\gamma_{2-2}$ is explicitly computed in 
the appendix \ref{sec:calculation} to give the result
\begin{equation}
\begin{split}
\tilde C_{A65} &= C_{-\theta} + C_{A65} + C_{A76}\, , \\
\tilde C_{A76} &= - C_{-\theta} \, , \\
\tilde C_{-\theta} &= - C_{A76},  
\end{split}
\label{eq:monod_2-2}
\end{equation}
while $C_{\alpha}^{1,2}$ and $C_{\beta}^{1,2}$ are left invariant.
Here $\tilde {C}_i$ denotes the transformed 2-cycle of $C_i$ after going 
along $\gamma_{2-2}$.
Similarly, the monodromies associated with $\gamma_{2-1}$ and
$\gamma_{2-3}$ can be computed. The result for $\gamma_{2-1}$ is
\begin{equation}
\begin{split}
\tilde C_{A65} &= C_{A76} + C_{A65} \, , \\
\tilde C_{A76} &= - C_{A76} \, , \\
\tilde C_{-\theta} &= C_{A76} + C_{-\theta} 
\end{split}
\label{eq:monod_2-1}
\end{equation}
and, for $\gamma_{2-3}$, the same monodromy as \eqref{eq:monod_2-2} 
is obtained. Thus, the monodromy for these three loops act only on the 
2-dimensional subspace of $H_2({\rm K3} \;; \Z)$ generated by 
$C_{A76}$ and $C_{-\theta}$, and are trivial 
on all the six generators of the visible $E_6$, all the eight generators
of the hidden $E_8$ and all the four 2-cycles $C_{\alpha,\beta}^{1,2}$ 
in the middle. These monodromies were generated essentially around 
the $(27 a_0 a_3^2 + 4 a_2^3) = 0$ locus.
This is also consistent with the expectation.

It is not difficult to see that these monodromy transformation 
on the visible $E_8$ 2-cycles are regarded as Weyl reflections of 
$\SU(3)_{\rm str} \subset E_8$. The monodromy $\rho(\gamma_{2-1})$ 
is a Weyl reflection generated by a root $C_{A76}$, $W_{C_{A76}}$, 
and $\rho(\gamma_{2-2})$ and $\rho(\gamma_{2-3})$ are 
$W_{C_{A76} + C_{-\theta}}$. 
When they are represented on a three-element basis, 
$(C_{A65}, C_{A65}+C_{A76}, C_{A65}+C_{A76}+C_{-\theta})$, they become 
\begin{equation}
\rho_{\mathbf{3}} (\gamma_{2-1}) =
\begin{pmatrix}
0 & 1 & 0 \\
1 & 0 & 0 \\
0 & 0 & 1
\end{pmatrix}
\label{eq:monod_2-1_matrix}
\end{equation}
\begin{equation}
\rho_{\mathbf{3}} (\gamma_{2-2}) = \rho_{\mathbf{3}} (\gamma_{2-3}) =
\begin{pmatrix}
0 & 0 & 1 \\
0 & 1 & 0 \\
1 & 0 & 0
\end{pmatrix}.
\label{eq:monod_2-2_matrix}
\end{equation}
Thus, they generate the permutation group $S_3$, the full Weyl group 
of $\SU(3)_{\rm str}$, just as expected in the gauge theory description 
without a factorization of the spectral surface.  

\subsubsection{Monodromy with factorization}
\label{sssec:w-factorization}

It is known that in the gauge theory description on 7+1 dimensions,
if we impose a factorization condition on a spectral surface,
then the structure group is reduced and an unbroken U(1) symmetry appears.
In the remainder of this subsection,
we will see that this fact can be described in terms of the 
monodromies of 2-cycles as well.

On the spectral surface \eqref{eq:rank3_spect}, let us impose the 
2+1 factorization condition \cite{TW-1, Caltech-0906}
\begin{equation}
 0 = a_0 \xi^3 + a_2 \xi + a_3 = (c_0 \xi^2 + c_1 \xi + c_2)(d_0 \xi + d_1)
\label{eq:2+1}
\end{equation}
for some sections $c_{0,1,2}$ and $d_{0,1}$ on $S_{GUT}$.
These sections have to satisfy a condition \cite{Caltech-0906}
\begin{equation}
 c_0 d_1 + c_1 d_0 = 0, 
\label{eq:vanish-a1}
\end{equation}
and we adopt a solution to this condition, $d_0 = c_0$ and 
$d_1 = - c_1$. Thus, 
\begin{align}
a_0 &= c_0^2 \, , \label{eq:factorization_cond1} \\
a_2 &= c_0 c_2 - c_1^2 \, ,\label{eq:factorization_cond2} \\
a_3 &= - c_1 c_2 \, .
\label{eq:factorization_cond3}
\end{align}
We now consider a family of elliptic K3 manifold parametrized 
by $(c_0, c_1, c_2, g_0)$, and study monodromy of 2-cycles 
for loops in this moduli space. 

The 8D gauge theory region in the $(c_0, c_1, c_2, g_0)$ space is 
characterized by 
\begin{equation}
 c_0 = c_{0,*} \epsilon^{1/2}_\eta, \quad 
 c_1 = c_{1,*} \epsilon_K \epsilon^{1/2}_\eta, \quad 
 c_2 = c_{2,*} \epsilon_K^2 \epsilon^{1/2}_\eta, 
\label{eq:scaling-rg-c}
\end{equation}
where parameters $c_{r,*} \in \C$ ($r=0,1,2$) and $g_0 \in \C$ are 
at most ${\cal O}(1)$. $\epsilon_K$ and $\epsilon_\eta$ are small 
but non-zero parameters as before. We further require that 
\begin{equation}
 \epsilon_K \ll |c_{0,*}| \sim {\cal O}(1).
\end{equation}
We will take a base point as 
\begin{equation}
 (c_0, c_1, c_2) = (c_{0,0}, c_{1,0}, c_{2,0}) \equiv 
  (1, -i \epsilon_K, \delta' \epsilon_K^2) \epsilon_{\eta}^{1/2},
\label{eq:reference_point_c}
\end{equation}
and $g_{0} = g_{0,0} = 1$; real positive values are used for small 
numbers $\epsilon_K$, $\epsilon_\eta$ and $\delta'$, so that this 
base point is mapped to the base point in the $(a_0, a_2, a_3, g_0)$ 
parameter space 
through (\ref{eq:factorization_cond1}--\ref{eq:factorization_cond3}).

The monodromy locus in the 8D gauge theory region in the $(c_0,c_1,c_2, g_0)$
parameter space is given by simply rewriting (\ref{eq:tilde_discriminant_lqp})
by (\ref{eq:factorization_cond1}--\ref{eq:factorization_cond3}).
It factorizes as
\begin{equation}
\tilde \Delta_{\mathrm{lower-quad}} = 4
 (c_0 c_2 - c_1^2)^3 (4 c_0 c_2 - c_1^2) (c_0 c_2 + 2 c_1^2)^2 \, .
\label{eq:tilde_discriminant_lqp_factorize}
\end{equation}

\begin{figure}[tb]
 \begin{center}
  \includegraphics[scale=0.3]{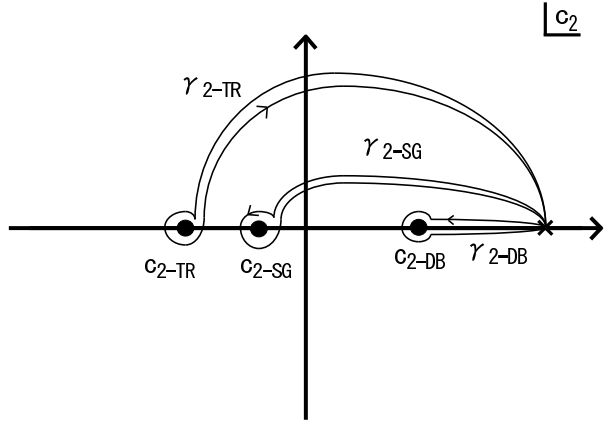}
  \caption{The monodromy locus $\tilde \Delta_{\mathrm{lower-quad}} = 0$ 
on the $c_2$-plane and independent loops around these points. 
The cross mark is the base point $c_2 = c_{2,0}$.}
  \label{fig:c2-plane_e_K}
 \end{center}
\end{figure}

If we focus on a constant $c_0$, $c_1$, $g_0$ slice (complex
$c_2$-plane) of the moduli space that contains the base point, 
then the monodromy locus $\tilde{\Delta}_{\mathrm{lower-quar}} = 0$
consists of three points; see Figure~\ref{fig:c2-plane_e_K}.
One of them $c_{2-TR}$ is a triple point, 
another $c_{2-SG}$ a single point, and the other 
$c_{2-DB}$ a double point; they correspond to the zero locus of 
the first, second and the last factor 
of (\ref{eq:tilde_discriminant_lqp_factorize}), respectively. 
Three independent non-trivial loops are contained in this slice 
(Figure~\ref{fig:c2-plane_e_K}), and we study the monodromy of 2-cycles 
for these loops. 

Since the loops $\gamma_{2-TR}, \gamma_{2-SG}$ and $\gamma_{2-DB}$ are 
mapped, respectively, to $\gamma_{2-A}, \gamma_{2-3}$ and
$(\gamma_{2-1})^2$ by 
(\ref{eq:factorization_cond1}--\ref{eq:factorization_cond3}) topologically, 
the monodromies are given by $\rho(\gamma_{2-TR}) = \rho(\gamma_{2-A}) =
{\rm id}.$,
$\rho(\gamma_{2-SG}) = \rho(\gamma_{2-3})$ and
$\rho(\gamma_{2-DB}) = (\rho(\gamma_{2-1}))^2 = {\rm id}.$.
Therefore, the full monodromy group of the 2-cycles for these loops 
is generated only by 
\begin{equation}
 \rho(\gamma_{2-SG}) = W_{C_{A76}+C_{-\theta}} \simeq \Z_2.
\label{eq:mod_gauge_fctr}
\end{equation}
The monodromy group is reduced from the full Weyl group $S_3$ of
$\SU(3)_{\rm str}$ to its $\Z_2$ subgroup, and the 
2-cycle $C_{A76}$ is monodromy invariant. 

The reduction of the monodromy group is a direct consequence of the 
factorization limit (\ref{eq:2+1}); the $(27 a_0 a_3^2 + 4 a_2^3) = 0$ 
component of the monodromy locus factorized as 
in (\ref{eq:tilde_discriminant_lqp_factorize}), and one of the
irreducible pieces become multiplicity two. That is essential in 
making the monodromy $\rho(\gamma_{c_{2-DB}})$ trivial. 
The remaining monodromy is generated essentially around 
$(c_1^2 - 4 c_0 c_2) = 0$, which is precisely the ramification 
locus of the 2-fold spectral cover in the 2+1 factorization limit. 
All these results obtained in terms of monodromy of 2-cycles in 
elliptic K3 manifold agree with the expectation from the gauge theory 
description on 7+1 dimensions. 

\subsection{The ``Full'' Monodromy Group}
\label{subsec:str_sing}

In the previous subsection, we focused on the $a_2$-plane in the 
8D gauge theory region and reproduced the expected monodromy group 
by looking directly at the monodromy of 2-cycles. 
Thus, we can use the monodromy of the 2-cycles to study physics/geometry 
that cannot be described precisely in the gauge theory description 
on 7+1 dimensions.

As discussed in section \ref{sec:effect_beyond_gauge_theory}, 
the gauge theory description on 7+1 dimensions have two independent 
difficulties. The problem A was that one has no choice but to drop 
terms that either are higher order in the $z'$ coordinate expansion, 
or have coefficients suppressed by $\epsilon_K$, in order 
to fit the geometry into an $E_8$ gauge theory. This approximation 
corresponds to using (\ref{eq:discriminant_lqp}) instead 
of (\ref{eq:discriminant}) by dropping higher order terms. 
Thus, this problem of the gauge theory description can be overcome 
by simply repeating the same monodromy analysis 
for (\ref{eq:discriminant}) by keeping higher order terms.

The other difficulty of the gauge theory description, the problem B, 
was how to formulate a region near the $a_0 = 0$ locus in $S_{GUT}$.
In order to study the geometry of an elliptic K3 manifold near 
the $a_0 = 0$ region of the moduli space, we only have to follow 
loops that step into the $a_0 \simeq 0$ region and calculate the monodromy, 
just like we did for loops that stay within the 8D gauge theory region. 

In order to study the full monodromy in the scaling 
region (\ref{eq:scaling-rg}) without staying strictly in the 
8D gauge theory region (\ref{eq:gauge-th-rg}), it is no longer 
possible to maintain only the lower quadratic terms from 
(\ref{eq:discriminant}). 
As $|a_{0,*}|$ becomes much smaller than 1 and comes closer 
to $|\epsilon_K^2|$, one of the two 7-branes in the 
$z \sim {\cal O}(\epsilon_K^6 \epsilon_\eta)$ region behaves 
as 
\begin{equation}
 z_i \simeq - \frac{4}{27}\frac{a_2^3}{a_0^2} \sim 
  {\cal O}\left(\epsilon_\eta \epsilon_K^6 \right) \times 
  \frac{1}{a_{0,*}^2}.
\label{eq:A7-behavior}
\end{equation}
This behavior in terms of the discriminant locus corresponds
to (\ref{eq:shoot-spec-surf}) in terms of the spectral surface.
At the same time, the two 7-branes in the 
$z \sim {\cal O}(\epsilon_\eta)$ region move as 
$z_i \sim a_0 = \epsilon_\eta a_{0,*}$. 
Thus, the three 7-branes come close to one another when 
$a_{0,*}$ is as small as $\epsilon_K^2$. Those 7-branes are 
no longer clearly separated into the groups of [A6 + A7] and 
[A8 + D]. We should use at least quartic polynomial part of 
(\ref{eq:discriminant}). 

Monodromy of 2-cycles are (potentially) generated only around a locus 
in the moduli space where more than one 7-branes come on top of 
the other(s). Thus, the monodromy locus on the moduli space is 
characterized by the discriminant locus $\tilde{\Delta} = 0$ of 
a polynomial $\Delta'(z) = 0$ in (\ref{eq:discriminant}) that 
determines the positions of the 7-branes. 
$\tilde{\Delta} = 0$ defines a divisor in the moduli space. 
Monodromy of 2-cycles should be calculated for loops on the moduli 
space that stay away from the monodromy locus. Therefore, the 
full monodromy group is a representation of the fundamental group 
of the moduli space without the monodromy divisor. 
There is no essential difficulty in approaching the 
$a_0 \simeq 0$ region, or incorporating higher order terms of 
the $\Delta'(z)$ polynomial.

As long as we stay within the scaling region (\ref{eq:scaling-rg}) 
with an $a''_0 = \epsilon_\eta \neq 0$ gauge, however, the problem 
becomes a little easier. The two 7-branes [A8$'$ + D$'$] stay within 
the $z \sim {\cal O}(\epsilon_\eta^{-1})$ region, and are frozen out
there;  we thus only need to follow the behavior of the four other 7-branes. 
We can now use 
\begin{equation}
\begin{split}
\Delta'_{\eta} &= \frac{27}{16} a_3^4
 + \left( \frac{27}{2} a_0 a_3^2 + 4 a_2^3 \right) z
 + \left( \frac{27}{2} a_3^2 g_0 + 27 a_0^2 + 12 a_2^2 f_0 \right) z^2 \\
 & \quad + (54 a_0 g_0 + 12 a_2 f_0^2) z^3
  + (4 f_0^3 + 27 g_0^2) z^4 \, 
\end{split}
\label{eq:discriminant_four}
\end{equation}
instead of (\ref{eq:discriminant}) or (\ref{eq:discriminant_lqp}).

The monodromy divisor in the moduli space is the discriminant 
locus of $\Delta'_\eta(z) = 0$:
\begin{equation}
\tilde \Delta_{\eta} = - 19683 A^3 B = 0, 
\label{eq:tilde_discriminant}
\end{equation}
where $A$ and $B$ are given by
\begin{align}
A &= 4 a_0 a_2 f_0 - a_3^2 f_0^2 - 4 a_2^2 g_0 \: , \label{eq:A} \\
B &= 4 a_0^3 a_2^3 + 27 a_0^4 a_3^2 + 4 a_0 a_2^5 f_0
 + 30 a_0^2 a_2^2 a_3^2 f_0 - a_2^4 a_3^2 f_0^2 - 24 a_0 a_2 a_3^4 f_0^2 \notag
 \\
  & \quad  + 4 a_3^6 f_0^3 - 4 a_2^6 g_0 - 36 a_0 a_2^3 a_3^2 g_0
   - 54 a_0^2 a_3^4 g_0 + 18 a_2^2 a_3^4 f_0 g_0 + 27 a_3^6 g_0^2
   \: . \label{eq:B}
\end{align}
$\tilde{\Delta}_{\mathrm{lower-quadr}}$ in (\ref{eq:tilde_discriminant_lqp}) 
is reproduced\footnote{
The triple point $a_{2-A}$ and the three points $a_{2-1,2,3}$ 
on the $a_2$-plane are obtained as the roots of $A = 0$ and $B=0$;
all of $a_0$, $a_3$ and $g_0$ are treated as fixed numbers here. 
Although $B$ is an order-six polynomial of $a_2$, the remaining three 
roots are not in the scaling region $a_{2,*} \lesssim {\cal O}(1)$.} 
by keeping only the leading order terms in $\epsilon_K$  
in \eqref{eq:A} and \eqref{eq:B}:
\begin{align}
A_{\mathrm{leading}} &= 4 a_0 a_2 f_0 \, , \\
B_{\mathrm{leading}} &= a_0^3 (4 a_2^3 + 27 a_0 a_3^2) \label{eq:B_leading} \,.
\end{align}

\begin{figure}[tb]
 \begin{center}
\begin{tabular}{cc}
  \includegraphics[scale=0.25]{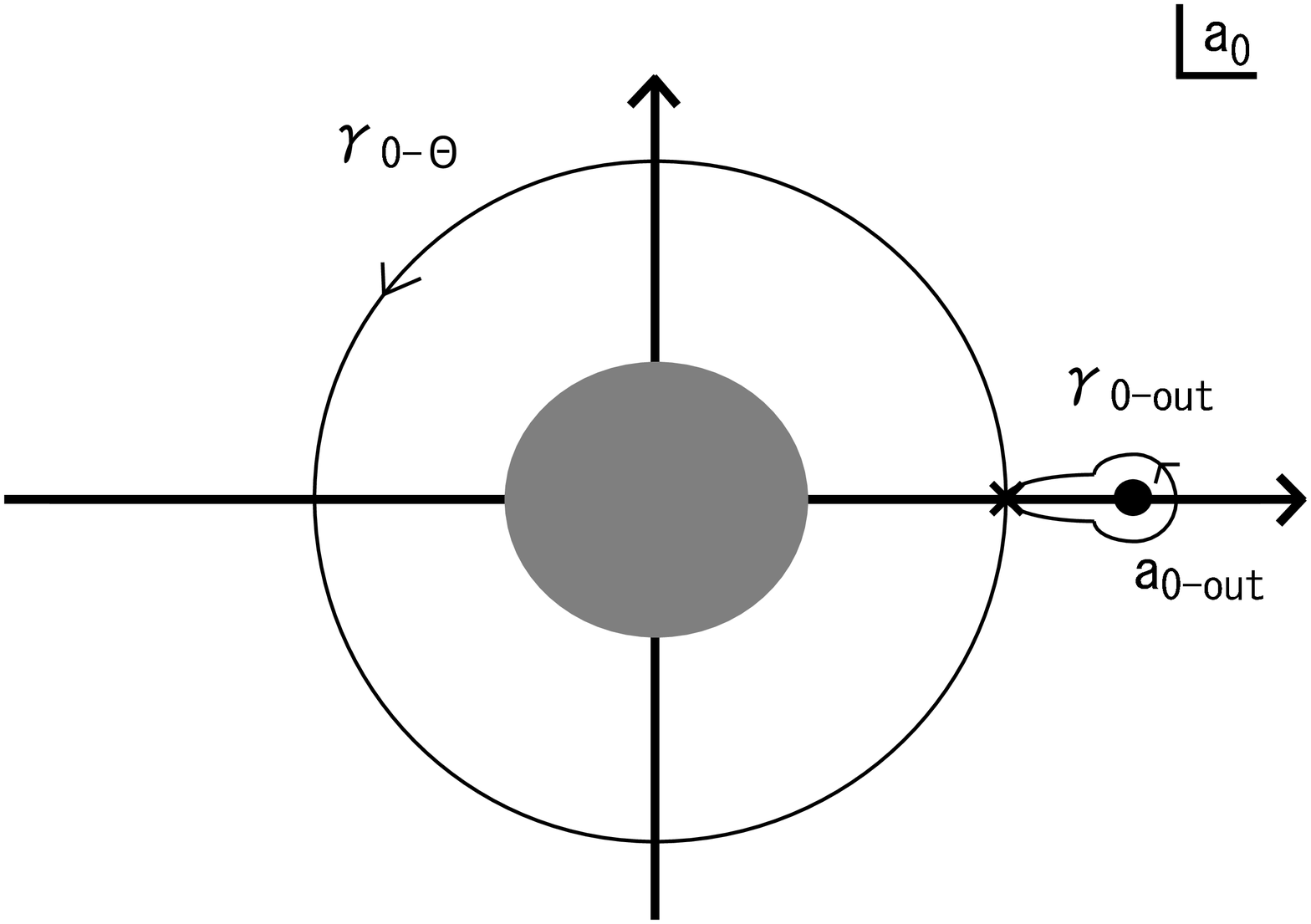} &  
  \includegraphics[scale=0.25]{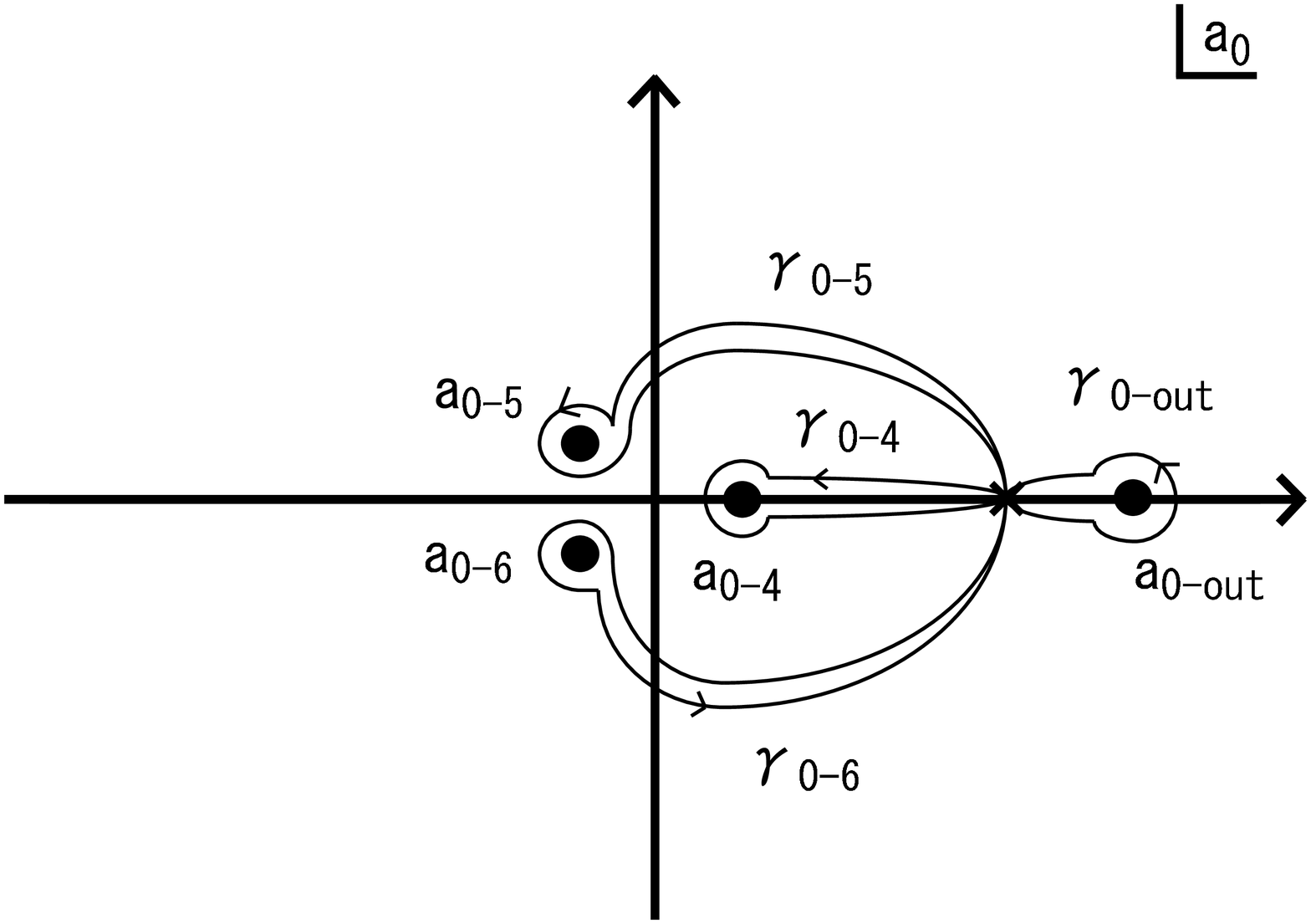} \\
 (a) & (b)
\end{tabular}
  \caption{The $B=0$ component of the monodromy locus consists of four points 
on the $a_0$-plane containing the base point (cross mark). Among them, 
only $a_{0-out}$ lies in the 8D gauge theory region 
$|\epsilon_K^2 \epsilon_\eta| \ll |a_0|$, 
and all others $a_{0-4,5,6} \sim {\cal O}(\epsilon_K^2 \epsilon_\eta)$ are 
located outside of the 8D gauge theory region (in a shaded region in (a)).  
}
\label{fig:a0-plane_e_K}
 \end{center}
\end{figure}

In order to see study the monodromy associated with 
the $a_0 \simeq 0$ region, it is useful to take a slice 
of the moduli space at constant $a_2$, $a_3$ and $g_0$ with 
a free $a_0$, and look at the distribution of the monodromy 
divisor in the complex $a_0$ plane.
We take a slice that contains the base point (\ref{eq:reference_point}). 
The $B = 0$ component of the monodromy locus appears in the 
$a_0$ plane as in Figure~ \ref{fig:a0-plane_e_K}.
%
There is only one monodromy locus $a_{0-out}$ in the 8D gauge theory region 
$|\epsilon_K^2 \epsilon_\eta| \ll |a_0|$, but three new types of
monodromy loci appear in the $a_0 \sim {\cal O}(\epsilon_K^2 \epsilon_\eta)$
region. 
We can think of various loops drawn in Figure~\ref{fig:a0-plane_e_K}; 
it is clearly a question of interest what the monodromy of 2-cycles 
will be for those loops.

Note that we do not have to consider the $A=0$ component of the
monodromy locus any more. In section \ref{subsec:mod_gauge}, we see
that the monodromy of the loop $\gamma_{2-A}$ is trivial. 
For any loops $\gamma_A$ in the moduli space (except $A=0$ and $B=0$)
that are homotopic to a loop of the form 
\begin{equation}
 	\gamma_{A}=\gamma_{B}^{-1}\circ\gamma_{2-A}\circ\gamma_{B}
\end{equation}
for some loop $\gamma_B$, $\rho(\gamma_A) = {\rm id}.$, because 
$\rho(\gamma_{2-A}) = {\rm id}.$. 
Thus, we focus only on the monodromy of the loops which go around the 
$B=0$ component (\ref{eq:B}) in the following. 

Among the loops in the $a_0$ plane in Figure~\ref{fig:a0-plane_e_K}, 
$\gamma_{0-out}$ and $\gamma_{0-\theta}$ stay within the 8D gauge theory 
region of the moduli space. Direct computation of the monodromy 
$\rho(\gamma_{0-out})$ shows that 
\begin{equation}
 \rho(\gamma_{0-out}) = W_{C_{A76}} = \rho(\gamma_{2-1}).
\label{eq:0-out}
\end{equation}
This is actually expected, because the loop $\gamma_{0-out}$ is
homotopic to $\gamma_{2-1}$ in the $(a_0, a_2, a_3, g_0)$ moduli space 
without the $B = 0$ monodromy locus. One can also see that 
$\gamma_{2-2} \sim \gamma_{0-\theta} \circ \gamma_{0-out} \circ 
\gamma_{0-\theta}^{-1}$ and 
$\gamma_{2-3} \sim \gamma_{0-\theta}^{-1} \circ \gamma_{0-out} \circ 
\gamma_{0-\theta}$, and hence the following relations  
\begin{eqnarray}
  \rho(\gamma_{0-\theta})\rho(\gamma_{0-out})\rho(\gamma_{0-\theta})^{-1}
 &=& \rho(\gamma_{2-2}) = W_{C_{A76}+C_{-\theta}}, \label{eq:0-theta1}\\
  \rho(\gamma_{0-\theta})^{-1}\rho(\gamma_{0-out})\rho(\gamma_{0-\theta})
 &=& \rho(\gamma_{2-3}) = W_{C_{A76}+C_{-\theta}}  \label{eq:0-theta2}
\end{eqnarray}
should hold true. We confirmed these relations by explicit computation
of the monodromy\footnote{\label{fn:0-theta}
The monodromy matrix $\rho(\gamma_{0-\theta})$
splits into a 2 by 2 block on {\rm Span}$_{\Z} \{C_{A76}, C_{-\theta}\}$ 
and a 4 by 4 block ${\rm Span}_{\Z} \{ C_{\alpha, \beta}^{1,2} \}$. 
It is a Weyl reflection $W_{C_{-\theta}}$ in the former. In the latter, 
$\tilde{C}_{\alpha, \beta}^{1} = C_{\alpha, \beta}^1$,  
$\tilde{C}_\alpha^2 = C_\alpha^2 - C_\beta^1$ and 
$\tilde{C}_\beta^2 = C_\beta^2 + C_\alpha^1$. The monodromy still splits  
between the visible $E_8$ sector and others for this loop 
$\gamma_{0-\theta}$.} 
$\rho(\gamma_{0-\theta})$. 
Other loops in Figure~\ref{fig:a0-plane_e_K} in the $a_0$ plane, 
which go away from the 8D gauge theory region of the moduli space, 
are not homotopic at least apparently to the loops whose monodromy 
we have already calculated. 

The full monodromy group of the 2-cycles is a representation of the 
fundamental group of the $(a_0, a_2, a_3, g_0)$ moduli space from which 
the $B=0$ monodromy locus is deleted, and this is what we are interested 
in. The monodromy group observed in the gauge theory description on 
7+1 dimensions, however, correspond to the representation of a subgroup 
generated only by loops that stay within the 8D gauge theory region 
of the moduli space. The monodromy group on the 2-cycles split into direct 
product of the one on ${\rm Span}_{\Z} \{C_{A76}, C_{-\theta}\}$ and 
the one on ${\rm Span}_{\Z} \{ C_{\alpha, \beta}^{1,2} \}$ at this 
level of analysis.
The loops such as $\gamma_{0-4,5,6}$ in Figure~\ref{fig:a0-plane_e_K}, 
however, may not be contained in this subgroup, and in general, the 
full monodromy group is larger than the expectation from the gauge 
theory description on 7+1 dimensions.  
Exactly the same thing can be said about the monodromy group 
of 2-cycles of a family of elliptic K3 manifold parametrized by 
$(c_0, c_1, c_2, g_0)$.

\subsection{The monodromy beyond the 8D gauge theory region}  
\label{subsec:monod_beyond_gauge}

We first show that the full monodromy group of the family
(\ref{eq:model}) with $(a_0, a_2, a_3, g_0)$ moduli space is 
indeed larger than the monodromy group expected from the gauge 
theory description on 7+1 dimensions. This is done by 
calculating the monodromy of the loops $\gamma_{0-4,5,6}$ in 
Figure~\ref{fig:a0-plane_e_K}; this is to probe physics and geometry 
of a region of small $a_0$, where the $E_8$ gauge theory description on 
7+1 dimensions breaks down. We then move on to study the monodromy group 
of the family for the factorization limit parametrized 
by $(c_0, c_1, c_2, g_0)$.

The monodromy for the loops $\gamma_{0,4,5,6}$ can be computed by 
the same method as in the preceding sections; the calculation for 
the loop $\gamma_{0-4}$ is demonstrated explicitly in the 
appendix \ref{sec:calculation}. It turns out that the 
monodromy of $\gamma_{0-4}$ is 
\begin{eqnarray}
\tilde{C}_{A65}&=&C_{A65},\;\;\;\;\;\; \tilde{C}_{A76}=C_{A76}+C_{-\theta}-C_{\alpha}^{1}, \nonumber \\
\tilde{C}_{\alpha}^{1}&=&C_{\alpha}^{1},\;\;\;\;\;\; \tilde{C}_{\alpha}^{2}=C_{\alpha}^{2}+C_{\alpha}^{1}-C_{-\theta}, \nonumber \\
\tilde{C}_{\beta}^{1}&=&C_{\beta}^{1},\;\;\;\;\;\; \tilde{C}_{\beta}^{2}=C_{\beta}^{2},
	\label{eq:0-4}
\end{eqnarray}
or equivalently, 
\begin{equation}
 \rho(\gamma_{0-4}) = 
  \left(
\begin{array}{cc|cc|c}
 1 &    &   &    & \\
 1 & -1 &   & -1 & \\
\hline
-1 &  2 & 1 &  1 & \\
   &    &   &  1 & \\
\hline
   &    &   &    & {\bf 1}_{2 \times 2}
\end{array}
  \right) \label{eq:0-4-mtrx}
\end{equation}
when we choose the basis as $(C_{A76}, C_{-\theta}, C_\alpha^1,
C_\alpha^2, C_\beta^1, C_\beta^2)$. The monodromy matrix 
$\rho(\gamma_{0-4})$ is trivial on the 2-cycles in the visible $E_6$ and 
hidden $E_8$. Similarly, the monodromy for $\gamma_{0-5}$ is given by 
	\begin{eqnarray}
	\tilde{C}_{A65}&=&C_{A65},\;\;\;\;\;\;\tilde{C}_{A76}=C_{A76}+C_{-\theta}-C_{\alpha}^{1}+C_{\beta}^{1},\nonumber \\
	\tilde{C}_{\alpha}^{1}&=&C_{\alpha}^{1},\;\;\;\;\;\; \tilde{C}_{\alpha}^{2}=C_{\alpha}^{2}+C_{\alpha}^{1}-C_{\beta}^{1}-C_{-\theta},\nonumber \\
	\tilde{C}_{\beta}^{1}&=&C_{\beta}^{1},\;\;\;\;\;\; \tilde{C}_{\beta}^{2}=C_{\beta}^{2}-C_{\alpha}^{1}+C_{\beta}^{1}+C_{-\theta},
	\label{eq:0-5}
	\end{eqnarray}
and, finally, the one for $\gamma_{0-6}$ is 
	\begin{eqnarray}
	\tilde{C}_{A65}&=&C_{A65},\;\;\;\;\;\;\tilde{C}_{A76}=C_{A76}+C_{-\theta}-C_{\beta}^{1},\nonumber\\
	\tilde{C}_{\alpha}^{1}&=&C_{\alpha}^{1},\;\;\;\;\;\; \tilde{C}_{\alpha}^{2}=C_{\alpha}^{2}, \nonumber \\
	\tilde{C}_{\beta}^{1}&=&C_{\beta}^{1},\;\;\;\;\;\; \tilde{C}_{\beta}^{2}=C_{\beta}^{2}+C_{\beta}^{1}-C_{-\theta}.
	\label{eq:0-6}
	\end{eqnarray}
The 2-cycles inside the $E_{8}$ root lattice and
$C_{\alpha,\beta}^{1,2}$  
are mixed up under the monodromy $\rho(\gamma_{0-4,5,6})$. 
This clearly shows that the loops $\gamma_{0-4,5,6}$ are not 
homotopic to the loops that stay within the 8D gauge theory region 
of the moduli space. More importantly, $\rho(\gamma_{0-4,5,6})$ 
should be added to the list of generators of the full monodromy 
group; the full monodromy group is no longer the product of the 
$\SU(3)_{\rm str}$ Weyl group $S_3$ on the visible $E_8$ and something 
acting on the four 2-cycles $C_{\alpha,\beta}^{1,2}$, but the full 
monodromy group is larger\footnote{
The only non-vanishing entries in the 4th and 6th rows of the 
monodromy matrices (\ref{eq:0-4}--\ref{eq:0-6}) are in the 
4th and 6th column, respectively, and are all ``$1$''.
This is an artifact of restricting the moduli space to 
the scaling region with $|\epsilon_\eta| \ll 1$. 
The period integrals over $C_{\alpha, \beta}^2$ are large and 
those for others small for $|\epsilon_\eta| \ll 1$. This is why 
$C_{\alpha, \beta}^{2}$ cannot mix into other 2-cycles. 
See also the appendix \ref{sec:Het}. 
When one considers the full monodromy group for a family without 
the restriction of $|\epsilon_\eta| \ll 1$, new generators will 
appear, and we expect that this special feature will disappear. }
than that.

The subject of our real interest is the monodromy group of 2-cycles 
at the factorization limit (\ref{eq:2+1}--\ref{eq:factorization_cond3}).
Although we have seen that the monodromy subgroup generated by loops 
in the constant $(c_0, c_1, g_0)$ slice in the 8D gauge theory region 
is the $\Z_2$ subgroup of the Weyl group $S_3$, there may be other 
generators in the full monodromy group, and the full monodromy group 
may be larger than $\Z_2$. To find out candidates for such a generator, 
let us take a constant $(c_1, c_2, g_0)$ slice of the moduli space and 
look at the complex $c_0$ plane. 

\begin{figure}[tb]
\begin{center}
\begin{tabular}{cc}
\includegraphics[scale=0.3]{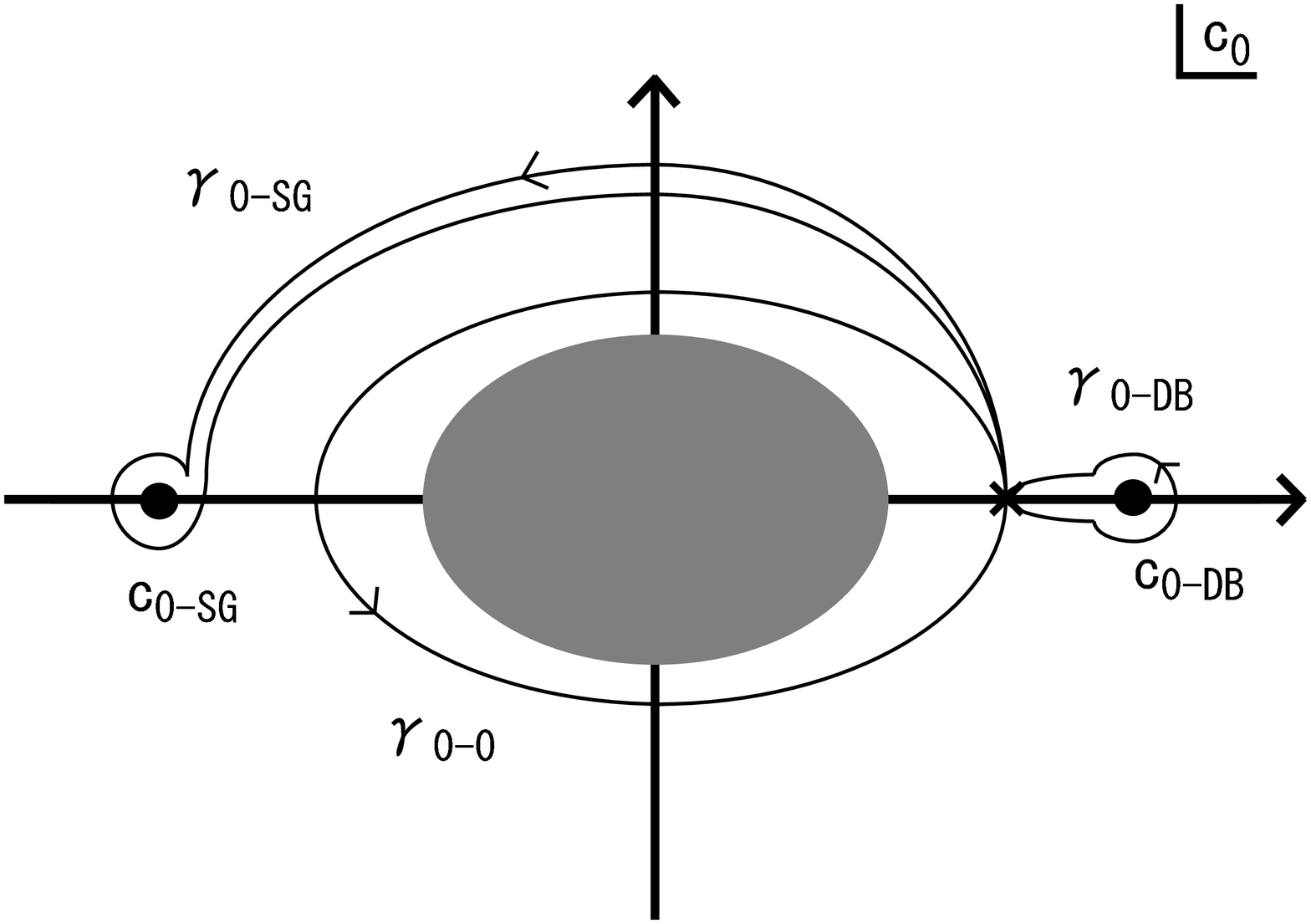} & 
\includegraphics[scale=0.3]{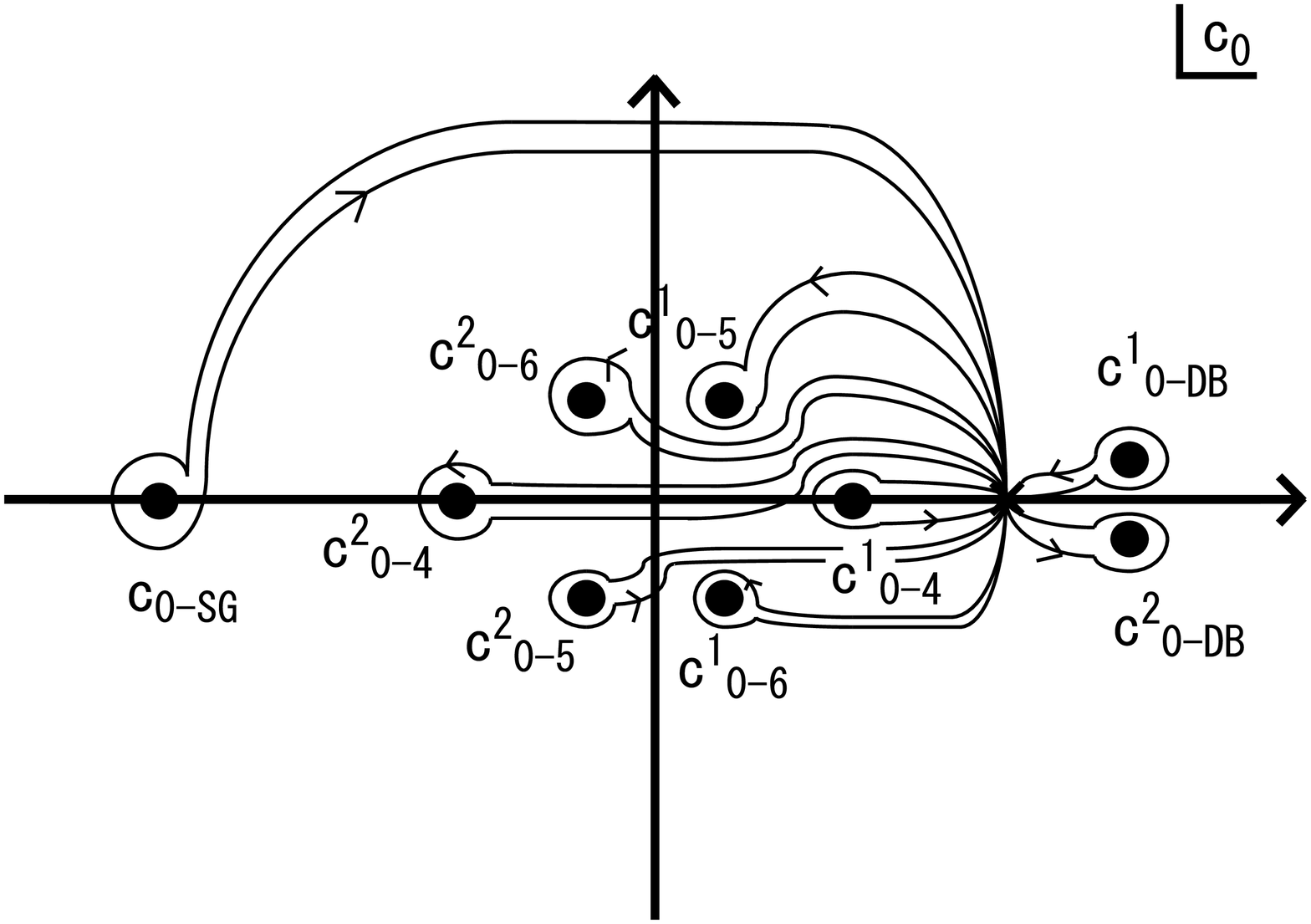} \\
(a) & (b)
\end{tabular}
\caption{The monodromy locus appearing in the constant $(c_1, c_2, g_0)$
 slice of the moduli space, and non-trivial loops that go around them. 
We have chosen a slice that contains the base 
point (\ref{eq:reference_point_c}).
The panel (a) is a coarse picture based on $B_{\rm leading} = 0$, where
higher order corrections in $\epsilon_K$ are ignored. The three loops, 
$\gamma_{0-DB}$, $\gamma_{0-SG}$ and $\gamma_{0-0}$ stay within the 
8D gauge theory region, away from the shaded region 
$|c_{0,*}| \lesssim \epsilon_K$. The panel (b) is a fine picture where 
higher order terms in $\epsilon_K$ are not dropped from $B = 0$. 
The $c_0 = 0$ monodromy point with the multiplicity 6 
in $B_{\rm leading} = 0$ (gray blob in the panel (a)) now splits 
into 6 distinct points; we can thus find six loops 
$\gamma_{0-4,5,6}^{1,2}$ in moduli space. The $c_{0-DB}$ monodromy point  
with multiplicity 2 in the panel (a) also splits into two points 
$c_{0-DB}^{1,2}$, and two loops $\gamma_{0-DB}^{1,2}$ are found.
}
\label{fig:c0-plane_e_K}
\end{center}
\end{figure}

The monodromy locus $B = 0$ in (\ref{eq:B}) should now be rewritten in 
terms of $c_0, c_1, c_2$. At the leading order in $\epsilon_K$, it
becomes
\begin{equation}
 B_{{\rm leading}} = (c_0)^6 (4 c_0 c_2 - c_1^2)(c_0 c_2 + c_1^2)^2,
\end{equation}
and hence it appears as in Figure~\ref{fig:c0-plane_e_K}~(a).
However, it was our motivation to study the geometry directly instead of 
gauge theory descriptions, to take account of higher order 
corrections in $\epsilon_K$, and also to study the region with small 
$c_0$. With all the higher order terms of $B = 0$ maintained, 
Figure~\ref{fig:c0-plane_e_K}~(b) is the precise picture of the 
monodromy locus appearing in the $c_0$ plane.
It is important to note that the double point $c_{0-DB}$ in 
Figure~\ref{fig:c0-plane_e_K}~(a) splits into two points,
$c_{0-DB}^{1,2}$ in Figure~\ref{fig:c0-plane_e_K}~(b), due 
to the higher order terms in $\epsilon_K$ in $B = 0$. 
%
%
The split in the $c_0$ plane is approximately 
%
\begin{equation}
	\Delta c_{0-DB}^{1,2}=\pm \frac{\sqrt{3f}}{2} c_{2} \sim
	 \mathcal{O}(\epsilon_{K}^{2}\epsilon_{\eta}^{\frac{1}{2}}), 
\end{equation}
which shows clearly that this is a higher order effect in $\epsilon_K$.
The split is certainly small, but still non-zero. 
Thus, the loops $\gamma^{1,2}_{0-DB}$ that go around either one of 
them can be separately non-trivial topologically in the moduli space; 
they may even become extra generators of the monodromy group. 
Each one of the $a_{0}=a_{0-4,5,6}$ monodromy locus points in the 
$a_0$ plane (Figure~\ref{fig:a0-plane_e_K}) splits into a pair 
$c_{0-4}^{1,2}, c_{0-5}^{1,2}, c_{0-6}^{1,2} \sim 
\mathcal{O}(\epsilon_{K}\epsilon_{\eta}^{\frac{1}{2}})$ 
in the $c_0$ plane, because of the factorization condition 
(\ref{eq:factorization_cond1}). 
There are six loops $\gamma_{0-4,5,6}^{1,2}$ that go around them 
in the $c_0$ plane, and they might also generate extra monodromy 
of the 2-cycles.

We found by numerical study that the motion of 7-branes in the $z$ plane 
along the loops $\gamma_{0-4,5,6}^{1,2}$ in the $c_{0}$-plane is 
topologically the same\footnote{This is easily guessed by the 
map (\ref{eq:factorization_cond1}--\ref{eq:factorization_cond3}).} 
as those along certain combinations of the loops $\gamma_{0-4,5,6}$ 
in the $a_{0}$-plane. Thus, the monodromy of the loops 
$\gamma^{1,2}_{0-4,5,6}$ in the $c_0$ plane is given by  
	\begin{eqnarray}
	\rho(\gamma_{0-4}^{1})&=&\rho(\gamma_{0-4}), \;\;\;\;\;\; \rho(\gamma_{0-4}^{2})=\rho(\gamma_{0-4}), \nonumber\\
	\rho(\gamma_{0-5}^{1})&=&\rho(\gamma_{0-5}), \;\;\;\;\;\; \rho(\gamma_{0-5}^{2})=\rho(\gamma_{0-4})\rho(\gamma_{0-5})\rho(\gamma_{0-4})^{-1}, \nonumber\\
	\rho(\gamma_{0-6}^{1})&=&\rho(\gamma_{0-6}), \;\;\;\;\;\; \rho(\gamma_{0-6}^{2})=\rho(\gamma_{0-4})^{-1}\rho(\gamma_{0-6})\rho(\gamma_{0-4}).
	\label{eq:mod_full}
	\end{eqnarray}
This clearly shows that the loops digging into the $c_0 \simeq 0$
region give rise to monodromy of 2-cycles that have not been observed 
in the gauge theory descriptions on 7+1 dimensions; 
2-cycles within the visible $E_8$ and those that are not are mixed up 
under the monodromy (\ref{eq:mod_full}).

An U(1) symmetry that survives all these monodromies correspond to 
a six-component row vector that remain invariant after multiplying 
any one of these monodromy matrices (\ref{eq:mod_gauge_fctr},
\ref{eq:mod_full}) from the right. There is none.
We therefore conclude that the U(1) symmetry remaining in the 
S[U(2)$\times$U(1)] Higgs bundle compactification is broken in the 
full geometry of F-theory compactification that has a region with 
$c_0 = 0$ on $S_{GUT}$. The problem B is indeed a problem.

To determine the monodromy associated with the loops
$\gamma_{0-DB}^{1,2}$, one only needs to note that the motion of 
7-branes in the $z$ plane along these loops 
are topologically the same as those along the loop $\gamma_{0-out}$ 
in the $a_{0}$-plane. Thus, 
\begin{equation}
	\rho(\gamma_{0-DB}^{1,2})=\rho(\gamma_{0-out})=W_{C_{A76}}.
	\label{eq:0-DB_1,2}
\end{equation}
These loops generate monodromy of 2-cycles in the visible $E_8$; 
combining this new generator and the one (\ref{eq:mod_gauge_fctr}) 
that we already know in the gauge theory description on 7+1 dimensions, 
the whole $S_3$ Weyl group is generated. 
We have thought that the monodromy of 2-cycles is reduced from 
$S_3$ to $\Z_2$ in the factorization limit of the spectral surface, 
but because of the $\epsilon_K$-suppressed higher order terms that 
were simply ignored in the 8D gauge theory description, actually 
the monodromy is not reduced from $S_3$.
We conclude that there is no unbroken U(1) symmetry with non-trivial 
components in the visible $E_8$ that survives the monodromy generated 
by both (\ref{eq:mod_gauge_fctr}) and (\ref{eq:0-DB_1,2}). 
The problem A is a problem indeed. 

To summarize,\footnote{
It is useful for sanity check to exploit the relations following 
from the homotopy equivalence of various loops in the moduli space, 
as we did in the $(a_0, a_2)$ moduli space in section \ref{subsec:str_sing}.
In the $(c_0, c_2)$ moduli space, we have relations  
\begin{eqnarray}
& & 
\gamma_{2-DB} \sim \gamma_{0-DB} \sim \gamma_{0-DB}^1 \circ
\gamma_{0-DB}^2, \label{eq:hom-1} \\
& & 
\gamma_{2-SG} \sim \gamma_{0-SG}, \label{eq:hom-2} \\
& & 
\gamma_{0-6} \circ \gamma_{0-4} \circ \gamma_{0-5} \sim \gamma_{0-\theta}, 
 \label{eq:hom-3}\\
& & 
\gamma_{0-6}^1 \circ \gamma_{0-5}^2 \circ \gamma_{0-4}^1 \circ 
\gamma_{0-4}^2 \circ \gamma_{0-6}^2 \circ \gamma_{0-5}^1 \sim
\gamma_{0-0}. \label{eq:hom-4}
\end{eqnarray}
Our results in section \ref{subsec:mod_gauge} and (\ref{eq:0-DB_1,2}) 
both consistently yield $[W_{C_{A76}}]^2 = [\rho(\gamma_{2-1})]^2$ 
in (\ref{eq:hom-1}).
Direct computation of $\rho(\gamma_{0-SG})$ indeed turned out to be 
the same as $\rho_{2-SG}$ in (\ref{eq:mod_gauge_fctr}). 
The homotopy relation (\ref{eq:hom-3}) in the monodromy matrices 
is confirmed directly by using the results (\ref{eq:0-4}--\ref{eq:0-6}) 
and that in footnote \ref{fn:0-theta}.
Finally, the product of (\ref{eq:mod_full}) in the order specified in 
the left-hand side of (\ref{eq:hom-4}) becomes 
$[\rho(\gamma_{0-6})\rho(\gamma_{0-4})\rho(\gamma_{0-5})]^2$, which 
is equal to $[\rho(\gamma_{0-\theta})]^2$, because of (\ref{eq:hom-3}). 
This should be the same as $\rho(\gamma_{0-0})$, because 
the loop $\gamma_{0-0}$ in the $c_0$ plane is mapped to 
$\gamma_{0-\theta} \circ \gamma_{0-\theta}$ in the $a_0$ plane 
through (\ref{eq:factorization_cond1}).
All these consistency checks as a whole provides confidence in the results 
of our calculation. } 
the full monodromy group contains at least 
(\ref{eq:mod_gauge_fctr}), 
(\ref{eq:mod_full}) and (\ref{eq:0-DB_1,2}) as generators, maybe more, 
when the spectral surface is in the factorization limit 
(\ref{eq:factorization_cond1}--\ref{eq:factorization_cond3}). 
There is no unbroken U(1) symmetry with non-trivial components 
in the visible $E_8$ under the full monodromy group. 
Thus, we cannot expect an unbroken U(1) symmetry in the low-energy 
effective theory in ruling out the dimension-4 proton decay operators. 

\subsection{Non-K3-fibred 4-fold $X_4$}
\label{subsec:non-K3}

We have so far 
used a Calabi--Yau 4-fold $X_4$ that is a K3-fibration on $S_{GUT}$. 
However, this is just for a concreteness. The monodromy analysis 
certainly needs to be phrased for individual cases for $X_4$'s that 
are not K3-fibration over $S_{GUT}$. But the essence of the problem A 
and B, and the essence of analysis remain the same. 
Let us take simplest non-K3 fibered models as examples: 
$B_3 = \P^3$, and $S_{GUT}$ is a quadratic ($d = 2$) or cubic ($d=3$) 
surface of $B_3$, the models discussed in \cite{BHV-2}.
First, $A_k$ in (\ref{eq:local-defeq}) are set to be \cite{Caltech-0904}
\begin{equation}
A_k = s^{k-1} \tilde{A}_k, \qquad 
\tilde{A}_{k} \in 
\Gamma(B_3; {\cal O}_{B_3}( - k K_{B_3}-(k-1)S_{GUT})), 
\label{eq:global-choice-Ak}
\end{equation}
where $s$ is a global holomorphic section of 
${\cal O}_{B_3}(S_{GUT}) \simeq {\cal O}_{\P^3}(d H)$ whose 
zero locus is the GUT divisor $S_{GUT}$. $a_{6-k}$ on $S_{GUT}$ in 
(\ref{eq:Ak-expnsn}) correspond to $\tilde{A}_k|_{S_{GUT}}$. 
Suppose that a point $p_0 \equiv [0:0:0:1] \in \P^3$ is not contained in 
$S_{GUT}$. The base manifold except this point, $\P^3 \backslash p_0$ 
is covered by three patches $\simeq \C^3$, which constitute 
${\cal O}_{\P^2}(1)$. Restriction of the original elliptic fibration 
over $B_3 = \P^3$ to that over ${\cal O}_{\P^2}(1)$, combined with 
a projection $\pi_0: {\cal O}_{\P^2}(1) \rightarrow \P^2$ defines 
a complex surface fibration 
\begin{equation}
\pi_0 \circ \pi_X: \pi^{-1}_X \left( {\cal O}_{\P^2}(1) \right) 
  \rightarrow \P^2. 
\label{eq:nonK3-fib}
\end{equation}
The fiber of this map is an elliptic fibration over $\C$, with 48 
discriminant points on the $\C$ plane. Although the number of 
$[p, q]$ 7-branes is not the same as in the case of an elliptic K3 
manifold, independent topological 2-cycles and their intersection form 
of the fiber complex surface can be worked out by using the techniques 
in \cite{jcn-intersection, jcn-KcMd}.
Monodromy can be studied for loops\footnote{Since 
$\pi_0|_{S_{GUT}}: S_{GUT} \rightarrow \P^2$ is a $d$-fold covering, 
one can pull-back the complex-surface fibration (\ref{eq:nonK3-fib}) from 
$\P^2$ to $S_{GUT}$. The monodromy analysis can then be phrased for 
loops in $S_{GUT}$.} in $\P^2$. 
When $d \neq 1$, other points $p_i$'s in $B_3 = \P^3$ should also 
be chosen so that the analysis for 
$\pi_i: \P^3 \backslash p_i \simeq {\cal O}_{\P^2} (1) \rightarrow \P^2$
is carried out and all the ramification locus of the projection 
$\pi_i|_{S_{GUT}}: S_{GUT} \rightarrow \P^2$ is covered by some of the 
analysis of $\P^3 \backslash p_i$.
Generalization from these examples to, e.g., toric $B_3$ (c.f. \cite{Yau}), 
is straightforward. 

The problem A arises from the difference between 
$a_{6-k}=\tilde{A}_k|_{S_{GUT}}$ and $\tilde{A}_k$, and the problem B arises 
essentially because of the local behavior of 2-cycles wherever $a_0$ 
vanishes. Thus, we expect similar phenomena also in the case of 
$B_3 = \P^3$, though we have not done the analysis. Since more than 
one points of $S_{GUT}$ are projected onto the same point in $\P^2$ for 
$d \neq 1$, 2-cycles in $E_8$ on a point of $S_{GUT}$ may mix with 2-cycles 
in $E_8$ on another point of $S_{GUT}$ in the context of problem B. 

\section{Consequences in Physics}
\label{sec:physics}

\subsection{U(1) Violating State-Mixing and Trilinear Couplings}
\label{subsec:M2brane}

We have studied a model of $E_8 \rightarrow E_6$ symmetry breaking with 
a spectral surface in the 2+1 factorization limit in the previous
section, as a toy model of $E_8 \rightarrow {\rm SU}(5)_{\rm GUT}$ 
symmetry breaking in a certain factorization limit. On the contrary 
to the expectation in gauge theory in 7+1 dimensions, the study shows 
that there is no monodromy-invariant 2-cycle, and hence there is no 
unbroken U(1) symmetry in the low-energy effective theory below 
the Kaluza--Klein scale. 
Without an unbroken U(1) symmetry, we generally expect that dimension-4 
proton decay operators will be generated. In this section, we discuss 
whether the dimension-4 proton decay operators are really generated from 
known interactions in string theory.

In F-theory, charged matter fields come from the degree of freedom of 
M2-branes wrapping on the 2-cycles corresponding to the representation 
of charged matters. 
Trilinear interactions are (likely to be) generated, if the sum of 2-cycles for 
the three fields is topologically trivial. 
Thus, by using this criterion, we can check whether dimension-4 proton
decay operators are generated or not.
	
\begin{figure}[tb]
\begin{center}
\includegraphics[scale=0.3]{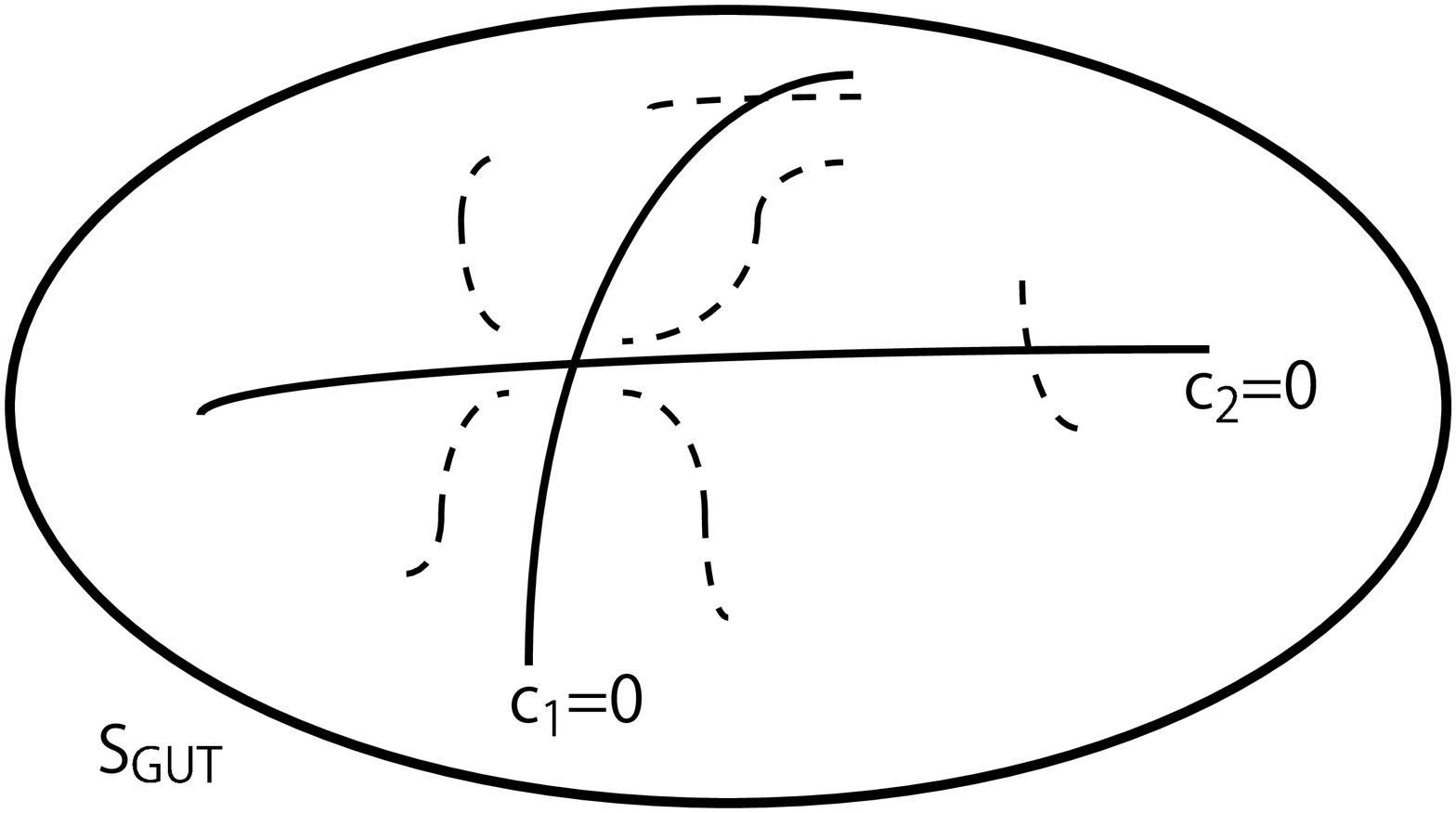}
  \caption{\label{fig:S-GUT-E6}A schematic picture of matter curves 
and monodromy loci on $S_{GUT}$. The ramification locus 
$c_1^2 - 4 c_2 c_0 \simeq 0$ and a locus $(2c_1^2 + c_2 c_0) \simeq 0$ 
with multiplicity 2 both approach the points of enhanced $E_8$ 
singularity. The multiplicity-2 branch, however, has a small split, and 
consists of two branches with multiplicity 1. 
Furthermore, all the three branches form a single irreducible component 
monodromy locus $B=0$ around $(c_1, c_2) \simeq (0,0)$. The monodromy 
locus $B = 0$ also intersect the matter curves at $(c_1,c_0) \simeq (0,0)$ 
and $(c_2,c_0) \simeq (0,0)$. }
\end{center}
\end{figure}

In the $E_8 \rightarrow E_6$ symmetry breaking model with the 2+1
factorization, the matter curve for $E_6$-{\bf 27} representation 
splits into two irreducible pieces; one is characterized by 
$c_2 = 0$, and the other by $d_1 = - c_1 = 0$. Based on the gauge theory 
description on $S_{GUT}$, one would expect that the charged matter 
fields in the $E_6 \times {\rm U}(1)$-${\bf 27}_{+1}$ representation 
are localized 
along the $c_2 = 0$ curve in $S_{GUT}$, and those in the ${\bf 27}_{-2}$
along the $c_1 = 0$ curve. Ignoring the monodromy that we studied in 
section \ref{subsec:monod_beyond_gauge}, the 2-cycle $(C_{A76}+C_{A65})$ 
vanishes along the $c_1 = 0$ curve, while either $C_{A65}$ or
$(C_{-\theta}+C_{A76}+C_{A65})$ does along $c_2 = 0$. 
$4 c_0 c_2 - c_1^2 = 0$ is the ramification locus of the 2-fold 
spectral cover, where monodromy $W_{C_{A76}+C_{-\theta}}$ exchanges $C_{A65}$ and 
$(C_{-\theta}+C_{A76}+C_{A65})$. See Figure~\ref{fig:S-GUT-E6}. 
The trilinear interaction 
$\Delta W = {\bf 27}_{+1} \cdot {\bf 27}_{+1} \cdot {\bf 27}_{-2}$ 
is generated at $E_8$ singularity enhancement points $c_1 = c_2 = 0$, 
because all the three 2-cycles vanish simultaneously there, and 
they satisfy 
\begin{equation}
 (C_{-\theta}+C_{A76}+C_{A65}) + C_{A65} + (C_{A76}+C_{A65}) \equiv 0
  \quad ({\rm mod~}E_6 ).
\label{eq:cycle-sum}
\end{equation}
This Yukawa coupling is invariant under the U(1) symmetry.

We know, however, that there are more monodromy among the 2-cycles.
Under the monodromy (\ref{eq:0-DB_1,2}), the 2-cycle $C_{A65}$ for 
${\bf 27}_{+1}$ turns into $(C_{A76}+C_{A65})$ for ${\bf 27}_{-2}$ and 
vice versa. The matter fields in the $E_6$-${\bf 27}$ representation 
cannot remain pure eigenstates of the U(1) symmetry. This mixing 
among states with different U(1) charges (and hence the monodromy) 
takes place around the monodromy locus 
\begin{equation}
 (c_{0,*})^4 (c_{0,*} c_{2,*} + 2 c_{1,*}^2)^2 \simeq 
 12 f_0 (\epsilon_K)^4 (c_{1,*}^2)^4, 
\end{equation}
which is a part of $B=0$ locus, just like the ramification locus 
$4c_0c_2-c_1^2 \simeq 0$.
Since this branch also approach the $c_1 = c_2 = 0$ point just like 
the ramification locus (see Figure~\ref{fig:S-GUT-E6}), we do not 
have a reason to believe that the mixing (monodromy) due to this 
branch cancels, while that of the ramification locus remain.
The mixing may be suppressed by some function of $\epsilon_K$, 
but we cannot take $\epsilon_K$ to be literally zero 
in (\ref{eq:eK-scale}, \ref{eq:coordinate-scale}, \ref{eq:scaling-rg}, 
\ref{eq:scaling-rg-c}) in $\SU(5)_{\rm GUT}$ models.
Therefore, all sorts of U(1) violating trilinear interactions 
$\Delta W = {\bf 27} \cdot {\bf 27} \cdot {\bf 27}$ will be 
generated, unless there is a cancellation. 

There are other groups of regions in $S_{GUT}$ where the matter curves 
$c_1 = 0$ or $c_2 = 0$ encounters the $B = 0$ monodromy locus.
One is $c_0 \simeq 0$ on $c_1 \simeq 0$ ($c_{2,*} \neq 0$), and the other 
is $c_0 \simeq 0$ along $c_2 \simeq 0$ ($c_{1,*} \neq 0$). The monodromy 
we studied in section \ref{subsec:monod_beyond_gauge} is relevant to 
the latter. 
The 2-cycle $(C_{-\theta}+C_{A76}+C_{A65})$ for ${\bf 27}_{+1}$ and 
$(C_{A76}+C_{A65})$ for ${\bf 27}_{-2}$ are exchanged under the
monodromy (\ref{eq:mod_full}) modulo 2-cycles $C_{\alpha, \beta}^{1,2}$.
Thus, the ``${\bf 27}_{+1}$'' fields will have non-vanishing 
$[C_{A76}+C_{A65}]$ (mod $C_{\alpha, \beta}^{1,2}$) component 
and vice versa. The mixing, however, will presumably be suppressed 
exponentially, because it is 2-cycles of finite size, rather than 
vanishing 2-cycles, that are exchanged under the monodromy at 
$c_0 \simeq 0$.

The U(1) symmetry violating Yukawa couplings ${\bf 27} \cdot {\bf 27}
\cdot {\bf 27}$ are generated only when the sum of three topological 
2-cycles vanish in $H_2({\rm K3}; \mathbb{Z})$, just like 
in (\ref{eq:cycle-sum}). Thus, just a single monodromy out of
(\ref{eq:mod_full}) acting on $(C_{-\theta}+C_{A76}+C_{A65})$ does not 
generate such a U(1) violating coupling ${\bf 27}_{}$; 
\begin{equation}
 (C_{-\theta}+C_{A76}+C_{A65}) + C_{A65}
  +(\tilde{C}_{-\theta}+\tilde{C}_{A76}+\tilde{C}_{A65}) \neq 0
\end{equation}
in $H_2({\rm K3}; \mathbb{Z})$ (mod $E_6$). After exploiting all the 
monodromies available in a 4-fold $X_4$ that is a K3 fibration on 
$S_{GUT}$,\footnote{Note, for example, that both $f_0$ and $g_0$ vanish 
at some points on $S_{GUT}$, where one cannot say $\epsilon_\eta$ is 
small. Although we did not study monodromy associated with non 
$|\epsilon_\eta| \ll 1$ region, generically we should expect such a
monodromy in a compact model.} however, all $C_{\alpha, \beta}^{1,2}$ 
will be mixed up and twisted over $S_{GUT}$, and eventually such a
U(1) violating coupling will be generated, although the coupling 
constant may be highly suppressed. In the case of non-K3 fibred $X_4$, 
$H_2({\rm K3}; \Z)$ just has to be replaced by $H_2$ of the 
complex-surface fibration of (\ref{eq:nonK3-fib}). 

The renormalizable proton decay operators are not the only consequence 
of the U(1) symmetry-breaking state mixing. Factorized spectral surface 
limit has been used for dimension-5 proton decay problem \cite{Caltech-0906}, 
or for the purpose of exploiting \cite{Hayashi-3, Cordova}
the idea of flavor structure in \cite{HV-08Nov, Harvard-nonC, 
Conlon, Marchesano}.
The discussion so far implies that it is impossible to confine 
$\bar{\bf 5}_H$ and ${\bf 5}_H$ to separate irreducible pieces of 
$\SU(5)_{\rm GUT}$ matter curves completely. This may not be a 
big problem, as opposed to the dimension-4 proton decay problem, 
however, because the dimension-5 proton decay problem only requires 
a small amount of suppression; complete separation is not necessary.
In the context of flavor structure of Yukawa couplings, the three copies 
of the visible sector matter fields $\bar{\bf 5}_M$ or 
${\bf 10}_M$ may not be able to have wavefunctions strictly in 
a single irreducible component of the factorized matter curves, even 
at the factorization limit of the spectral surface. This means 
that the up-type Yukawa matrix of the low-energy effective theory below 
the Kaluza--Klein scale receives contributions from more than 
one point of enhanced $E_6$ singularity \cite{Hayashi-3, Cordova}. If the 
state mixing is small, then this mixing is not terribly bad, and may 
even contribute in generating smaller eigenvalues; the flavor structure
generated in this way is not necessarily similar to the one we observe 
in the Standard Model, however. 

It should be remembered, however, that our analysis employed a 2+1 
factorization in the $E_8 \rightarrow E_6$ symmetry breaking. 
Thus, it will not be a terribly bad guess to expect similar results 
for the 4+1 and 3+2 factorization in the $E_8 \rightarrow \SU(5)_{\rm GUT}$ 
symmetry breaking \cite{TW-1, Tsuchiya, Caltech-0906}. 
When another type of factorization (or monodromy subgroup of $S_5$) is 
employed, as in \cite{Harvard-nu, Harvard-E8}, separate study is necessary, 
especially because it is not clear how the problem A will look like 
in such a factorization limit. 

\subsection{Loopholes in This Argument}
\label{subsec:loophole}

The discussion so far hints that the dimension-4 proton decay operators 
are generated in the factorized spectral surface scenario. There are
some loopholes in the argument, however. At the end of this article, 
here, we list up some loopholes that come to our minds. The list can 
also be taken as possible ways to save the factorized spectral surface
solution. 

First of all, the argument so far only showed that 
\begin{itemize}
 \item there is no unbroken 
U(1) symmetry (apart from exceptional cases that we discuss later) in 
the low-energy effective theory that we hope would exclude the
dimension-4 proton decay operators, and 
 \item the picture of interactions as recombination of M2-branes without
       a change in the total topology does not exclude the U(1)
       violating operators. 
\end{itemize}
It is not that we have calculated the coefficients of such operators, 
and in fact, we do not even have a theoretical framework to calculate 
the coefficients within F-theory. A gauge theory on 7+1 dimensions 
cannot handle this. A possible direction is to exploit the Type
IIB--M-theory duality or F-theory--Heterotic duality; dual descriptions 
may be used to see whether cancellation mechanism is likely to exist, 
or to make an estimate of the coefficients. That will tell us how small 
$\epsilon_K$ should be. 
Such a study is beyond the scope of this article, however.

If one literally sets $\epsilon_K = 0$, instead of fine-tuning it to be 
sufficiently small, then $S_{GUT}$ becomes a locus of $E_8$ singularity. 
If a vector bundle on $S_{GUT}$ has a structure group that is smaller 
than $\SU(5)_{\rm str} \times \U(1)_Y$ so that an extra U(1) factor is 
contained, then an unbroken U(1) symmetry remains in the low-energy 
effective theory and prevent proton decay. The chirality of various 
charged matter fields in this case, however, are determined simply 
by intersection numbers of first Chern classes of those bundles and 
$K_{S_{GUT}}$ \cite{TW-1, DW-1, BHV-1, Hayashi-1}, and existence of 
exotics is predicted easily \cite{TW-1, BHV-2, Caltech-0906}.
It is also obvious that theories of flavor structure like those 
in \cite{Hayashi-2, HV-08Nov, Font, Harvard-nonC, Hayashi-3, Cordova} 
are not applied to such a case, because only $A_4$ singularity 
is assumed along $S_{GUT}$.

Secondly, one will notice that our analysis in section \ref{sec:mod} 
is based on a 
choice (\ref{eq:factorization_cond1}--\ref{eq:factorization_cond3}) of 
the solution to (\ref{eq:vanish-a1}). The condition (\ref{eq:vanish-a1}), 
however, can be solved in the form of 
\begin{equation}
 c_0 = p s, \quad d_1 = - q r, \quad c_1 = p r, \quad d_0 = q s 
\end{equation}
for some global holomorphic sections 
\begin{eqnarray}
& & s \in \Gamma(S_{GUT}; {\cal O}_S(\eta_0)), \quad 
 p \in \Gamma(S_{GUT}; {\cal O}_S(\eta'_1)), \quad 
 q \in \Gamma(S_{GUT}; {\cal O}_S(\eta'_2)), \\
& & r \in \Gamma(S_{GUT}; {\cal O}_S(\eta_0 + K_{S_{GUT}}));
\end{eqnarray}
here, $\eta'_{1,2}$ and $\eta_0$ are some divisors on $S_{GUT}$, and 
$c_2 \in \Gamma(S_{GUT}; {\cal O}_S(\eta'_1 + \eta_0 + 2 K_{S_{GUT}}))$.
What we studied in section \ref{sec:mod} is the monodromy associated 
with $s=0$ locus. 

What if the global section $s$ does not have a zero locus?
This is possible if the line bundle ${\cal O}_S(\eta_0)$ is trivial; 
$s$ can now take a constant non-zero value over the entire $S_{GUT}$. 
Suppose that this topological condition is satisfied. If we further 
assume that the divisor $K_{S_{GUT}}$ is not effective, as in del Pezzo 
surfaces and Hirzebruch surfaces, then there is no non-trivial 
global holomorphic section $r$. We have to set $r=0$.
This means that 
\begin{equation}
c_1 = d_1 = 0 
\end{equation}
over the entire $S_{GUT}$, which is nothing but the case we already
listed as the rank-5 GUT scenario (ii) in Introduction. The dimension-4 
proton decay problem can be solved completely in this scenario, because 
of an unbroken (or spontaneously broken) U(1) symmetry in the low-energy 
effective theory; an extra 2-cycle is along $S_{GUT}$, and a semi-local 
geometry of this form along $S_{GUT}$ is sufficient in ensuring the proton 
stability \cite{TW-1}. The singularity along $S_{GUT}$ is either $D_5$ or $A_5$ 
in this case, and theories of flavor structure \cite{Hayashi-2, HV-08Nov, 
Font, Harvard-nonC, Hayashi-3} are not applied here. Exotic-matter 
free conditions have also been studied for rank-5 GUT 
scenarios \cite{BHV-2, Chung}.

If the divisor $K_{S_{GUT}}$ is effective, on the other hand, 
we can introduce a different set of topological conditions: all of 
line bundles ${\cal O}_S(\eta_0)$ and ${\cal O}_S(\eta'_{1,2})$ are  
trivial, so that there is no zero locus in $s$, $p$ and $q$.
This is possible for an effective $K_{S_{GUT}}$, because the matter 
curves belong to topological classes of effective divisors. 
Now there is no $a_0 = 0$ locus, and at least the problem B is gone, 
under this set of topological conditions. 

An obvious loophole in the argument that follows (\ref{eq:spec-surf-1})
is that the coefficient of the highest degree term $a_{5-k}$ may not 
have a zero locus. We have already exploited a case that $a_0$
is constant and non-zero, and a remaining alternative is a case\footnote{
\label{fn:ex}
As an example, one can consider an elliptic fibration 
$\pi_X : X_3 \rightarrow B_2 = \P^2$, and take $S_{GUT}$ to be the zero 
locus of a homogeneous function $s$ of degree 4 on $B_2= \P^2$. 
The line bundle for $\tilde{A}_4$ is trivial on $B_2$, and hence 
$a_2 = \tilde{A}_4|_{S_{GUT}}$ has to be constant on $S_{GUT}$. 
One will also see that $\tilde{A}_6 = 0$ everywhere on $B_2$. 
The divisor $K_{S_{GUT}}$ on $S_{GUT}$ is effective. }
where $a_2$ is constant and non-zero everywhere on $S_{GUT}$. 
This is possible only when $K_{S_{GUT}}$ is effective on $S_{GUT}$, 
and $a_0 \in \Gamma(S_{GUT}; {\cal O}_S(- 2 K_{S_{GUT}}))$ vanishes 
as a consequence.  An $E_7$ gauge theory\footnote{$E_7$ is a minimal choice  
in obtaining all the matter fields and their trilinear interactions of 
the supersymmetric standard Models \cite{TW-1, Bourjaily-1, Harvard-E8}. 
Thus, it is not an option for realistic GUT models to assume 
that $a_3$ is constant and non-zero over $S_{GUT}$.} is well-defined over the 
entire $S_{GUT} \times \R^{3,1}$. There is no mixing between 2-cycles 
in $E_7$ and those outside, and the problem~B is gone in this case. 
What is more intriguing in this model is that there is no need to impose 
a factorization condition like (\ref{eq:2+1}) by hand. The commutant 
of $\SU(5)_{\rm GUT}$ in $E_7$ is $\U(3)$. The decomposition 
\begin{eqnarray}
 {\rm Res}^{E_7}_{\U(3) \times \SU(5)_{\rm GUT}} {\bf adj.} & = & 
 ({\bf adj.}, {\bf 1}) + ({\bf 1}, {\bf adj.}) \nonumber \\
 & + &  
 \left[ ({\bf 3}, \overline{\bf 10}) + (\wedge^2 {\bf 3}, {\bf 5}) 
        + (\wedge^3 {\bf 3}, \bar{\bf 5}) \right] + {\rm h.c.}
\end{eqnarray}
shows that there are two different kinds of charged matter fields 
in the $\SU(5)_{\rm GUT}$-$\bar{\bf 5}$ representation. For a fully 
generic 3-fold spectral cover globally defined on $S_{GUT}$, 
$({\bf 3},\overline{\bf 10})+(\bar{\bf 3},{\bf 10})$ matter fields 
are localized along the $a_5=0$ locus, while  
$(\wedge^2 {\bf 3}, {\bf 5})+{\rm h.c.}$ and 
$(\wedge^3 {\bf 3}, \bar{\bf 5})+{\rm h.c.}$ matter fields are localized 
along the $(-a_2 a_5 + a_4 a_3)=0$ and $a_3=0$ loci, respectively. 
This factorization / decomposition automatically takes place under this 
topological condition.\footnote{We thank Cumrun Vafa for discussion.}
Because of the commutation relation of the $E_7$ Lie algebra, 
$(\wedge^2 {\bf 3}, {\bf 5})$ is identified with the origin of the 
up-type Higgs multiplet. If $(\wedge^3 {\bf 3}, \bar{\bf 5})$ and 
$(\wedge^2 \bar{\bf 3}, \bar{\bf 5})$ are identified with 
$\bar{\bf 5}_H$ and $\bar{\bf 5}_M$, respectively,\footnote{The exotic 
free condition and successful doublet--triplet splitting cannot be 
realized simultaneously in this identification, however \cite{Caltech-0906}. 
If the two components are identified in the opposite way with 
$\bar{\bf 5}_H$ and $\bar{\bf 5}_M$, the interaction 
(\ref{eq:Yukawa-in-E7}) gives rise to the trilinear interaction 
$\Delta W = S \bar{{\bf 5}}_H {\bf 5}_H$ of the next-to-minimal supersymmetric 
Standard Models, and a candidate for the right-handed neutrinos is lost.} 
then the moduli 
$({\bf adj.}, {\bf 1})$ for the $\U(3)$ spectral surface may become 
right-handed neutrinos \cite{Harvard-nu, Tsuchiya}. 
The up-type Yukawa couplings and down-type/charged lepton Yukawa couplings 
originate from 
\begin{equation}
 \Delta W = \left(\wedge^4 {\bf 8}\right)^{MNPQ} \cdot 
  {\bf adj.}^{I}_{\; H} \cdot \left(\wedge^4 {\bf 8}\right)^{HJKL} \; 
   \epsilon_{IJKLMNPQ}
\end{equation}
in the language of $\mathfrak{su}(8) \subset \mathfrak{e}_7$, and 
the neutrino Yukawa couplings from \cite{TW-1}
\begin{equation}
 \Delta W = \tr {}_{\mathfrak{su}(8)} 
    \left( {\bf adj.}, [ {\bf adj.}, {\bf adj.} ] \right). 
\label{eq:Yukawa-in-E7}
\end{equation}
It is worthwhile to study the monodromy for this case, e.g, using the example 
in footnote \ref{fn:ex}, to see how the problem A will look like, because 
the ``problem A'' may be quite different in nature, or may even be absent, 
in such a case without a need for imposing factorization condition by hand. 

Finally, there are also loopholes that resort to continuous fine-tuning 
of complex structure parameters. Flux compactifications should ultimately 
explain such a tuning. At the level of gauge theory description on 7+1 
dimensions, we introduced a parametrization of the complex structure 
moduli $a_0$, $a_2$ and $a_3$ by $c_{0,1,2}$ as 
in (\ref{eq:factorization_cond1}--\ref{eq:factorization_cond3}), 
so that the spectral surface factorizes in the new parametrization. 
But the ultimate goal is to reduce the monodromy of 2-cycles so that 
an unbroken U(1) symmetry is maintained in the low-energy effective 
theory.\footnote{Strictly speaking, we do not need a full U(1) symmetry, 
or equivalently a monodromy-invariant 2-cycle. If a torsion
component survives the monodromy, then some selection rule will 
remain, as we discussed in section \ref{subsec:M2brane}. 
} Thus, the generalized version of the solution is to consider a 
parametrization of the complex structure moduli $(a_{0,2,3}, f_0, g_0, a''_0)$ 
of $X_4$ so that the polynomial $B$ in (\ref{eq:B}) is factorized 
appropriately in the new parameters. The geometry for compactification 
should be given by holomorphic sections from $S_{GUT}$ to the new 
parameter space. 
It will be straightforward to work out the expression for ``$B$'' 
in the case of $E_8 \rightarrow {\rm SU}(5)_{\rm GUT}$ symmetry breaking. 

In the case of $E_8 \rightarrow E_6$ symmetry breaking, the polynomial $B$ 
should be factorized so that the split between 
$c^1_{2-DB}$--$c^1_{0-DB}$ and $c^2_{2-DB}$--$c^2_{0-DB}$ disappears and they 
form a monodromy locus of multiplicity 2. 
The monodromy (\ref{eq:0-DB_1,2}) is gone in this limit; 
this is a fine-tuning solution to the problem A. 

A hint for a fine-tuning solution to the problem B comes from 
Heterotic dual description. In Heterotic string language, one can 
tune complex structure moduli so that the elliptic fibred Calabi--Yau 3-fold 
admits a non-trivial section; moduli of spectral surfaces of $V_1$ and $V_2$ 
may be tuned so that the line bundle ${\rm det} \; V_1$ corresponds to 
the non-trivial section.\footnote{We thank Ron Donagi for discussion.} 
In other words, this is to consider the factorization limit 
of (\ref{eq:Het-spec-surf}) instead of the factorization limit 
of (\ref{eq:5-fold}, \ref{eq:rank3_spect}). There still remains a problem 
of determining the $\epsilon_\eta$ suppressed corrections, because the 
spectral surface picture relies on supergravity + $E_8 \times E_8$ 
super Yang--Mills approximation in the Heterotic description (see 
the appendix \ref{sec:Het} for more). In F-theory language, this 
tuning, we expect, will correspond to factorization of $B$ so that 
all the six branches $c^{1,2}_{0-4,5,6}$ somehow come on top of one another 
in the new parametrization space to form a monodromy locus of 
multiplicity 6. The monodromy, then, is $\rho(\gamma_{0-0})$, which we 
know is trivial on the $E_8$ part of the 2-cycle. We do not have a 
concrete picture of how to modify the parametrization (\ref{eq:factorization_cond1}--\ref{eq:factorization_cond3}) systematically to obtain the new 
parametrization, however.

  \section*{Acknowledgements}  

TW thanks Ron Donagi, Sergei Galkin, Joseph Marsano, Cumrun Vafa and
Martijn Wijnholt for stimulating discussion, useful comments and 
communications. This work started when two of the authors (TK and TW) 
were staying at YITP, Kyoto University, during a program 
``Branes, Strings and Black Holes,'' September--November, 2009.  
TW thanks Caltech theory group and KITP, UC Santa Barbara, for hospitality, 
where he stayed during the final stage of this project. 
This work was supported in part 
by JSPS Research Fellowships for Young Scientists (HH), 
by a Grant-in-Aid \#19540268 from the MEXT of Japan (TK), 
by Global COE Program ``the Physical Sciences Frontier'', MEXT, Japan (YT), 
by WPI Initiative, MEXT, Japan and by the NSF under Grant 
No. PHY05-51164 (TW).


\appendix

\section{Some examples of computing the monodromy of 2-cycles}
\label{sec:calculation}

In sections \ref{subsec:mod_gauge}--\ref{subsec:monod_beyond_gauge}, 
we need to study monodromy of 2-cycles of an elliptic fibered K3 manifold 
for loops in the moduli space of the elliptic K3; this monodromy 
group is used in this article, to find out whether an unbroken 
U(1) symmetry remains in the effective theory and the dimension-4 
proton decay operators (\ref{eq:dim-4}) are ruled out.
In order to calculate the monodromy of 2-cycles, we first 
need to identify 2-cycles in an elliptically fibered K3 
surface that constitute a basis of $H_2({\rm K3} ; \Z)$.
Second, we analyze the changes of the 2-cycles when we move along 
a loop in the moduli space. 

We are not interested so much in the fiber class and base class of 
the elliptic fibration, because they do not correspond to a U(1) 
vector field or a U(1) symmetry. When an M2-brane is wrapped on 
one of other topological 2-cycles of an elliptic K3 manifold, it 
is interpreted as a string junction configuration on the base manifold 
$\P^1$. The first task, therefore, is to identify ``independent'' 
string junction configurations, which has already been done completely 
in \cite{jcn-KcMd}. In section \ref{subsec:indep-2cycle}, we briefly 
review the results of \cite{jcn-KcMd}, while explaining details of 
our conventions that are used in the calculation in section
\ref{subsec:sample-calc}.

String junction configurations on $\P^1$ are easier to deal with 
(for string theorist) than topological 2-cycles in an elliptic 
K3 manifold. Thus, we calculate monodromy of topological 2-cycles by 
following the configuration of string junctions along a loop in 
the moduli space of the elliptic K3. 
We demonstrate the technique of the computation of the monodromy 
explicitly for some loops as examples in section \ref{subsec:sample-calc}.

The appendix \ref{sec:calculation} constitutes nearly a third of 
this article, but it is a technical note in nature. The main text 
will be readable without reading the details of this appendix. 

\subsection{Independent 2-cycles in the language of string junctions}
\label{subsec:indep-2cycle}

Let us consider a 7-brane configuration in a complex $z$ plane, where 
$[1, 0]$ 7-branes (called A-branes), $[1, -1]$ 7-brane (called
B-branes), $[1, 1]$ (C-branes) and $[3, -1]$ 7-branes (we call them 
``$D$''-branes) are line up from left to right as 
\begin{equation}
  A^8 B C^2 D \; A^8 B C^2 D,
\label{eq:E9E9}
\end{equation}
and branch cuts run from all of those 7-branes to the infinity in the 
$z$-plane in the positive imaginary direction. This is the 
$\hat{E}_9 \hat{E}_9$ configuration in \cite{jcn-KcMd}, which contains 
two sets of 7-brane configuration $[A^7 B C^2]$ for $E_8$.
We assign names to these 7-branes. The eight A-branes on the left 
are called $A8$--$A1$ from left to right, and the two C-branes 
on the left called $C1$ and $C2$ from left to right (see 
Figure~\ref{fig:2-cycles}). The B-brane and ``D''-brane on the left 
are simply called $B$ and $D$. The twelve 7-branes on the right 
are named similarly, with an extra $'$, such as $A8'$, $A7'$ etc.
We will see shortly that the configuration of $[p, q]$ 7-branes 
and branch cuts can be made topologically the same as this for 
an explicit choice of a base point in the elliptic K3 moduli space.

We have adopted a convention that a $(p, q)$ string corresponds to 
an M2-brane wrapped on $(p \alpha + q \beta)$ cycle of the 
$T^2$-fiber.\footnote{Here, the topological 1-cycles of $T^2$, $\alpha$ 
and $\beta$ are assumed to have a intersection form 
$\vev{\alpha, \beta} = - \vev{\beta, \alpha} = 1$. 
This follows the convention of \cite{Denef, GZ}.  
References \cite{Schwarz, GHZ, jcn-KcMd}, on the other hand,
define a $(p^{\prime}, q^{\prime})$ string as an M2-brane wrapping 
on $(p^{\prime}\alpha-q^{\prime}\beta)$-cycle of a torus. 
While the $(r, s)$ charge are re-labeled as in (\ref{eq:SL2Z-mx-A}) 
in the convention of \cite{Denef, GZ} when crossing a cut of a 
$[p, q]$ 7-brane in an anti-clockwise direction, the
\begin{eqnarray}
\left(\begin{array}{c}
r^{\prime}\\
s^{\prime}
\end{array}
\right)=\left(
\begin{array}{cc}
1&0\\
0&-1
\end{array}
\right)\left(
\begin{array}{c}
r\\
s
\end{array}
\right)
\end{eqnarray}
charge in the convention of \cite{GHZ, jcn-KcMd, Schwarz} changes 
by the monodromy matrix given by
\begin{eqnarray}
M^{\prime}_{p^{\prime}, q^{\prime}}
\equiv \left(
\begin{array}{cc}
1&0\\
0&-1
\end{array}
\right)M_{p=p^{\prime}, q=-q^{\prime}}\left(
\begin{array}{cc}
1&0\\
0&-1
\end{array}
\right)=\left(
\begin{array}{cc}
1+p^{\prime}q^{\prime} & -p^{\prime 2}\\
q^{\prime 2} & 1-p^{\prime}q^{\prime}
\end{array}
\right)
\label{eq:SL2Z-mx-B}
\end{eqnarray}
when crossing a branch cut of a $[p', q']$ 7-brane in the 
anti-clockwise direction. 
Certainly the last expression of $M'_{p', q'}$ happens to look 
like an inverse matrix of $M_{p, q}$ with $p$ and $q$ simply replaced 
(not rewritten!) by $p'$ and $q'$. But $M'_{p', q'}$ in
(\ref{eq:SL2Z-mx-B}) and $M_{p,q}$ in (\ref{eq:SL2Z-mx-A}) should be 
regarded as physically equivalent ${\rm SL}(2; \Z)$ monodromy matrices 
in different conventions. }
The $(r, s)$ charge of a string undergoes the monodromy as in 
\begin{eqnarray}
\left(
\begin{array}{c}
r\\
s
\end{array}
\right) \rightarrow
M_{p, q}\left(
\begin{array}{c}
r\\
s
\end{array}
\right)\equiv
\left(
\begin{array}{cc}
1-pq & p^{2}\\
-q^{2} & 1+pq
\end{array}
\right) \left(
\begin{array}{c}
r\\
s
\end{array}
\right)
\label{eq:SL2Z-mx-A}
\end{eqnarray}
when the string crosses a branch cut for a $[p, q]$ 7-brane
in anti-clockwise direction. 

Any string junctions on $\P^1$ correspond to closed 2-cycles of an
elliptic K3. We can pull out any junction configurations to 
the negative imaginary direction by continuous deformation, so that 
the configuration does not cross a branch cut; string creation process 
should be used if necessary. By deforming junction configurations 
in this way, we can express junction configurations by 24 integers
$N^i$, where $i=A8, A7, \cdots, A1, B, \cdots, C1', C2', D'$ run over 
the twenty four 7-branes; $N^i$ are the numbers of $(p_i,q_i)$ strings 
coming out of the $i$-th 7-brane whose charge is $[p_i, q_i]$. 
For example, the junction configuration (and the 2-cycle) named 
$C_{A76}$ is characterized by $N^{A7} = 1$, $N^{A6} = -1$, and 
$N^i = 0$ for all other 7-branes. $C_{BCD}$ corresponds to 
$N^{B} = N^{C1} = N^{C2} = 1$, $N^{D} = -1$, and $N^i = 0$ for all 
the other 7-branes. See Figure~\ref{fig:2-cycles}.
We denote 22 independent closed 2-cycles by \cite{jcn-KcMd}
\begin{itemize}
 \item visible $E_8$: $C_{A76} \sim C_{A21}, C_{ABC}$ and $C_{C12}$ (see
       Figure~\ref{fig:2-cycles}), 
 \item hidden $E_8$: $C'_{A76} \sim C'_{A21}, C'_{ABC}$ and $C'_{C12}$ 
      that are the same string junction configuration as the
       corresponding ones in visible $E_8$, except that their endpoints 
      are on the 7-branes $A8'$--$A1'$, $B'$, $C1'$, $C2'$ and $D'$, 
 \item others: $C_{A87}$, $C_{BCD}$, $C'_{A87}$, $C'_{BCD}$, $C_{AA'}$
       and $C_{DD'}$ (see Figure~\ref{fig:2-cycles}).
\end{itemize}

Among the 22 closed 2-cycles listed above, however, 2 linear 
combinations 
\begin{equation}
 C_{BCD} + C'_{BCD} \quad {\rm and} \quad  
-C_{A87}+-C_{A87}^{\prime}+C_{-\theta}+C_{-\theta}^{\prime}
\end{equation}
can expressed as a boundary of 3-dimensional cells; here, 
$C_{-\theta}$ is defined by (\ref{eq:C-theta-def}), and 
$C'_{-\theta}$ is its obvious $'$ version.
We thus drop $C'_{BCD}$ and $C'_{-\theta} - C'_{A87}$, and 
use $C_{BCD}$, $C_{-\theta}$, $C_{AA'}$, $C_{DD'}$ and the 16 
2-cycles of the visible and hidden $E_8$'s as representatives\footnote{
\label{fn:K3-RES}
The homology group of a rational elliptic surface (also known as
``$dP_9$'') can be expressed in a similar manner. Any linear
combinations of $C_{BCD}$ and $C_{-\theta} - C_{A87}$ become the
boundary, and hence $H_2(dP_9; \Z)$ is generated by the eight 
visible $E_8$ 2-cycles and the fiber and base classes. In the homology
group of a rational elliptic surface, $[C_{A87}] = [C_{-\theta}]$. 
This relation, however, does not hold for an elliptic K3 manifold.} 
of the 20 independent topological 2-cycles 
in $H_2({\rm K3} ; \Z)$ \cite{jcn-KcMd}.

For an appropriate choice of a basis of $H_2({\rm K3} ; \Z)$, the 
intersection form becomes \cite{Aspinwall}
\begin{eqnarray}
-C(E_{8}) \oplus\left(
\begin{array}{cc}
0&1\\
1&0
\end{array}
\right) \oplus \left(
\begin{array}{cc}
0&1\\
1&0
\end{array}
\right)
\oplus-C(E_{8}^{\prime}),
\label{eq:K3-int-form}
\end{eqnarray}
where $C(E_{8})$ and $C(E_{8}^{\prime})$ denotes the Cartan matrix of 
$E_{8}^{(\prime)}$ respectively, with an extra 2 by 2 block for the 
fiber and base classes. The eight 2-cycles of the visible $E_8$ and 
those of the hidden $E_8$ can be used as the first and last eight
elements of the basis. The remaining four elements of a basis 
for the intersection form above can be chosen as 
the following four linear combinations \cite{jcn-KcMd}:
%
\begin{eqnarray}
	C_{\alpha}^{1}=-C_{A87}+C_{-\theta},&\;& 
        C_{\alpha}^{2}=-C_{A87}+C_{-\theta}+C_{AA^{\prime}},
   \label {eq:basis1} \\
	C_{\beta}^{1}=C_{BCD},&\;&
	C_{\beta}^{2}=C_{BCD}+C_{DD^{\prime}}. 
    \label{eq:basis2}
\end{eqnarray}
$\vev{C_\alpha^1, C_\alpha^2} = 1$, $\vev{C_\beta^1, C_\beta^2} = 1$,
and all other intersection numbers among the four 2-cycles vanish.

\subsection{Some examples of the monodromy}
\label{subsec:sample-calc}

\subsubsection{The 7-brane configuration at the base point}

In section \ref{subsec:sample-calc}, we present the practical procedure 
of calculation of monodromy of 2-cycles for some loops in the moduli 
space of elliptic K3 manifold (\ref{eq:model}). 
We have chosen a base point as in (\ref{eq:reference_point}) in the 
moduli space. 
\begin{figure}[tb]
\begin{center}
\includegraphics[scale=0.25]{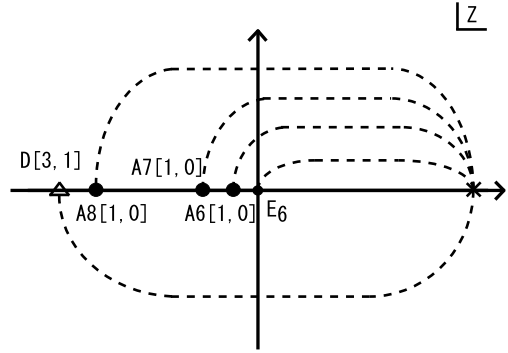}
\caption{The 7-brane configuration at the base point
 (\ref{eq:reference_point}). The dotted curves represent branch
 cuts. The approximate location of $A6$ 7-brane is $z_{A6}\sim
 -\frac{27}{64}\frac{a_{3}^{4}}{a_{2}^{3}}\sim
 \mathcal{O}(\epsilon_{K}^{6}\epsilon_{\eta}^{1}\delta^{4})$ and
 $z_{A7}\sim
 -\frac{4}{27}\frac{a_{2}^{3}}{a_{0}^{2}}\sim\mathcal{O}(\epsilon_{K}^{6}\epsilon_{\eta})$ for $A7$ 7-brane [see also (\ref{eq:A7-behavior})]. $A8$ and $D$ 7-branes are at $z\sim \mathcal{O}(\epsilon_\eta)$. }
\label{fig:7-branes_at_ref_pt}
\end{center}
\end{figure}
The 7-branes configuration on the base $\P^1$ at this base point is 
shown in Figure~\ref{fig:7-branes_at_ref_pt}. Since we introduced 
a small parameter $\delta$ in the definition of the base point, 
the elliptic K3 manifold for the base point must realize a hierarchical 
symmetry breaking $E_8 \rightarrow E_7 \rightarrow E_6$. We thus 
assign the name A6 to the 7-brane that is closest to the $E_6$ point. 
The branch cuts and the $[p, q]$ charges of the 7-branes are 
given as in Figure~\ref{fig:7-branes_at_ref_pt}. 
This arrangement is certainly the way anticipated in (\ref{eq:E9E9}),  
but it is not trivial whether the cut configuration and charge
assignment in the figure are correct for the specific choice of 
the complex structure at the base point. 
We confirmed that this is a right choice, by examining the 1-cycles 
of the $T^2$ fiber to degenerate and monodromy of the complex structure 
of the fiber at these 7-branes.

\subsubsection{The Monodromy of the loop $\gamma_{2-A}$}

First, we consider the monodromy around the loop $\gamma_{2-A}$. 
This loop goes around a triple point of the monodromy locus. 
The loop $\gamma_{2-A}$ is depicted in the Figure \ref{fig:a2-plane_e_K}. 
We decompose the loop $\gamma_{2-A}$ into three paths in calculating 
the monodromy.
\begin{enumerate}
\item approaching $a_{2}=a_{2-A}$ from the base point from the right (path 1),
\item going around $a_{2}=a_{2-A}$ in the anti-clockwise direction (path
      2), and finally, 
\item going back to the base point from $a_{2}=a_{2-A}$ (path 3).
\end{enumerate}
The 2-cycles at the base point change when evaluated after they go along path 1, path 2 and path 3.
We can read the monodromy matrix from the change of the 2-cycles.

{\bf Path 1}: The movement of 7-branes corresponding to the path 1 
is shown in the first part of the Table \ref{tb:2-A_path1}. 
After completing the path 1, we rearrange the branch cuts as a
preparation for the path 2, by two steps as in the second and third part 
of the Table \ref{tb:2-A_path1}. The changes of 2-cycles (string
junction) as well as the change in the $[p, q]$ charges of various
7-branes during the rearrangement of the branch cuts are also shown 
in the table below the corresponding figures (Table
\ref{tb:2-A_path1}). As we already explained in 
section \ref{subsec:indep-2cycle}, string junction configurations are 
always deformed continuously so that they do not cross any one of 
branch cuts. The numbers in the table show the number of 
$[p_i, q_i]$ strings coming out of a $[p_i, q_i]$ 7-brane.

\begin{longtable}{c}

\includegraphics[scale=0.25]{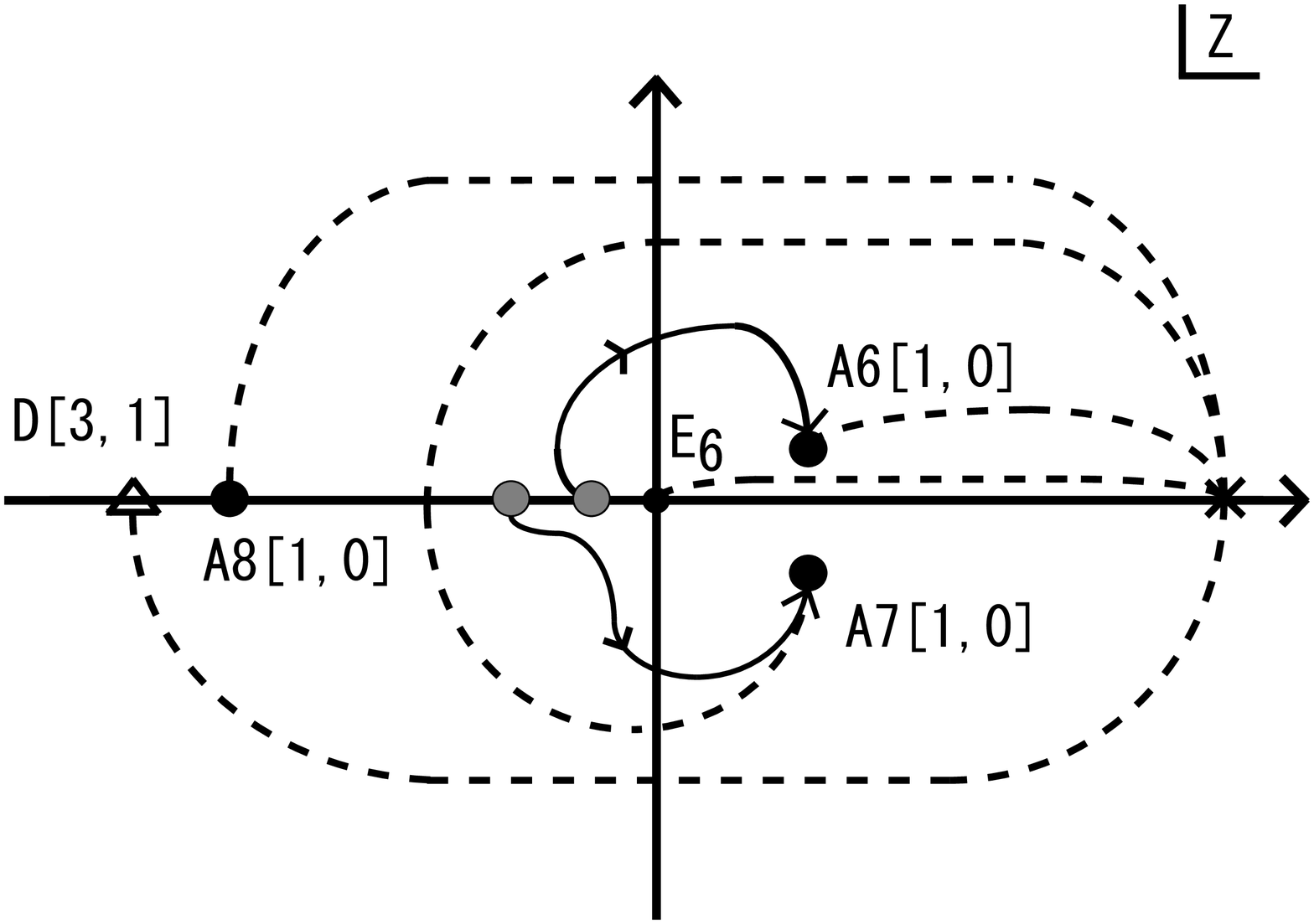}

\\ \\

\begin{tabular}{|c|c|c|c|c|c|c|c|c|c|}
\hline
&A8 [1,0] & A7 [1,0] & A6 [1,0] & A5 [1,0] & A4-A1 [1,0] & B [1,-1] & C1,2 [1,1] & D[3,1]\\
\hline
$C_{A76}$&0&1&-1&0&0&0&0&0\\
\hline
$C_{A65}$&0&0&1&-1&0&0&0&0\\
\hline
$C_{BCD}$&0&0&0&0&0&1&1&-1\\
\hline
\end{tabular}

\\ \\

$\downarrow$ step 1

\\ \\

Passing $A6+E_{6}$ 7-branes through\\
the branch cuts of $A7$ 7-branes from right to left.

\\ \\

\includegraphics[scale=0.25]{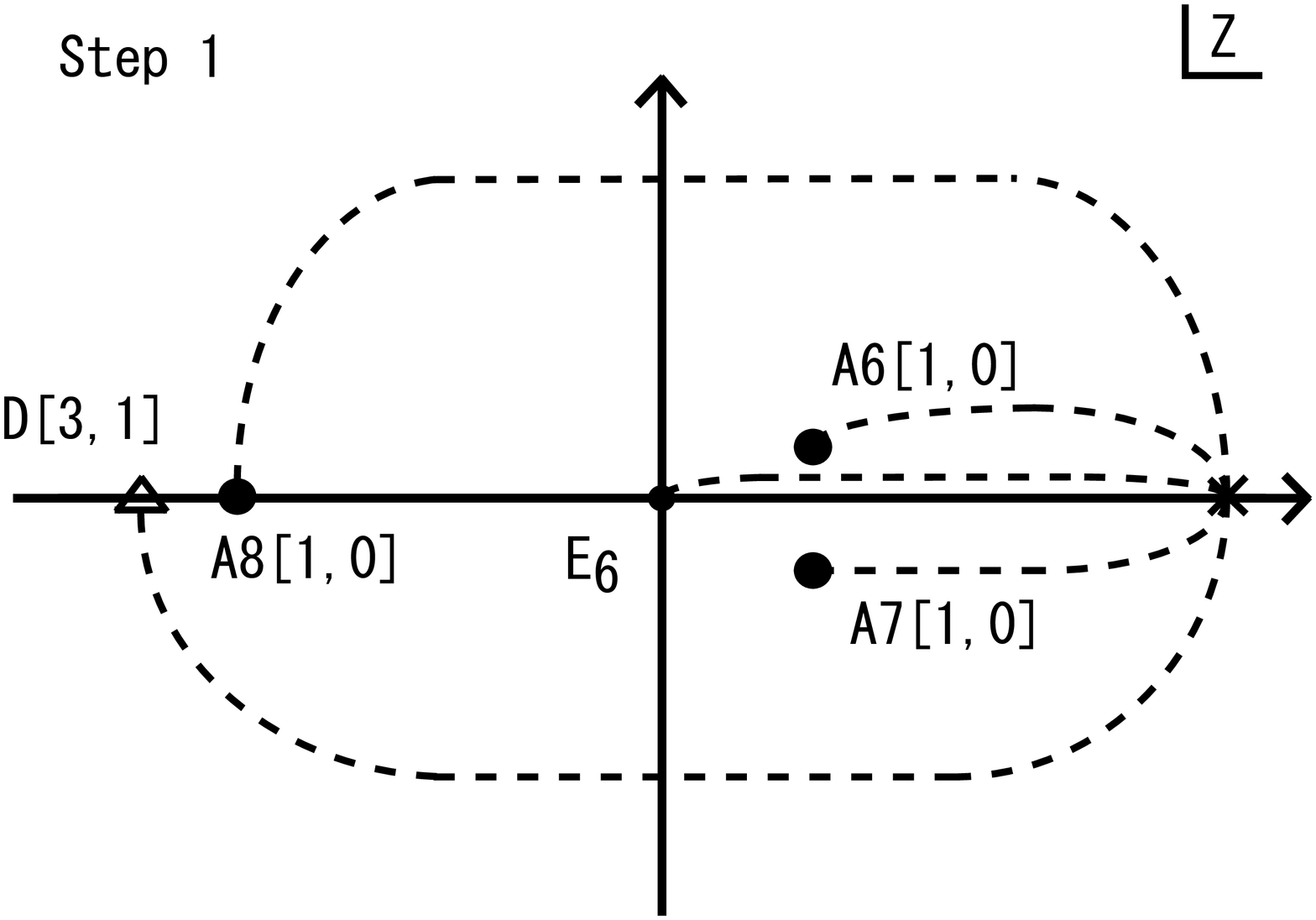}

\\ \\

\begin{tabular}{|c|c|c|c|c|c|c|c|c|c|}
\hline
&A8 [1,0] & A6 [1,0] & A5 [1,0] & A4-A1 [1,0] & B [0,-1] & C1,2 [2,1] & A7 [1,0]& D[3,1]\\
\hline
$C_{A76}$&0&-1&0&0&0&0&1&0\\
\hline
$C_{A65}$&0&1&-1&0&0&0&0&0\\
\hline
$C_{BCD}$&0&0&0&0&1&1&-1&-1\\
\hline
\end{tabular}

\\ \\

$\downarrow$ step 2

\\ \\

Passing $A6$ 7-branes through \\
the branch cuts of $E_{6}$ 7-branes from left to right.

\\ \\

\includegraphics[scale=0.25]{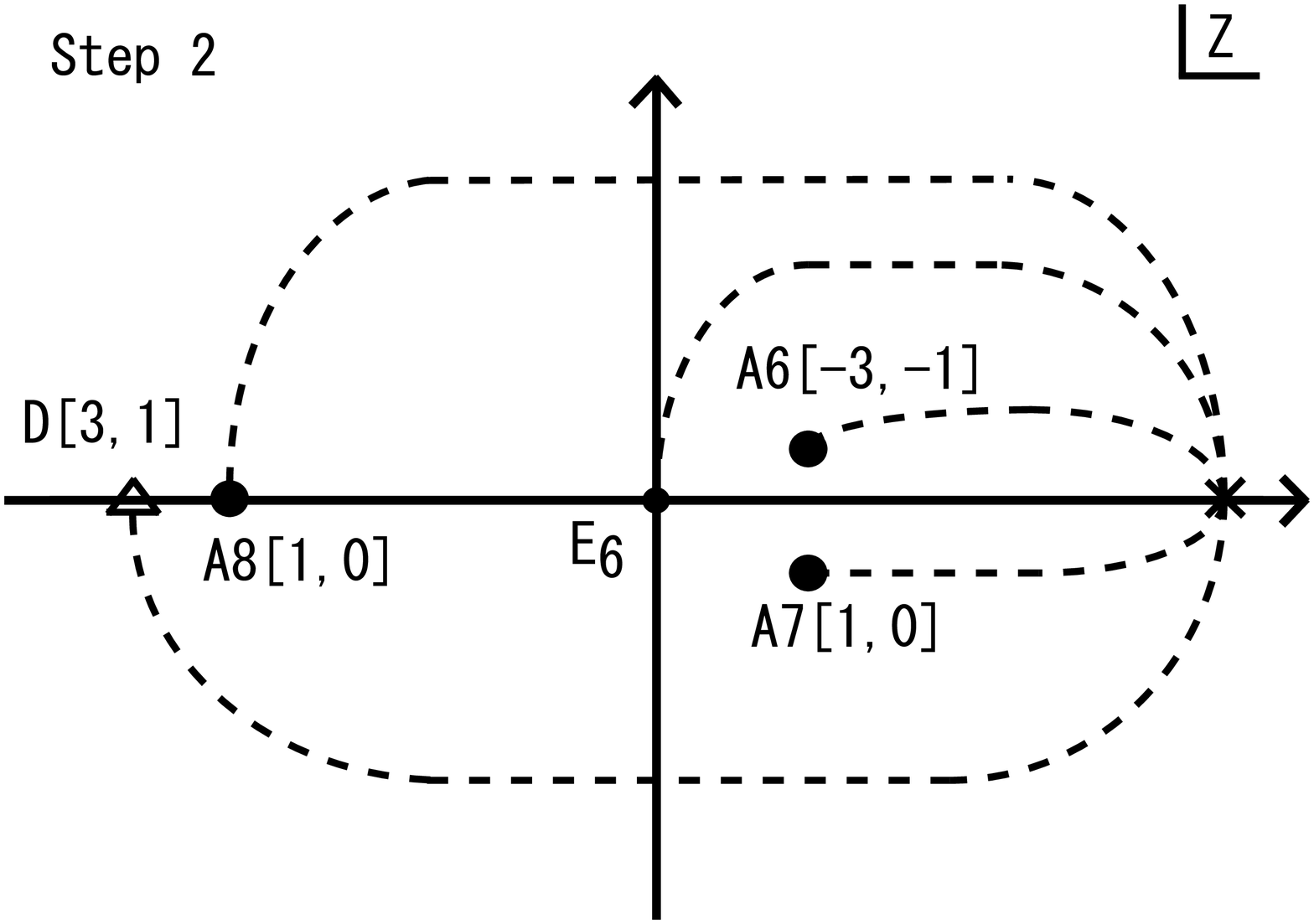}

\\ \\

\begin{tabular}{|c|c|c|c|c|c|c|c|c|c|}
\hline
&A8 [1,0] & A5 [1,0] & A4-A1 [1,0] & B [0,-1] & C1,2 [2,1] & A6 [-3, -1] & A7 [1,0]& D[3,1]\\
\hline
$C_{A76}$&0&0&0&-1&-1&-1&1&0\\
\hline
$C_{A65}$&0&-1&0&1&1&1&0&0\\
\hline
$C_{BCD}$&0&0&0&1&1&0&-1&-1\\
\hline
\end{tabular}
\\ \\
\caption{
The first figure shows the motion of 7-branes along the path
 1. The location of 7-branes before the movement is depicted by the gray
 circles. The next two figures show how to rearrange branch cuts. The
 changes of 2-cycles (string junctions) of the process are shown below
 the corresponding figures.  We denotes the $[p, q]$ charge next to the
 name of 7-branes. 
}
\label{tb:2-A_path1}
\end{longtable}

{\bf Path 2}:  During the path 2, the $A6$ 7-brane and $A7$ 7-brane
mutually rotate around the other by $3\pi$, as in 
Table~\ref{tb:2-A_path2}. We have rearranged the branch cuts at the 
end of path 1, so that they do not cross any branch cuts except the 
cuts of themselves.
Note, however, that $A6$ and $A7$ 7-branes are not mutually local 
after the rearrangement of the branch cuts. Therefore we have to take 
a close look during the path 2 at the changes of $[p, q]$ charges 
of the $A6$ and $A7$ 7-branes, and at the changes of string junction 
configurations that have end points on $A6$ or $A7$. 
For every $\pi$ rotation, either $A6$ or $A7$ has to cross the branch 
cut of the other. Thus, we need to trace the changes of junction 
configurations for every $\pi$ rotation. Table~\ref{tb:2-A_path2} 
shows the results.
From the Table \ref{tb:2-A_path2}, the 2-cycles do not change after 
the $3\pi$ rotation. 
		
\begin{longtable}{c}

\includegraphics[scale=0.25]{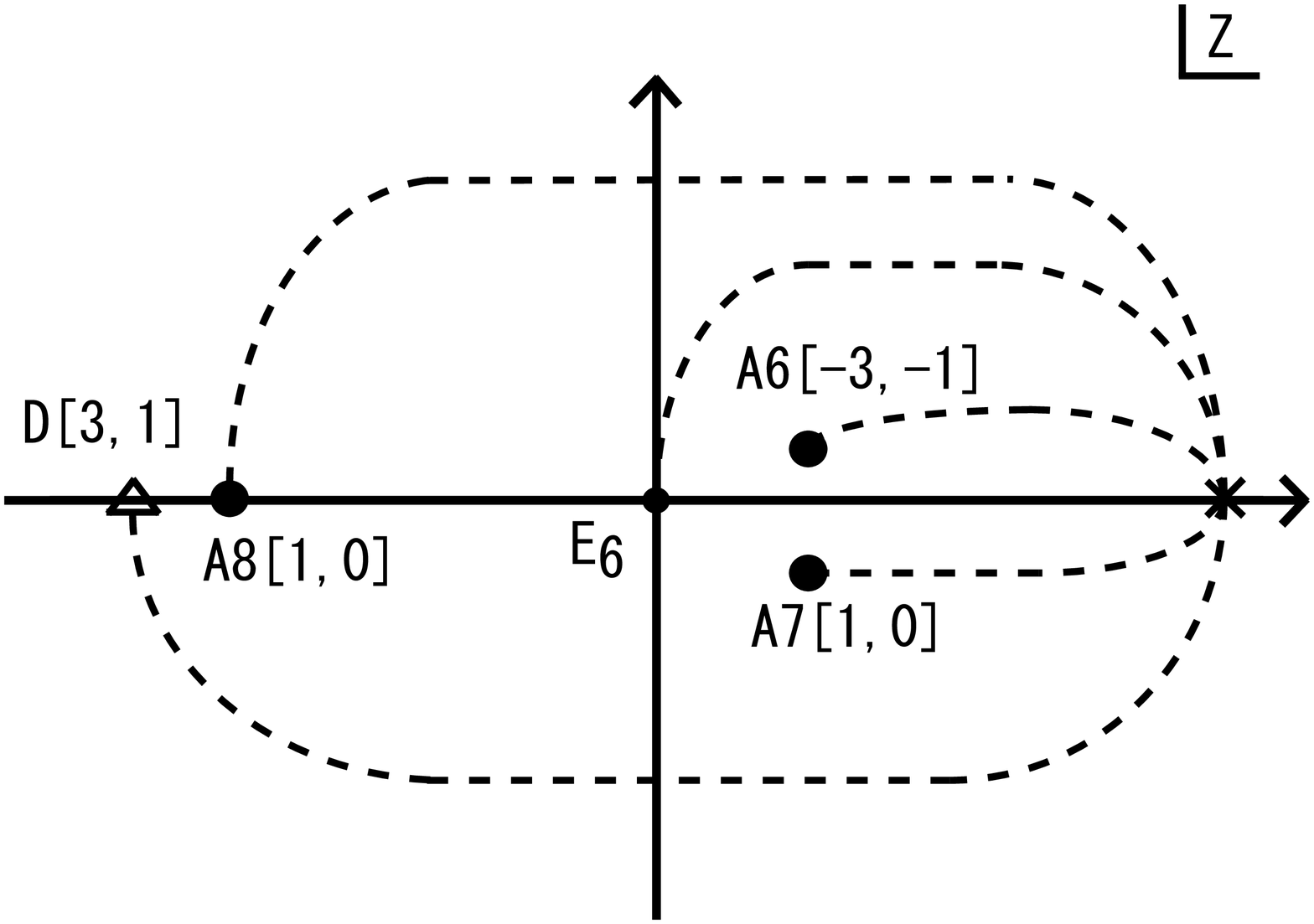}

\\ \\

\begin{tabular}{|c|c|c|c|c|c|c|c|c|c|}
\hline
&A8 [1,0] & A5 [1,0] & A4-A1 [1,0] & B [0,-1] & C1,2 [2,1] & A6 [-3, -1] & A7 [1,0]& D[3,1]\\
\hline
$C_{A76}$&0&0&0&-1&-1&-1&1&0\\
\hline
$C_{A65}$&0&-1&0&1&1&1&0&0\\
\hline
$C_{BCD}$&0&0&0&1&1&0&-1&-1\\
\hline
\end{tabular}

\\ \\
$\downarrow$ [$0 \rightarrow \pi$ rotation]

\\ \\ 

\includegraphics[scale=0.25]{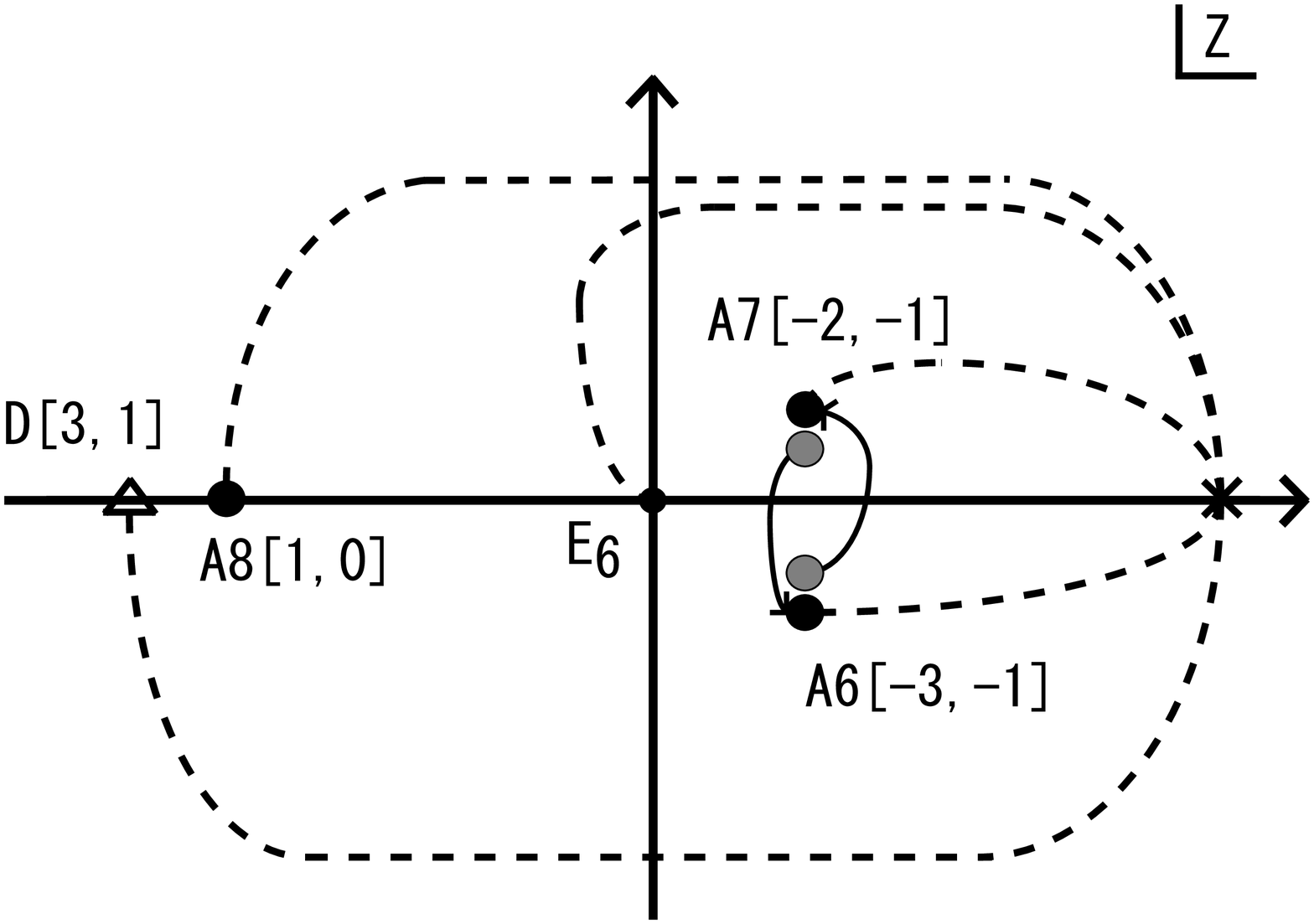}

\\ \\

\begin{tabular}{|c|c|c|c|c|c|c|c|c|c|}
\hline
&A8 [1,0] & A5 [1,0] & A4-A1 [1,0] & B [0,-1] & C1,2 [2,1] & A7 [-2, -1] & A6 [-3,-1]& D[3,1]\\
\hline
$C_{A76}$&0&0&0&-1&-1&1&-2&0\\
\hline
$C_{A65}$&0&-1&0&1&1&0&1&0\\
\hline
$C_{BCD}$&0&0&0&1&1&-1&1&-1\\
\hline
\end{tabular}

\\ \\

$\downarrow$ [$\pi \rightarrow 2\pi$ rotation]

\\ \\

\includegraphics[scale=0.25]{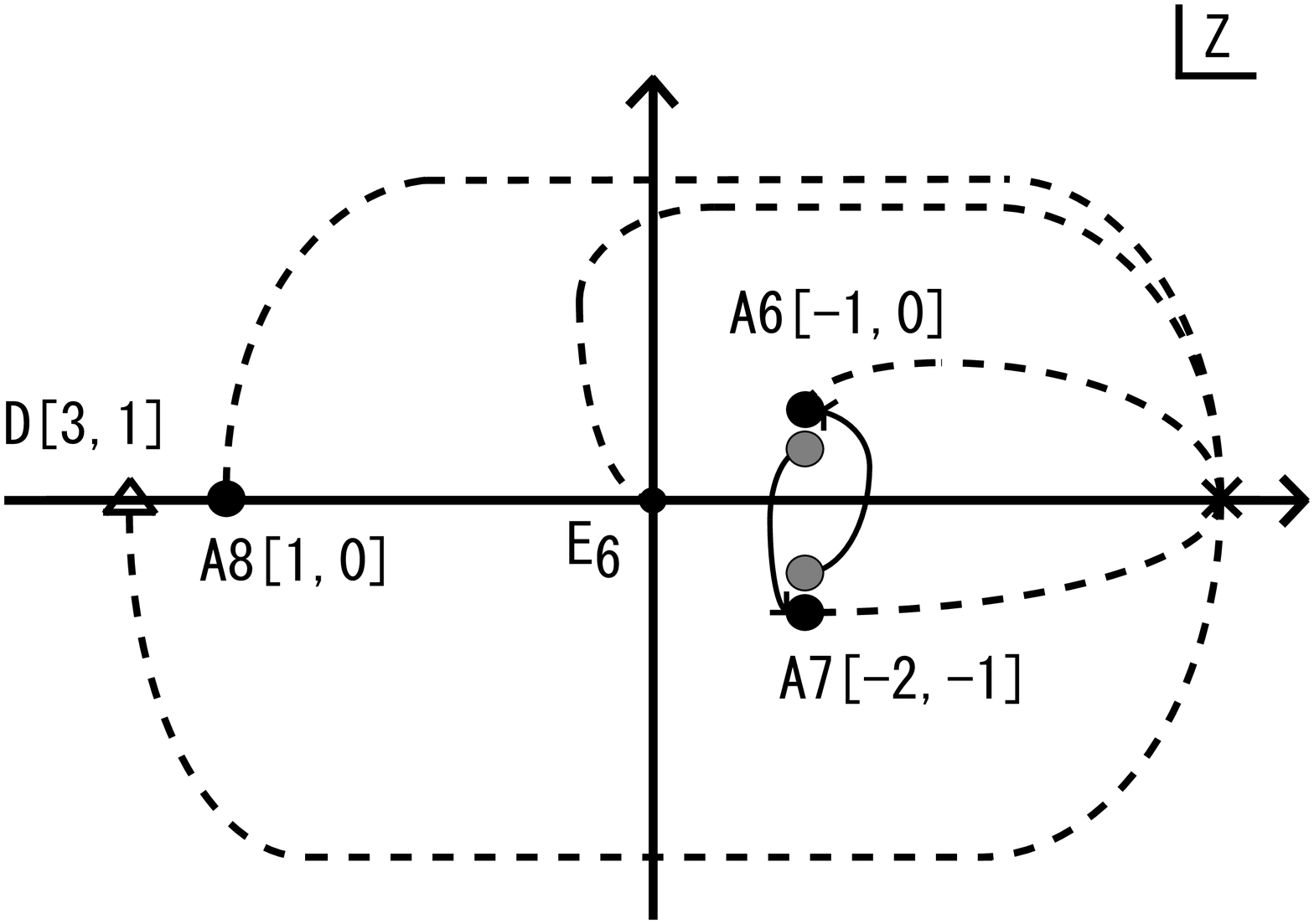}

\\ \\

\begin{tabular}{|c|c|c|c|c|c|c|c|c|c|}
\hline
&A8 [1,0] & A5 [1,0] & A4-A1 [1,0] & B [0,-1] & C1,2 [2,1] & A6 [-1, 0] & A7 [-2,-1]& D[3,1]\\
\hline
$C_{A76}$&0&0&0&-1&-1&-2&-1&0\\
\hline
$C_{A65}$&0&-1&0&1&1&1&1&0\\
\hline
$C_{BCD}$&0&0&0&1&1&1&0&-1\\
\hline
\end{tabular}

\\ \\

$\downarrow$ [$2\pi \rightarrow 3\pi$ rotation]

\\ \\

\includegraphics[scale=0.25]{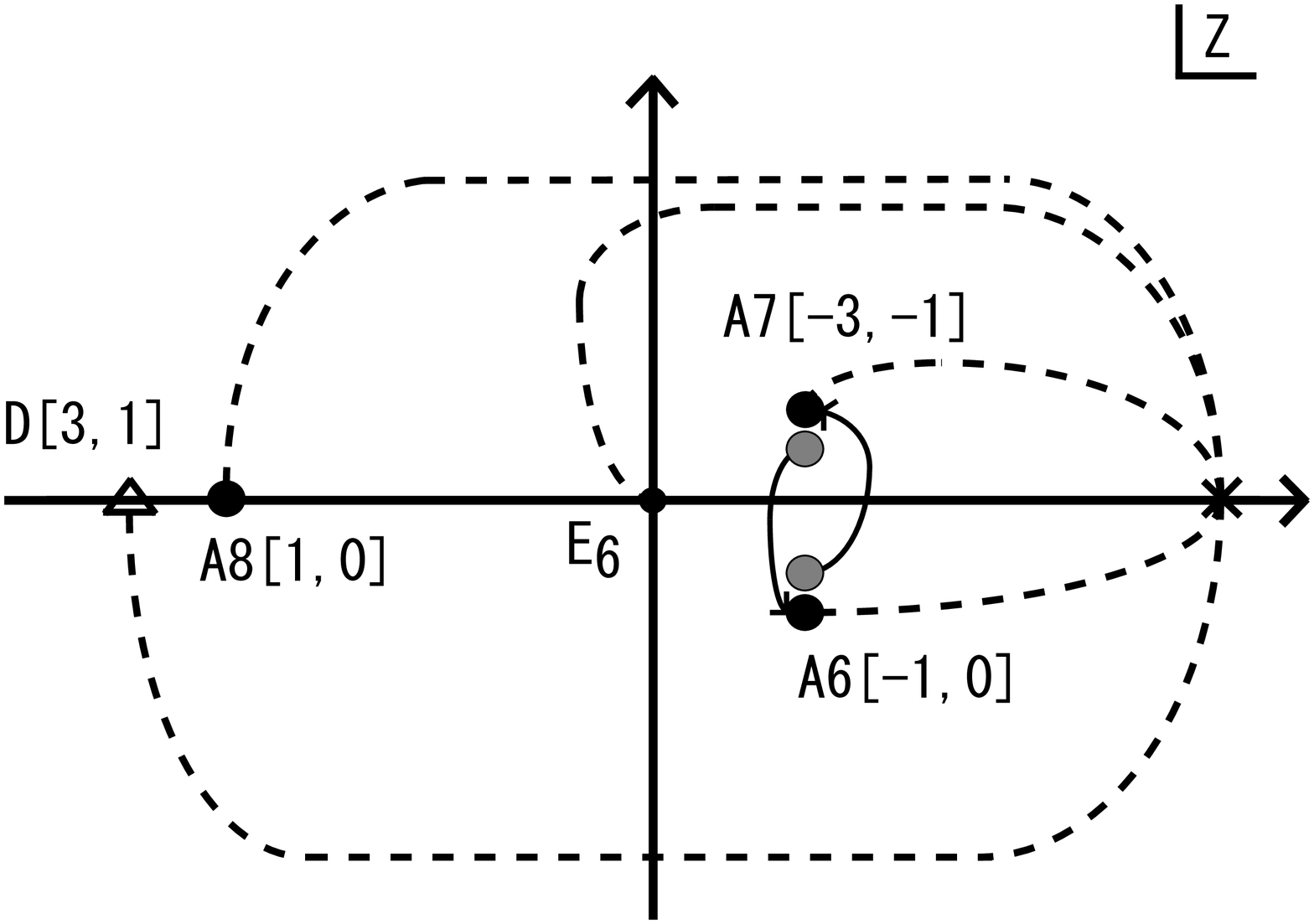}

\\ \\

\begin{tabular}{|c|c|c|c|c|c|c|c|c|c|}
\hline
&A8 [1,0] & A5 [1,0] & A4-A1 [1,0] & B [0,-1] & C1,2 [2,1] & A7 [-3, -1] & A6 [-1,0]& D[3,1]\\
\hline
$\tilde{C}_{A76}$&0&0&0&-1&-1&-1&-1&0\\
\hline
$\tilde{C}_{A65}$&0&-1&0&1&1&1&0&0\\
\hline
$\tilde{C}_{BCD}$&0&0&0&1&1&0&1&-1\\
\hline
\end{tabular}







\\ \\
\caption{The motion of 7-branes and the changes of the 2-cycles along the path 2. $\tilde{C}$s represent the 2-cycles after $3\pi$ rotation.
}
\label{tb:2-A_path2}
\end{longtable}

{\bf Path 3}: 
The path 3 simply follows the path 1 in the opposite direction. 
Before going back to the base point, however, we rearrange the branch
cuts in a backward direction of the Table \ref{tb:2-A_path1} with 
$A6$ and $A7$ exchanged. After that, we go back to the base point 
along the path 3, and 7-branes move along the same path as in the 
first part of Table \ref{tb:2-A_path1} in the opposite direction 
without crossing any branch cuts; this is depicted in the second part 
of the Table \ref{tb:2-A_path3}.

\begin{longtable}{c}

\\ \\

\includegraphics[scale=0.25]{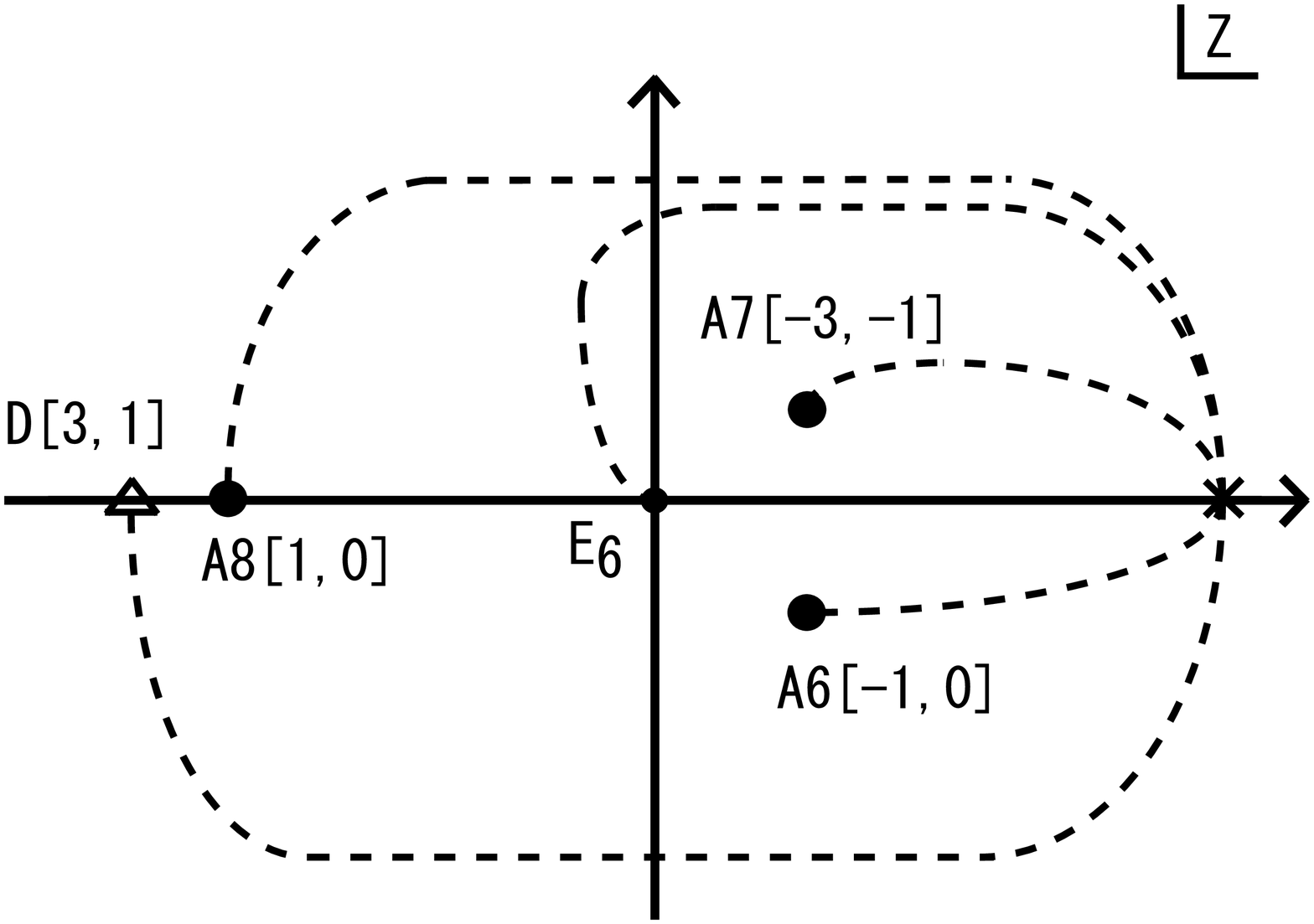}

\\ \\

\begin{tabular}{|c|c|c|c|c|c|c|c|c|c|}
\hline
&A8 [1,0] & A5 [1,0] & A4-1 [1,0] & B [0,-1] & C1,2 [2,1] & A7 [-3, -1] & A6 [-1,0]& D[3,1]\\
\hline
$\tilde{C}_{A76}$&0&0&0&-1&-1&-1&-1&0\\
\hline
$\tilde{C}_{A65}$&0&-1&0&1&1&1&0&0\\
\hline
$\tilde{C}_{BCD}$&0&0&0&1&1&0&1&-1\\
\hline
\end{tabular}

\\ \\

$\downarrow$ 

\\ \\
Rearrangement of branch cuts and going along the path 3.
\\ \\

\includegraphics[scale=0.25]{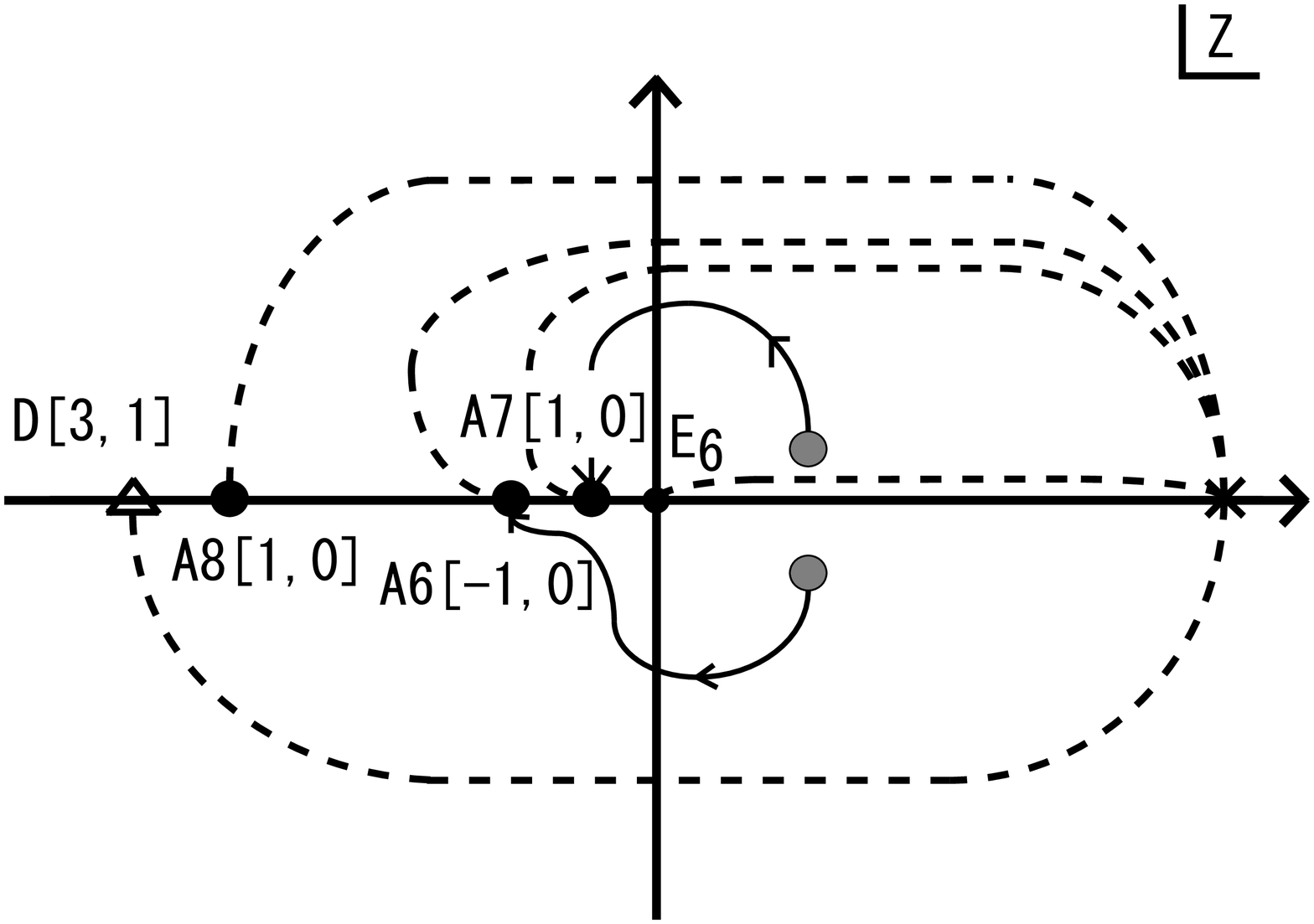}

\\ \\

\begin{tabular}{|c|c|c|c|c|c|c|c|c|c|}
\hline
&A8 [1,0] & A6 [-1,0] & A7 [1,0] & A5 [1,0] & A4-1 [1,0] & B [1,-1] & C1,2 [1,1] & D[3,1]\\
\hline
$\tilde{C}_{A76}$&0&-1&-1&0&0&0&0&0\\
\hline
$\tilde{C}_{A65}$&0&0&1&-1&0&0&0&0\\
\hline
$\tilde{C}_{BCD}$&0&0&0&0&0&1&1&-1\\
\hline
\end{tabular}

\\ \\

$\downarrow$ 

\\ \\

Changing the bases into the ones before the rotation

\\ \\

\includegraphics[scale=0.25]{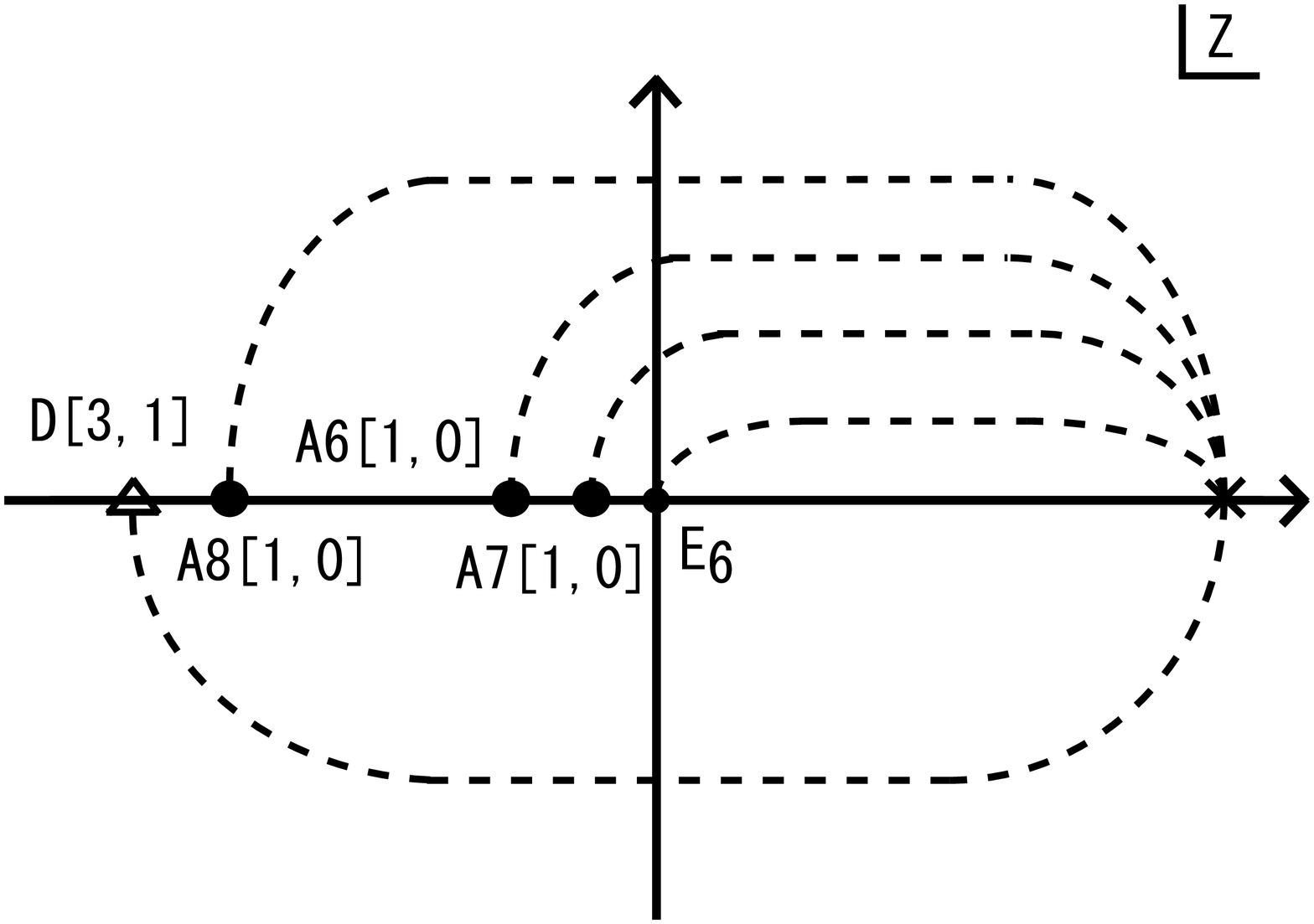}

\\ \\

\begin{tabular}{|c|c|c|c|c|c|c|c|c|c|}
\hline
&A8 [1,0] & A6 [1,0] & A7 [1,0] & A5 [1,0] & A4-1 [1,0] & B [1,-1] & C1,2 [1,1] & D[3,1]\\
\hline
$\tilde{C}_{A76}$&0&1&-1&0&0&0&0&0\\
\hline
$\tilde{C}_{A65}$&0&0&1&-1&0&0&0&0\\
\hline
$\tilde{C}_{BCD}$&0&0&0&0&0&1&1&-1\\
\hline
\end{tabular}

\\ \\

\caption{The motion of 7-branes and the changes of 2-cycles along the path 3.}
\label{tb:2-A_path3}

\end{longtable}

Comparing the first table of Table \ref{tb:2-A_path1} with the last
table of Table \ref{tb:2-A_path3}, we find that the 2-cycles do not
change at the end of the whole process. Thus, the monodromy of the 
2-cycles is trivial for the loop $\gamma_{2-A}$.

\subsubsection{The Monodromy of the loop $\gamma_{2-2}$}

\begin{figure}[!h]
\begin{center}
\includegraphics[scale=0.25]{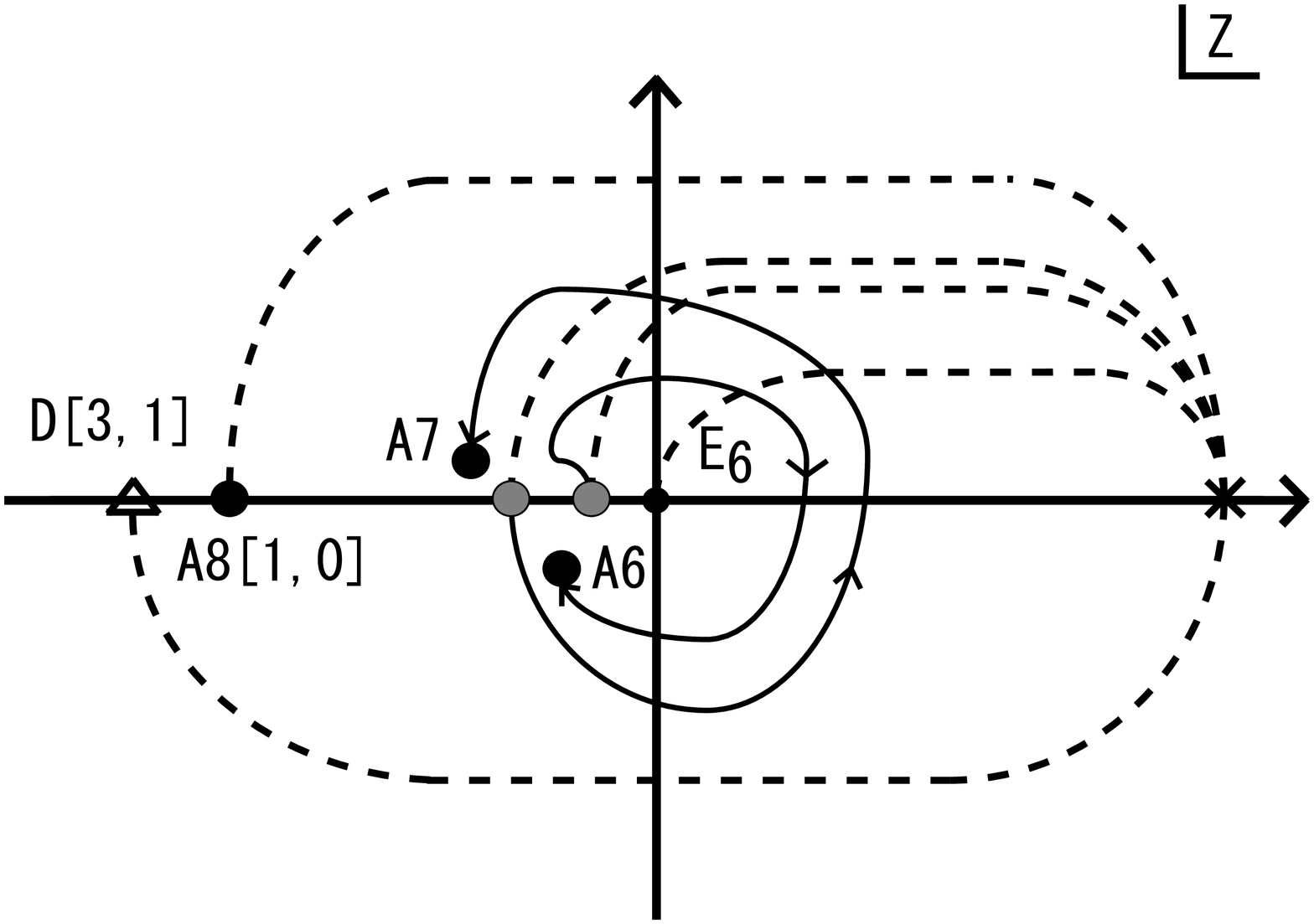}
\end{center}
\caption{The motion of $A6, A7$ 7-branes along the path 1. Since the motion of $A8, D$ 7-branes are not relevant to the monodromy of this case, we do not write it in the figure.}
\label{fig:2-2}
\end{figure}

Let us follow the motion of 7-branes when $a_2$ varies along the loop
$\gamma_{2-2}$.
First, let us separate the loop $\gamma_{2-2}$ into three pieces;
\begin{enumerate}
  \item a path from the base point to the right of $a_{2-2}$ (path 1).
  \item a loop around $a_{2-2}$ (path 2).
  \item a path which is the reverse of the first path (path 3).
\end{enumerate}
When $a_2$ varies along the first path,
7-branes move as shown in Figure \ref{fig:2-2}.

When $a_2$ varies along the second path,
the A6 7-brane's position and the A7 7-brane's position exchange with each other.

Next let us consider to rearrange the branch cuts of 7-branes
before the exchange of A6 and A7
so that both 7-branes do not cross any branch cuts except for themselves' during
the exchange. This process is collected in the Table \ref{tab:2-cycles_before_ex}. It shows that the configuration of 7-branes and how the 2-cycles change by the above rearrangement.

\begin{longtable}{c}
\includegraphics[scale=0.25]{ref_pt.eps}

\\ \\

  \begin{tabular}{|c|c|c|c|c|c|c|c|c|} \hline
           &  A8[1,0]
           &  A7[1,0]
           &  A6[1,0]
           &  A5[1,0]
           &  A4-A1[1,0]
           &  B[1,-1]
           &  C1, C2[1,1]
           &  D[3,1]
           \\\hline
 $C_{A65}$ &  0  &   0  &  1   & $-1$ & 0            & 0 &   0    & 0 \\\hline
 $C_{A76}$ &  0  &   1  & $-1$ & 0    & 0            & 0 &   0    & 0 \\\hline
 $C_{A87}$ &  1  & $-1$ &  0   & 0    & 0            & 0 &   0    & 0 \\\hline
 $C_{BCD}$ &  0  &   0  &  0   & 0    & 0            & 1 &  1    & $-1$ \\
 \hline
  \end{tabular}

\\ \\
$\downarrow$ step 1

\\ \\

A6 and E$_6$ go across the cut of A7 from right to left

\\ \\

\includegraphics[scale=0.25]{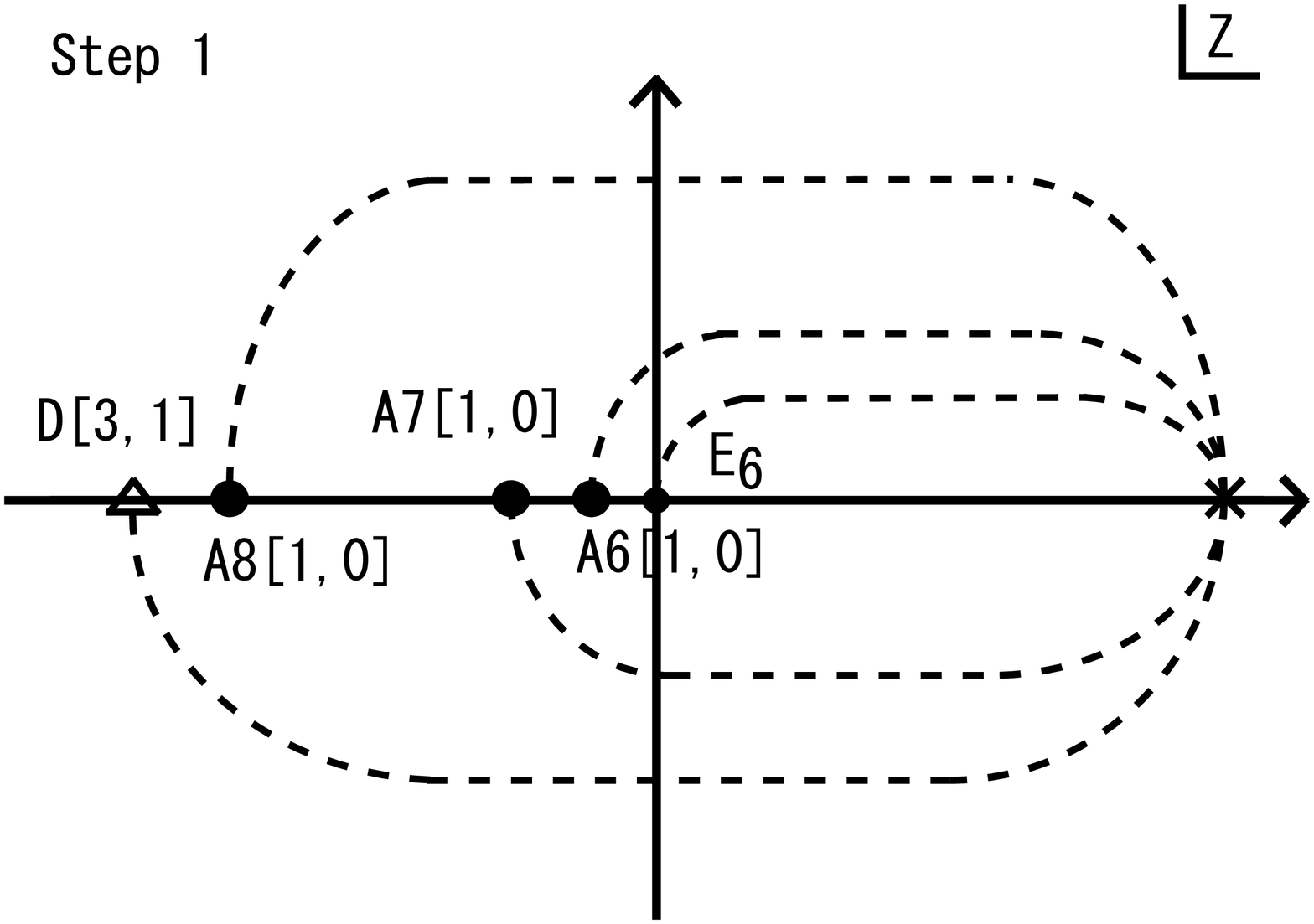}

\\ \\

  \begin{tabular}{|c|c|c|c|c|c|c|c|c|} \hline
           &  A8[1,0]
           &  A6[1,0]
           &  A5[1,0]
           &  A4-A1[1,0]
           &  B[0,-1]
           &  C1, C2[2,1]
           &  A7[1,0]
           &  D[3,1]
           \\\hline
 $C_{A65}$ &  0   & 1    & $-1$ & 0            & 0 & 0      &   0  & 0 \\\hline
 $C_{A76}$ &  0   & $-1$ & 0    & 0            & 0 & 0      &   1  & 0 \\\hline
 $C_{A87}$ &  1   & 0    & 0    & 0            & 0 & 0      & $-1$ & 0 \\\hline
 $C_{BCD}$ &  0   & 0    & 0    & 0            & 1 & 1      &  0   & $-1$\\
 \hline
  \end{tabular}

\\ \\
$\downarrow$ step 2

\\ \\

A7 goes across the cut of A6 and E$_6$ from right to left.

\\ \\

\includegraphics[scale=0.25]{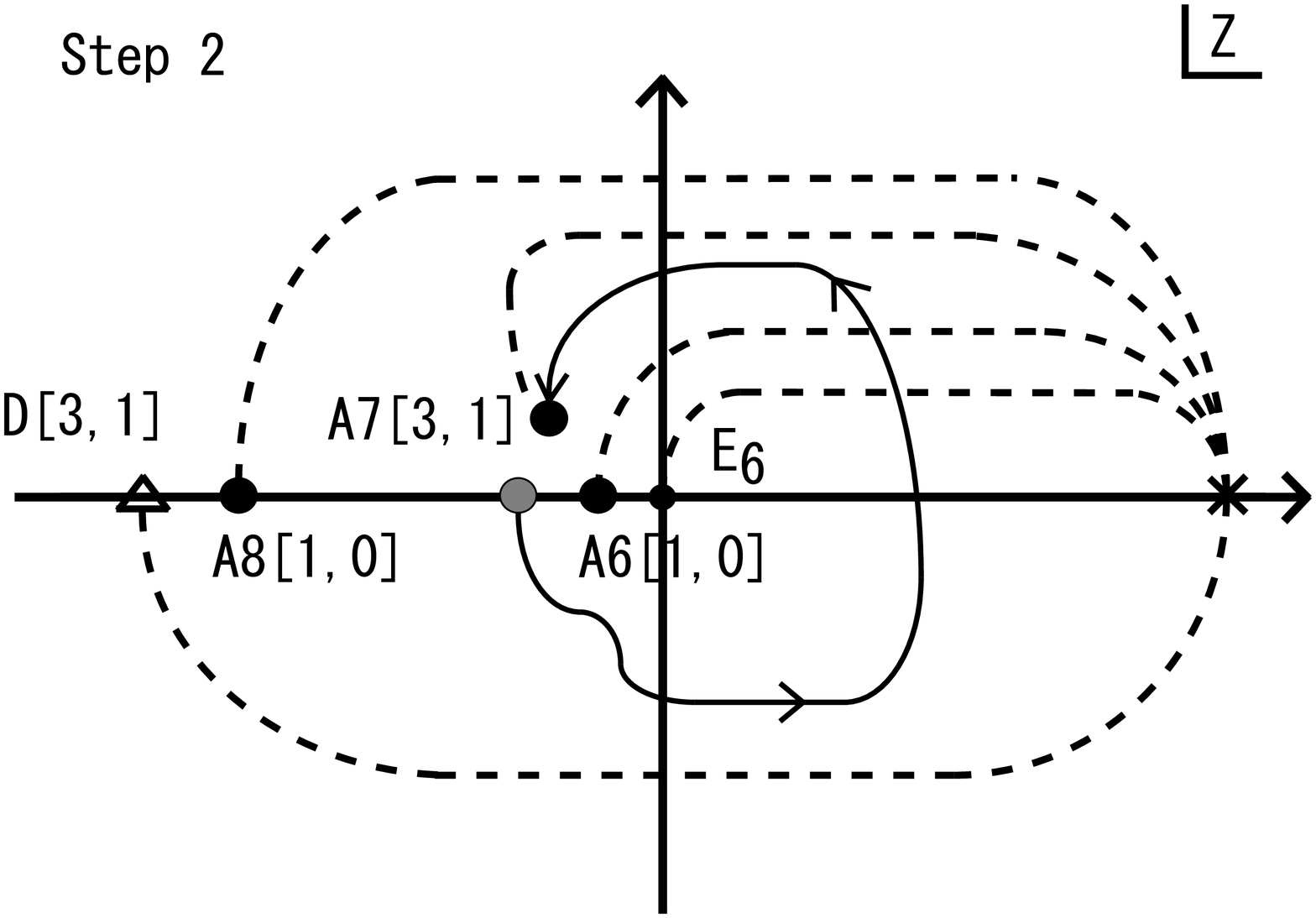}

\\ \\

  \begin{tabular}{|c|c|c|c|c|c|c|c|c|} \hline
           &  A8[1,0]
           &  A7[3,1]
           &  A6[1,0]
           &  A5[1,0]
           &  A4-A1[1,0]
           &  B[0,-1]
           &  C1, C2[2,1]
           &  D[3,1]
           \\\hline
 $C_{A65}$ &  0  &   0  &  1   & $-1$ & 0            & 0 &   0    & 0 \\\hline
 $C_{A76}$ &  0  &   1  & $-2$ & $-1$ & $-1$         & 3 &   1    & 0 \\\hline
 $C_{A87}$ &  1  & $-1$ &  1   & 1    & 1            &$-3$& $-1$  & 0 \\\hline
 $C_{BCD}$ &  0  & $-1$ &  1   & 1    & 1            &$-2$&  0    & $-1$ \\
 \hline
  \end{tabular}

\\ \\
$\downarrow$ step 3

\\ \\

A6 goes across the cut of E$_6$ from left to right.

\\ \\

\includegraphics[scale=0.25]{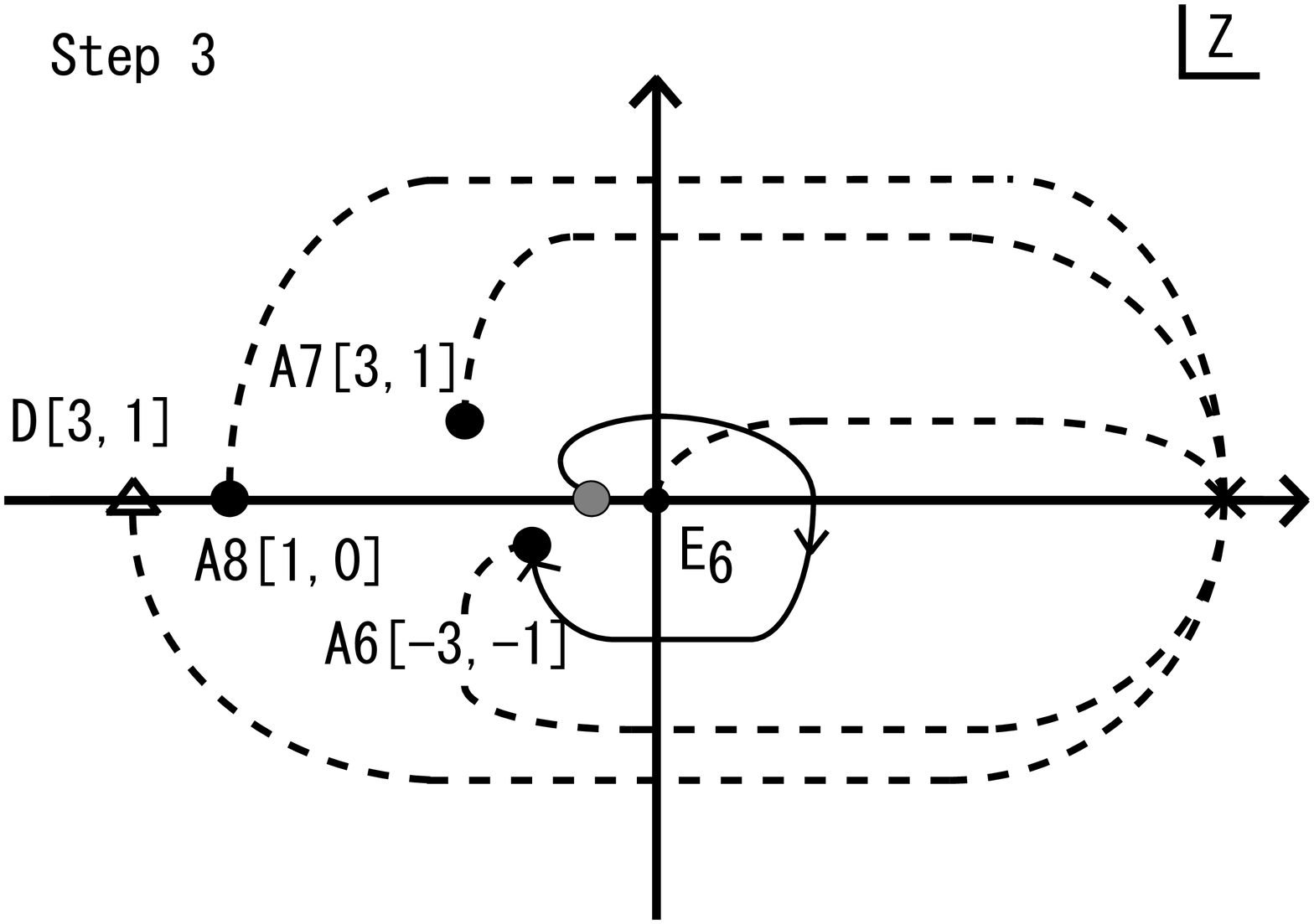}

\\ \\

  \begin{tabular}{|c|c|c|c|c|c|c|c|c|} \hline
           &  A8[1,0]
           &  A7[3,1]
           &  A5[1,0]
           &  A4-A1[1,0]
           &  B[0,-1]
           &  C1, C2[2,1]
           &  A6[-3,-1]
           &  D[3,1]
           \\\hline
 $C_{A65}$ &  0  &   0  & $-1$ & 0            & 1 &   1    & 1  & 0 \\\hline
 $C_{A76}$ &  0  &   1  & $-1$ & $-1$         & 1 & $-1$   &$-2$& 0 \\\hline
 $C_{A87}$ &  1  & $-1$ & 1    & 1            &$-2$&  0    & 1  & 0 \\\hline
 $C_{BCD}$ &  0  & $-1$ & 1    & 1            &$-1$&  1    & 1  & $-1$ \\
 \hline
  \end{tabular}

\\ \\
$\downarrow$ step 4

\\ \\

E$_6$ goes across the cut of A6 from left to right.

\\ \\

\includegraphics[scale=0.25]{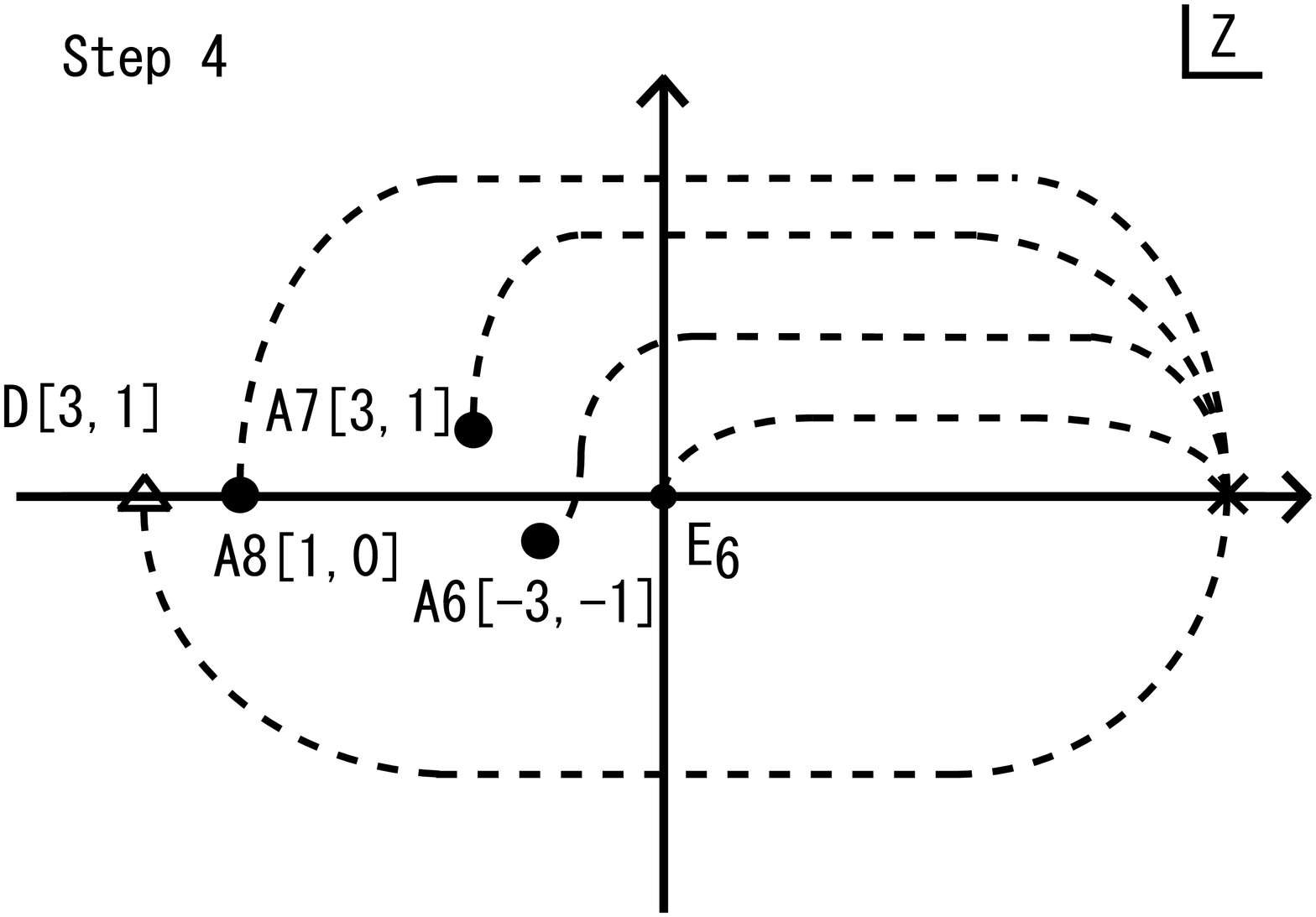}

\\ \\
  \begin{tabular}{|c|c|c|c|c|c|c|c|c|} \hline
           &  A8[1,0]
           &  A7[3,1]
           &  A6[-3,-1]
           &  A5[4,1]
           &  A4-A1[4,1]
           &  B[9,2]
           &  C1, C2[-1,0]
           &  D[3,1]
           \\\hline
 $C_{A65}$ &  0  &   0  &  1   & $-1$ & 0            & 1 &   1    & 0 \\\hline
 $C_{A76}$ &  0  &   1  & $-2$ & $-1$ & $-1$         & 1 & $-1$   & 0 \\\hline
 $C_{A87}$ &  1  & $-1$ &  0   &   1  & 1            &$-2$&  0    & 0 \\\hline
 $C_{BCD}$ &  0  & $-1$ &  1   &   1  & 1            &$-1$&  1    & $-1$ \\
 \hline
  \end{tabular}
\\ \\
 \caption{The motion of 7-branes and the changes of the 2-cycles along the path1.}
 \label{tab:2-cycles_before_ex}
\end{longtable}

After the step 4, A7's position and A6's position exchange with each other.
Note that the A7 charge and the A6 charge are the same up to sign.
Hence we do not need to rearrange the branch cuts in this exchange process.
We get the Table \ref{tab:after_ex} after the exchange.

\begin{longtable}{c}
 \includegraphics[scale=0.25]{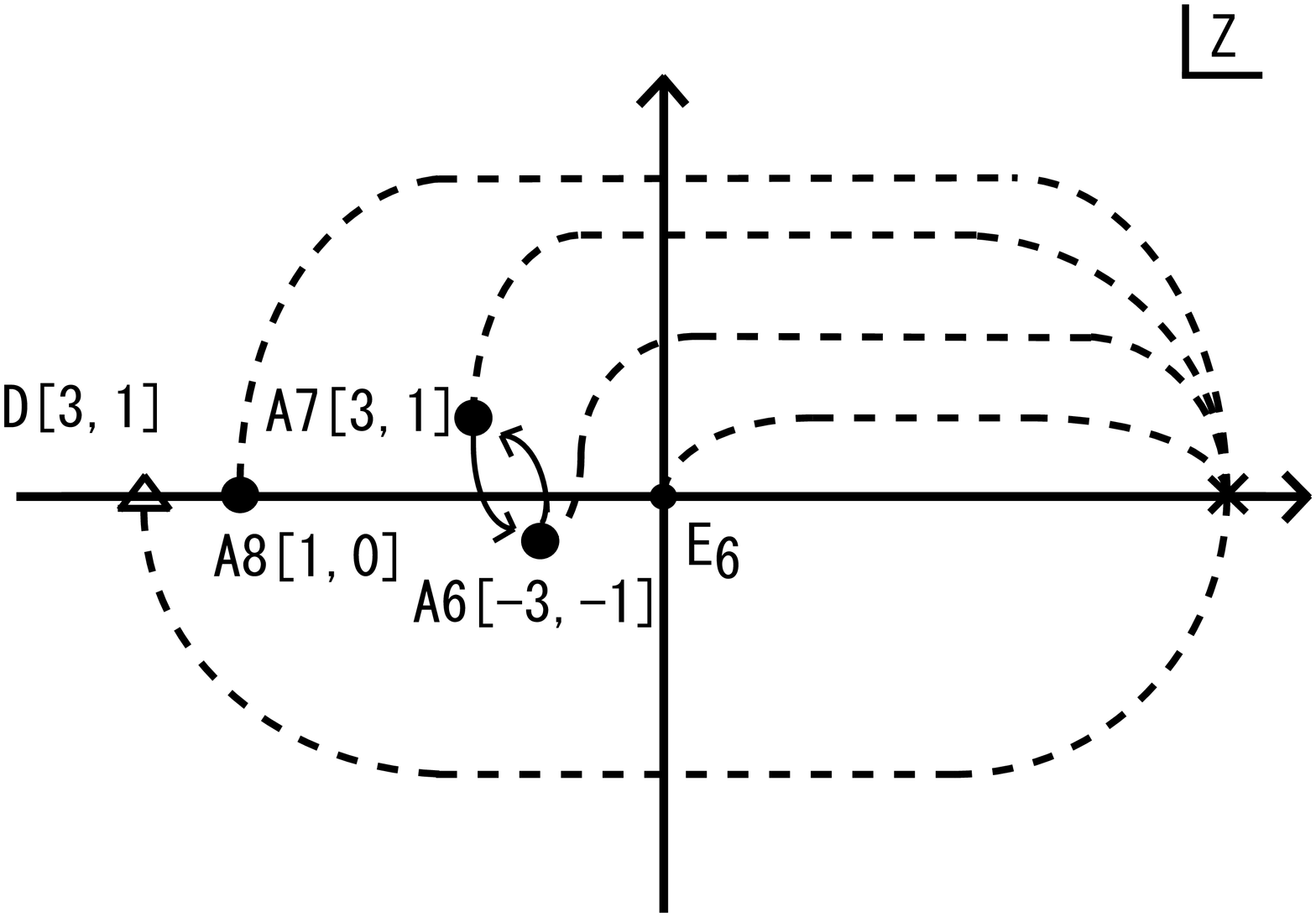}
 
 \\ \\
  \begin{tabular}{|c|c|c|c|c|c|c|c|c|} \hline
           &  A8[1,0]
           &  A6[-3,-1]
           &  A7[3,1]
           &  A5[4,1]
           &  A4-A1[4,1]
           &  B[9,2]
           &  C1, C2[-1,0]
           &  D[3,1]
           \\\hline
 $\tilde C_{A65}$ &  0  &   1  &  0   & $-1$ & 0            & 1 &   1    &
 0\\\hline
 $\tilde C_{A76}$ &  0  & $-2$ &  1   & $-1$ & $-1$         & 1 &
 $-1$   & 0 \\\hline
 $\tilde C_{A87}$ &  1  &   0  & $-1$ &   1  & 1            &$-2$&  0    & 0
 \\\hline
  $\tilde C_{BCD}$ &  0  &   1  & $-1$ &   1  & 1            &$-1$&  1    & $-1$
 \\
 \hline
  \end{tabular}
  
  \\ \\
   \caption{The figure shows the motion of 7-branes during the path 2 and the table represents the changes of 2-cycles after the path 2}
 \label{tab:after_ex}
\end{longtable}

We follow the above 4 steps backward, then we get the second part of the Table \ref{tab:back}.
The A6 charge and the A7 charge are $[-1,0]$.
Since their charge is $[1,0]$ originally, we make their charges into the the original charges $[1,0]$.
Then the second part of the Table \ref{tab:back} becomes last part of the Table \ref{tab:back}.

\begin{longtable}{c}
\includegraphics[scale=0.25]{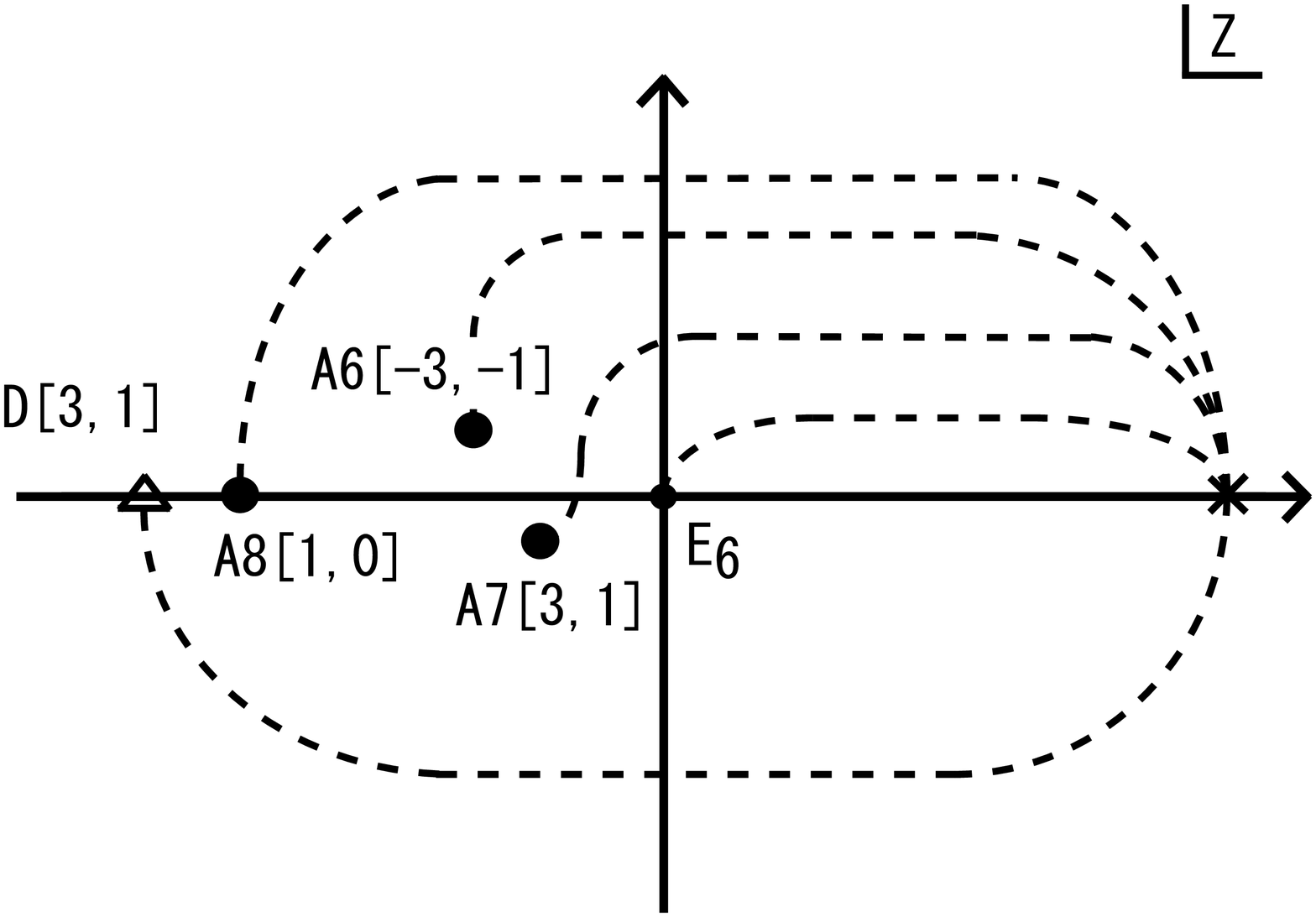}

\\ \\

 \begin{tabular}{|c|c|c|c|c|c|c|c|c|} \hline
           &  A8[1,0]
           &  A6[-3,-1]
           &  A7[3,1]
           &  A5[4,1]
           &  A4-A1[4,1]
           &  B[9,2]
           &  C1, C2[-1,0]
           &  D[3,1]
           \\\hline
 $\tilde C_{A65}$ &  0  &   1  &  0   & $-1$ & 0            & 1 &   1    &
 0\\\hline
 $\tilde C_{A76}$ &  0  & $-2$ &  1   & $-1$ & $-1$         & 1 &
 $-1$   & 0 \\\hline
 $\tilde C_{A87}$ &  1  &   0  & $-1$ &   1  & 1            &$-2$&  0    & 0
 \\\hline
  $\tilde C_{BCD}$ &  0  &   1  & $-1$ &   1  & 1            &$-1$&  1    & $-1$
 \\
 \hline
  \end{tabular}
  
  \\ \\
  
  $\downarrow$
  
  \\ \\
  
  Going back the four steps of the Table \ref{tab:2-cycles_before_ex}
  
  \\ \\
  
  \includegraphics[scale=0.25]{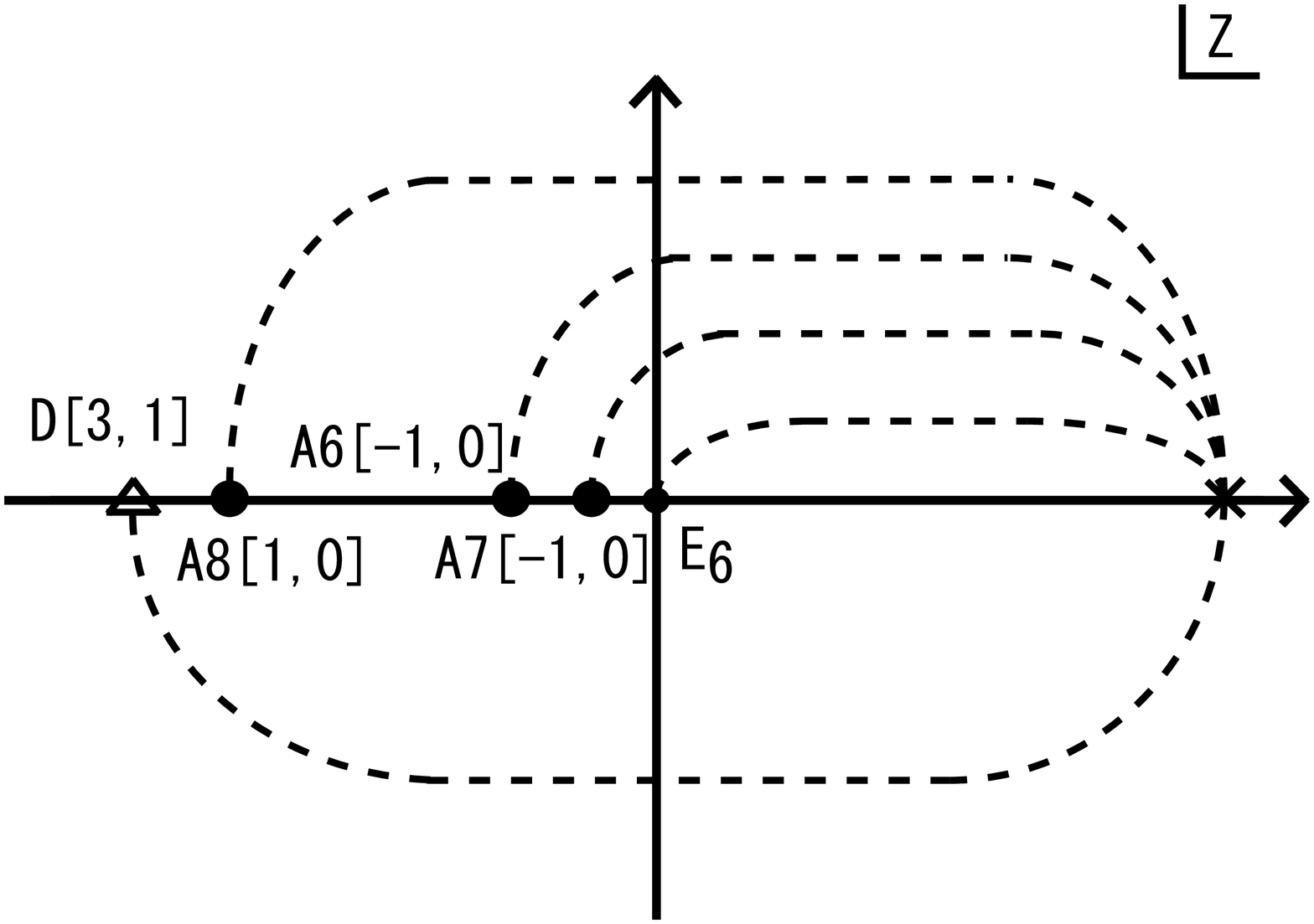}
  
  \\ \\
  \begin{tabular}{|c|c|c|c|c|c|c|c|c|} \hline
           &  A8[1,0]
           &  A6[-1,0]
           &  A7[-1,0]
           &  A5[1,0]
           &  A4-A1[1,0]
           &  B[1,-1]
           &  C1, C2[1,1]
           &  D[3,1]
           \\\hline
 $\tilde C_{A65}$ &  0  &   1  &  1   & $-2$ & $-1$         & 4 &   2    & 0
 \\\hline
 $\tilde C_{A76}$ &  0  & $-2$ & $-1$ &   1  & 1            &$-4$&
 $-2$  & 0 \\\hline
 $\tilde C_{A87}$ &  1  &   0  & $-2$ &   1  & 1            &$-4$& $-2$  & 0
 \\\hline
 $\tilde C_{BCD}$ &  0  &   0  &  0   &   0  & 0            & 1  &  1    & $-1$
 \\
 \hline
  \end{tabular}
  
  \\ \\
  $\downarrow$
  
  \\ \\
  
  Changing the bases into the ones before the rotation
  
  \\ \\
  
  \includegraphics[scale=0.25]{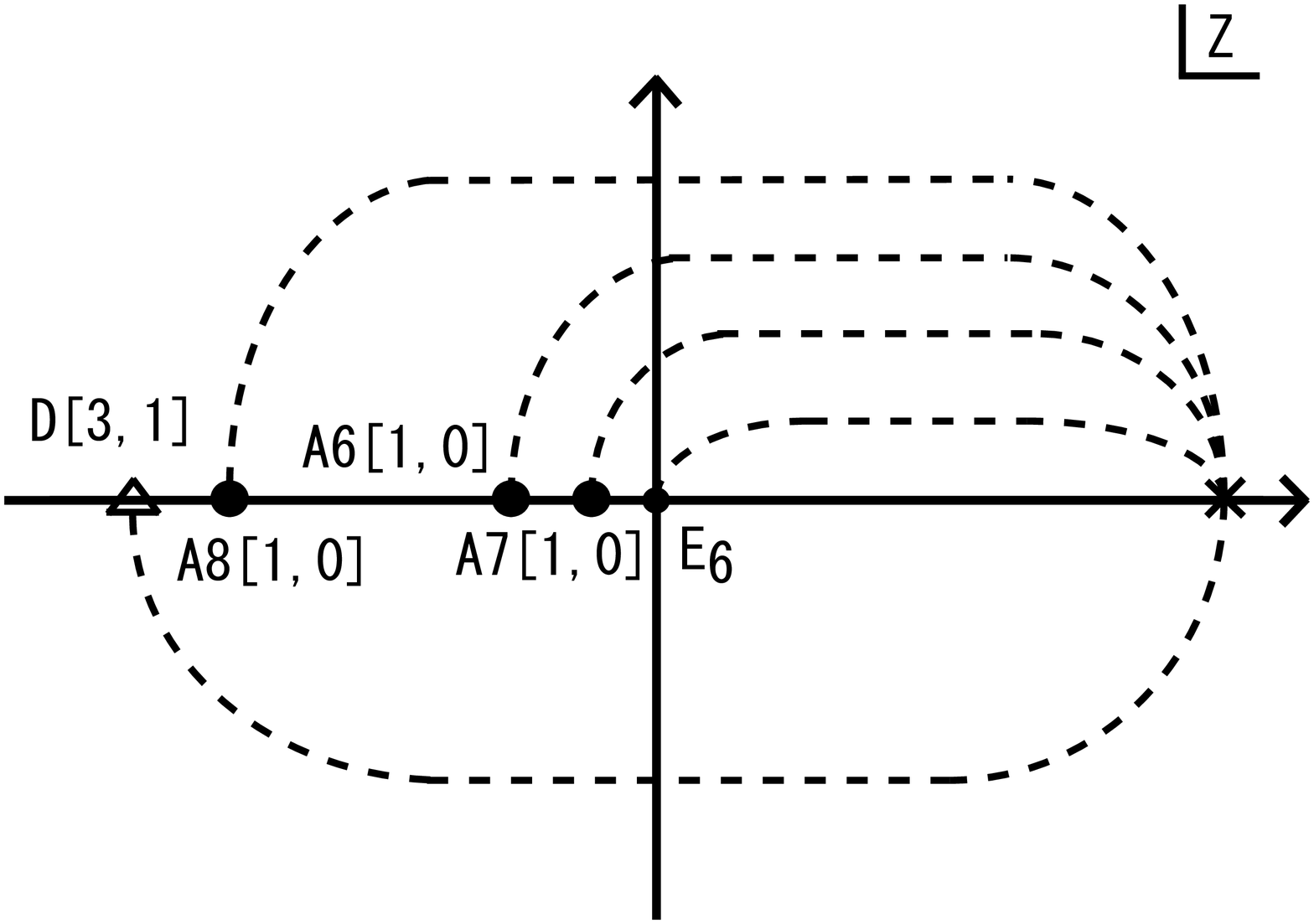}
  
  \\ \\

  \begin{tabular}{|c|c|c|c|c|c|c|c|c|} \hline
           &  A8[1,0]
           &  A6[1,0]
           &  A7[1,0]
           &  A5[1,0]
           &  A4-A1[1,0]
           &  B[1,-1]
           &  C1, C2[1,1]
           &  D[3,1]
           \\\hline
 $\tilde C_{A65}$ &  0  & $-1$ & $-1$ & $-2$ & $-1$         & 4 &   2    &
 0\\\hline
  $\tilde C_{A76}$ &  0  &   2  &  1   &   1  & 1            &$-4$& $-2$  & 0
  \\\hline
   $\tilde C_{A87}$ &  1  &   0  &  2   &   1  & 1            &$-4$& $-2$ & 0
   \\\hline
    $\tilde C_{BCD}$ &  0  &   0  &  0   &   0  & 0            & 1  &  1    & $-1$\\
 \hline
  \end{tabular}
  
  \\ \\
\label{tab:back}
 \\ \\
 \caption{The configurations of 7-branes and the changes of 2-cycles when we go 4 steps in the Table \ref{tab:2-cycles_before_ex} in a backward direction. The last part shows the change of the bases into the ones at the base point.}
\end{longtable}

From the first table of Table \ref{tab:2-cycles_before_ex} and
the last table of the Table \ref{tab:back}, we get
\begin{align}
\tilde C_{A65} &= C_{-\theta} + C_{A65} + C_{A76} \label{eq:2-2_1}\\
\tilde C_{A76} &= - C_{-\theta} \label{eq:2-2_2}\\
\tilde C_{-\theta} &= - C_{A76} \label{eq:2-2_3}\\
\tilde C_{A87} &= - C_{-\theta} - C_{A76} + C_{A87}\label{eq:2-2_4} \, .
\end{align}
where the equalities of the equations (\ref{eq:2-2_1}) $\sim$ (\ref{eq:2-2_4}) should be understood up to the boundaries of the 3-dimensional cells. Note that $C_{\alpha}^1$, $C_{\alpha}^2$, $C_{\beta}^1$ and $C_{\beta}^2$
, which are the outside of $E_8$ are invariant.
On the other hand, this monodromy acts on the 2-cycles inside $E_8$
as the Weyl reflection by $C_{A76} + C_{-\theta}$.

\subsubsection{The monodromy of the loop $\gamma_{0-4}$}

\begin{figure}[!h]
\begin{tabular}{cc}

\begin{minipage}{0.5\hsize}
\begin{center}
\includegraphics[scale=0.25]{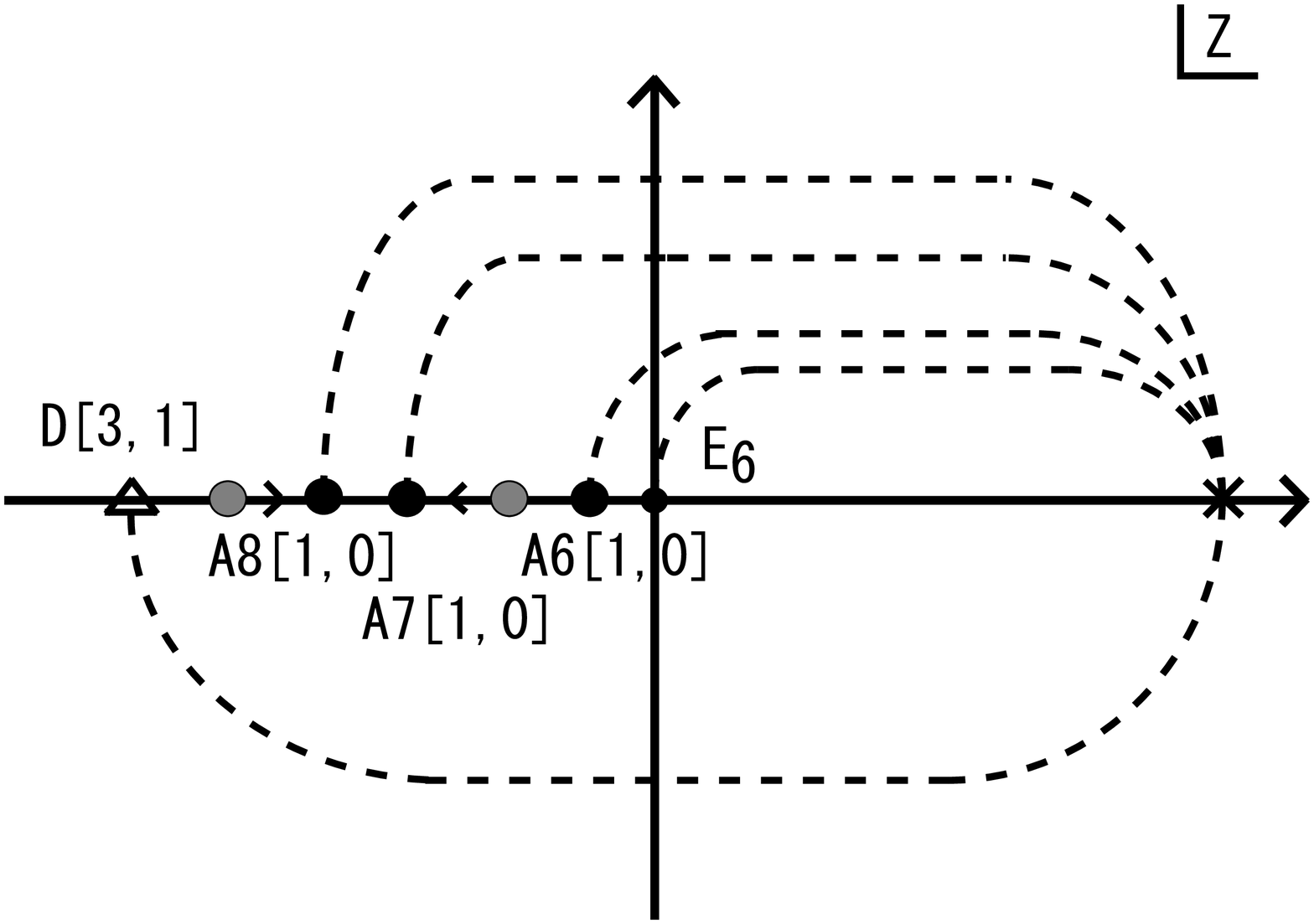}
\end{center}
\end{minipage}

\begin{minipage}{0.5\hsize}
\begin{center}
\includegraphics[scale=0.25]{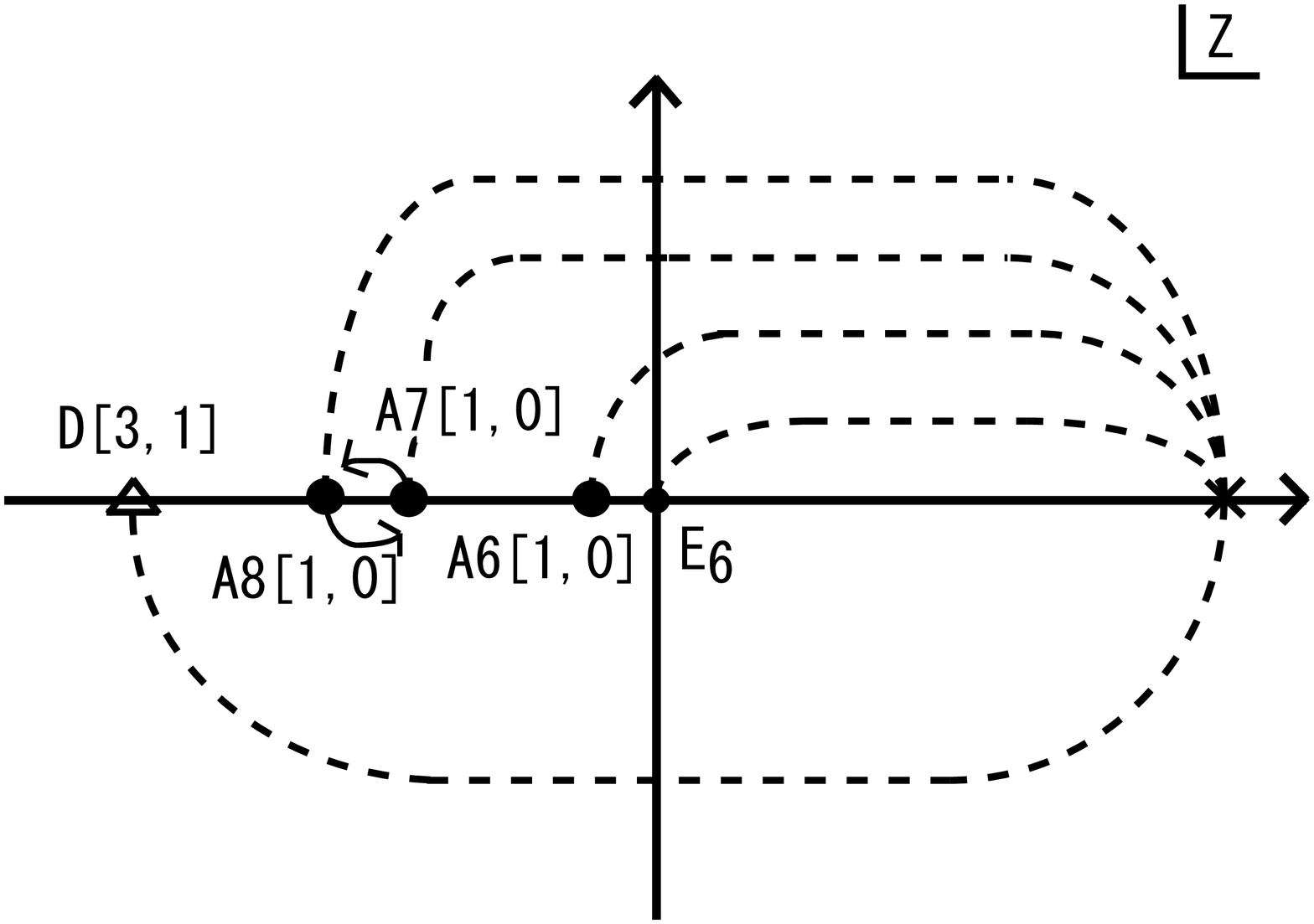}
\end{center}
\end{minipage}

\end{tabular}
\caption{The left figure shows the motion of 7-branes when we approach the right of $a_{0}=a_{0-4}$ singularity. The right figure shows the motion of 7-brane when we go round $a_{0}=a_{0-4}$ singularity in an anti-clockwise direction. }
\label{fig:0-4}
\end{figure}

When $a_0$ varies along the loop $\gamma_{0-4}$, 
the exchange of A7 and A8 only happens
(see Figure \ref{fig:0-4}).

Hence we get
\begin{align}
\tilde C_{A76} &= C_{A76} + C_{A87} \\
\tilde C_{A87} &= - C_{A87} \\
\tilde C_{AA'} &= - C_{A87} + C_{AA'}.
\end{align}

Note that $C_{A87}$ mixes into the 2-cycles inside $E_8$.
This means that the monodromy of the loop $\gamma_{0-4}$ causes
an effect beyond $E_8$ gauge theory.

$C_{\alpha}^2$ and $C_{\beta}^2$ do not mixes with $C_{\alpha}^1$, $C_{\beta}^1$
and the 2-cycles in $E_8$.
We can understand this structure due to the $\epsilon_{\eta}$ scaling.

\section{Monodromy around $a_0 = 0$ Locus in Light of Heterotic Dual}
\label{sec:Het}

It is instructive to see what the monodromy of 2-cycles 
(\ref{eq:0-4}--\ref{eq:0-6}) correspond to in Heterotic 
dual description. These monodromy matrices were obtained by 
a brute force calculation. Heterotic dual description tells 
us that this is a very natural result. 

The duality map between F-theory and Heterotic string theory 
\cite{Vafa-F} is now well-understood. The moduli space of 
F-theory compactification on an elliptic K3 and that of 
Heterotic string compactification on $T^2$ are both 
$O(2) \times O(18) \backslash O(2,18; \R) / O(2,18; \Z)$. 
For the purpose of describing the action of the monodromy group 
$O(2,18; \Z)$ explicitly, however, let us begin by reminding 
ourselves of the basic things about the duality map, while 
setting up notation. 

The complex structure moduli of elliptic fibered K3 manifold 
for F-theory compactification is described (e.g., \cite{Aspinwall}) 
by 20 complex numbers $\Pi_i$ ($i = 1, \cdots, 20$) satisfying 
\begin{equation}
 \Pi_i C^{ij} \Pi_j = 0, \qquad \Pi_i C^{ij} \Pi_j^* = 2.
\end{equation}
Multiplying an arbitrary complex phase on all the $\Pi_i$'s 
corresponds to the $O(2)$ redundancy of the moduli space. 
The moduli parameters $\Pi_i$ are obtained from a defining 
equation of an elliptic fibered K3 manifold as period integrals 
\begin{equation}
 \Pi_i = \int_{C_i} \Omega^{(2,0)}
\end{equation}
of a holomorphic (2,0) form satisfying 
$\int_{K3} \Omega \wedge \overline{\Omega} = 2$. 
$C^{ij}$ is the inverse matrix of the intersection form of 
2-cycles of a K3 other than the zero section and the generic 
fiber class. We choose the first four 2-cycles to be 
$C_\alpha^1$, $C_\alpha^2$, $C_\beta^1$ and $C_\beta^2$ in this 
order, and use those in the first and last rows of Table~\ref{tab:2-cycles} 
as the 16 other 2-cycles. 
For an elliptic fibered K3-manifold given by (\ref{eq:model}), 
for example, 
\begin{eqnarray}
 \Pi_{C_\alpha^1} \sim \Pi_{C_\beta^1} & \sim & 
       \frac{1}{\sqrt{\ln (1/\epsilon_\eta)}}, \\
 \Pi_{C_\alpha^2} \sim \Pi_{C_\beta^2} & \sim & 
       \sqrt{\ln (1/\epsilon_\eta)}, \\
 \Pi_{C_{A76}} \sim \Pi_{C_{-\theta}} \sim \Pi_{C'_{-\theta}} & \sim &
       \frac{1}{\sqrt{\ln (1/\epsilon_\eta)}} \epsilon_K.
\end{eqnarray}

The Narain moduli of Heterotic string compactifications on a $T^2$
are described by 20 complex parameters
\begin{equation}
 \Pi_i = \sqrt{\frac{\alpha'}{2}} 
   \left(k_{\hat{8}R} + i k_{\hat{9}R} \right)_i =  
  \sqrt{\frac{\alpha'}{2}} 
   \frac{1}{\tau_2}(-i \tau k_{8R} + i k_{9R})_i
\end{equation} 
for $i=1,\cdots 20$; $(k_{mR})_i$ ($m=8,9$) above determine the 
momenta $k_{8R}$ and $k_{9R}$ in the right-moving sector by 
$k_{mR} = (k_{mR})_i n^i$, where $(n_i)^T$ is a charge vector 
\begin{equation}
 (n^i)^T = (n_8, - w^8, n_9, -w^9, n^{i=5,\cdots,20})^T; 
\label{eq:charge-vct}
\end{equation} 
$(n_8, n_9)$ specify the Kaluza--Klein excitation level, and 
$(w^8, w^9)$ are the winding numbers. $n^i \in \Z$'s ($i=5,\cdots, 20$)
generate root lattice\footnote{One can use a basis so that the 
last $16 \times 16$ part of the inverse of the intersection form is 
just a unit matrix. The last 16 components of the charge vector in this 
basis, $n^I$ ($I = 5, \cdots, 20$), and $n^i$ ($i=5, \cdots, 20$) that we use 
in the main text, are related by $n^I = q^I_{\; i} n^i$; integer valued 
$q^I_{\; i}$'s specify the roots of $E_8$ corresponding to $C_i$.} 
of $E_8 \times E_8$.
$k_{8R}$ and $k_{9R}$ are parametrized by 16 Wilson lines 
$(-i\tau A_8 + i A_9)^i \in \C$ ($i=5,\cdots,20$), complex 
structure $\tau = \tau_1 + i \tau_2$ of the torus, $R$ and $B_{89}$; 
explicit expressions are found in (11.6.17) 
of \cite{Polchinski}. 

With all the conventions fixed as above, the duality map is 
to identify $\Pi_i$'s of the two descriptions. One can see 
by comparing $\Pi_i$'s calculated on both sides that 
\begin{equation}
 \frac{R}{\sqrt{\alpha'}} \quad ({\rm Het}) \leftrightarrow 
 \sqrt{\ln (1/\epsilon_\eta)} \quad ({\rm F-theory}),
\end{equation}
as is well known \cite{MV2, FMW}, and 
\begin{equation}
 \frac{1}{\tau_2} R (-i \tau A_8 + i A_9)^i \quad ({\rm Het}) 
  \leftrightarrow  {\rm fcn}\left(\frac{a_r}{a_0}\right) \sim \epsilon_K 
   \quad ({\rm F-theory}),
\end{equation}
where we assumed that $a_0$ may scale as $\epsilon_\eta$, but 
$a_{0,*}$ in (\ref{eq:scaling-rg}, \ref{eq:gauge-th-rg}) 
remains ${\cal O}(1)$ and non-zero. 
Thus the choice of complex structure moduli with a small $\epsilon_K$ 
in (\ref{eq:scaling-rg}) and $a_{0,*} \sim {\cal O}(1)$ in F-theory 
corresponds to Wilson lines in Heterotic string theory in $T^2$ that 
are parametrically smaller than the Kaluza--Klein scale of the $T^2$ 
compactification. Thus, the $T^2$ Kaluza--Klein states decouple 
and a gauge theory on {\it 7+1} dimensions is obtained also 
in the Heterotic description. The ``8D gauge theory region'' of 
F-theory moduli space corresponds to 8D gauge theory on the Heterotic side.

Both F-theory compactified on an elliptic K3 manifold and Heterotic 
string theory compactified on $T^2$ have 20 U(1) vector fields on 
7+1 dimensions. The charge vector $(n^i)^T$ that appeared 
in (\ref{eq:charge-vct}) specifies\footnote{In F-theory (M-theory is more 
appropriate, though), such an object corresponds to an M2-brane 
wrapped on a 2-cycle $C_i n^i$.} charges of an object under these 20 U(1) 
vector fields. In the convention we adopted above, the last 16 U(1) 
vector fields are those in $E_8 \times E_8$. In Heterotic string 
compactifications, the vector fields corresponding to the charges 
$n_{8,9}$ are Kaluza--Klein vector fields from the metric on 9+1 
dimensions, and those to $w^{8,9}$ are those from the $B$-field.
The monodromy group $O(2, 18; \Z)$ acts from the right 
on all of 2-cycles $(C_i)$, moduli parameters $(\Pi_i)$, and on 
20 independent U(1) vector fields.

Having prepared all above, we are now ready to interpret the 
monodromy matrix (\ref{eq:0-4}) in Heterotic dual description. 
Since we considered monodromy in F-theory that appears even in 
$|\epsilon_\eta| \ll 1$, it must be a monodromy in Heterotic 
string theory that appear even in $R/\sqrt{\alpha'}$, that is, 
in a region of moduli space where supergravity + super Yang--Mill 
approximation can be used. In particular, the compactification 
can be described in terms of Calabi--Yau 3-fold and a stable 
vector bundle on it. The bundle is described by a spectral surface. 
The spectral surface in the Heterotic dual \cite{FMW, Donagi-Het-spec-surf} 
is given by \cite{KMV-BM, CD, DW-1, Hayashi-1, Hayashi-2}
\begin{equation}
 a_0 + a_2 x + a_3 y + \cdots = 0,
\label{eq:Het-spec-surf}
\end{equation}
using the same coefficients $a_{0,2,3, \cdots}$ as in F-theory 
compactifications.\footnote{The overall normalization of
$a_{0,2,3,\cdots}$ and $a''_{0,2,3,\cdots}$ carries 
information---$\epsilon_\eta$---in F-theory compactifications, but 
this information are ignored in the defining equation of the 
spectral surface (\ref{eq:Het-spec-surf}) in Heterotic string, as 
this information is now carried by the volume of $T^2$ fiber 
$R^2/\alpha'$.
}
$(x,y)$ are the coordinates of the elliptic fiber for the Heterotic
string compactification. When $a_0 \neq 0$ and 
$a_r/a_0 \sim \epsilon_K^r$ for $r = 2,3,\cdots$, the spectral surface 
is near the zero section. In a neighbourhood of a base manifold where 
$a_0$ becomes zero, however, two roots of (\ref{eq:Het-spec-surf}) 
move away from the zero section, and behave as specified 
by (\ref{eq:shoot-spec-surf}), which indicates that a Weyl reflection 
takes place within $E_8$. 

Noting that the monodromy matrix (\ref{eq:0-4-mtrx}) 
can be decomposed into 
\begin{equation}
 \left(\begin{array}{cc|cc|c}
  1 & & & & \\  & -1 & & -1 & \\ \hline 
    & 2 & 1 & 1 & \\ & & & 1 & \\ \hline 
    & & & &
   {\bf 1}_{2 \times 2}   
       \end{array}\right) = 
 \left(\begin{array}{cc|cc|c}
  1 & & & & \\  & -1 & & & \\ \hline 
    & & 1 & & \\ & & & 1 & \\ \hline 
    & & & &
   {\bf 1}_{2 \times 2}   
       \end{array}\right)  
 \left(\begin{array}{cc|cc|c}
  1 & & & & \\ & 1 & & 1 & \\ \hline 
    & 2 & 1 & 1 & \\ & & & 1 & \\ \hline 
    & & & &
   {\bf 1}_{2 \times 2}   
       \end{array}\right);  
\end{equation}
here, we used a basis $(2C_{A76}+C_{-\theta}, C_{-\theta}, C_\alpha^1, 
C_\alpha^2, C_\beta^1, C_\beta^2)$, changing the first element of the basis 
from $C_{A76}$, so that the first two elements of this basis are dual 
to the Cartan elements $\diag(2, -1, -1)$ and $\diag(0, 1, -1)$ acting 
on $(C_{A65}, C_{A76}+C_{A65}, C_{-\theta}+C_{A76}+C_{A65})$.  
The first matrix on the right hand side is $W_{C_{-\theta}}$, a Weyl 
reflection in $E_8$.

The second matrix above is not hard to understand, either. 
It is satisfactory that non-winding states ($w^{8,9} = 0$) remain 
non-winding states; since we have chosen $|\epsilon_\eta| \ll 1$ and
hence $R/\sqrt{\alpha'} \gg 1$, supergravity / super Yang--Mills modes 
on 9+1 dimensions should not mix with winding states. 
When we take $|\epsilon_K| \ll 1$, the spectral surface 
scan the elliptic fiber once near a zero locus\footnote{This is very 
natural because the $n$-fold spectral cover given by
(\ref{eq:Het-spec-surf}) belongs to a topological class 
$|n K_S + \eta|$, and $a_0 \in \Gamma(S; {\cal O}(\eta))$.} of $a_0$.
This means that the Wilson line associated with the Weyl reflection 
varies by of order the Kaluza--Klein scale (i.e., 
$R (A_{\hat{8}}+iA_{\hat{9}} )^I$ changes by ${\cal O}(1)$). This 
Wilson line is identified with the original one modulo gauge 
transformation whose parameter depends on the $T^2$ direction. 
Thus, the Kaluza--Klein excitation number $n_8$ (the 3rd row) 
becomes $\tilde{n_8} = n_8 + 2 n^{-\theta}$ in this monodromy matrix. 
Once this coefficient $2$ is fixed, then there is no freedom left 
in the 4th column, because the intersection form needs to remain 
invariant. 

This monodromy involves field identification modulo gauge
transformation depending on the $T^2$ direction; there is no way 
describing this phenomenon in a gauge theory on {\it 7+1} dimensions!
This intuitive understanding of the monodromy matrix also explains 
why some U(1) symmetry charges within $E_8$ mix up with U(1) symmetries 
of Kaluza--Klein vector fields. 

One would not bother about all these things when one considers, say, 
a Heterotic string compactification on a Calabi--Yau 3-fold with a
vector bundle whose structure group is $\SU(5)_{\rm str}$. 
All the four U(1)'s in the Cartan of $\SU(5)_{\rm str}$ are broken 
completely, because of the ramification behavior of the spectral
surface. All the four Kaluza--Klein vector fields from metric and 
$B$-field in 7+1 dimensions do not remain in the massless spectrum 
in 3+1 dimensions. 
Since none of those U(1) vector fields and U(1) symmetries 
available at microscopic level is relevant to low-energy physics, 
nobody cares how they are mixed up. 

In the factorized spectral surface scenario in the context of 
dimension-4 proton decay problem, however, one wants to keep a 
U(1) symmetry out of a four independent U(1)'s in ${\rm SU}(5)_{\rm str}$. 
For this purpose, one has to make sure that at least one U(1) 
symmetry survives the monodromy that would potentially mix up 
all of U(1)'s in $E_8$ as well as those associated with Kaluza--Klein 
vector fields. 

%

\end{document}